\documentclass[a4paper,fleqn,usenatbib]{mnras}

\usepackage{newtxtext,newtxmath}
\usepackage[T1]{fontenc}
\usepackage{ae,aecompl}
\usepackage{graphicx}	% Including figure files
\usepackage{amsmath}	% Advanced maths commands
\usepackage{amssymb}	% Extra maths symbols

\usepackage{longtable}
\usepackage{xspace}

\title[SNe Ia spectropolarimetry]{Linear spectropolarimetry of 35 Type Ia Supernovae with VLT/FORS: An analysis of the Si II line polarization}

\author[A. Cikota et al.]{Aleksandar Cikota$^{1,2}$\thanks{E-mail: acikota@lbl.gov},
Ferdinando Patat$^{2}$,
Lifan Wang$^{3}$,
J. Craig Wheeler$^{4}$,
\newauthor Mattia Bulla$^{5}$,
 Dietrich Baade$^{2}$,
Peter H\"{o}flich$^{6}$,
Stefan Cikota$^{7}$,
\newauthor Alejandro Clocchiatti$^{8}$, 
Justyn R. Maund$^{9}$,
 Heloise F. Stevance$^{9}$,
Yi Yang$^{10}$
\\
$^{1}$Physics Division, Lawrence Berkeley National Laboratory, 1 Cyclotron Road, Berkeley, CA 94720, USA\\
$^{2}$European Southern Observatory, Karl-Schwarzschild-Str. 2, 85748 Garching b. M\"{u}nchen, Germany\\
$^{3}$Department of Physics, Texas A\&M University, College Station, TX 77843, USA\\
$^{4}$Department of Astronomy, University of Texas at Austin, Austin, TX 78712-1205, USA\\
$^{5}$Oskar Klein Centre, Department of Physics, Stockholm University, SE 106 91 Stockholm, Sweden\\
$^{6}$Department of Physics, Florida State University, Tallahassee, FL 32306$-$4350, USA\\
$^{7}$University of Zagreb, Faculty of Electrical Engineering and Computing, Department of Applied Physics, Unska 3, 10000 Zagreb, Croatia\\
$^{8}$Institute of Astrophysics, Universidad Cat\'{o}lica de Chile, and Millennium Institute of Astrophysics, Santiago, Chile\\
$^{9}$Department of Physics and Astronomy, University of Sheffield, Hicks Building, Hounsfield Road, Sheffield S3 7RH, U.K.\\
$^{10}$Department of Particle Physics and Astrophysics, Weizmann Institute of Science, Rehovot 76100, Israel
}

\date{Accepted XXX. Received YYY; in original form ZZZ}

\pubyear{2019}

\begin{document}
\label{firstpage}
\pagerange{\pageref{firstpage}--\pageref{lastpage}}
\maketitle

\begin{abstract}
Spectropolarimetry enables us to measure the geometry and chemical structure of the ejecta in supernova explosions, which is fundamental for the understanding of their explosion mechanism(s) and progenitor systems. We collected archival data of 35 Type Ia Supernovae (SNe\,Ia), observed with FORS on the Very Large Telescope at 127 epochs in total. We examined the polarization of the \ion{Si}{ii} $\lambda$6355\,\AA\ line (p$_{\rm \ion{Si}{ii}}$) as a function of time which is seen to peak at a range of various polarization degrees and epochs relative to maximum brightness. We reproduced the $\Delta$m$_{15}$--p$_{\rm \ion{Si}{ii}}$ relationship identified in a previous study, and show that subluminous and transitional objects display polarization values below the $\Delta$m$_{15}$--p$_{\rm \ion{Si}{ii}}$ relationship for normal SNe\,Ia. We found a statistically significant linear relationship between the polarization of the \ion{Si}{ii} $\lambda$6355\,\AA\ line before maximum brightness and the \ion{Si}{ii} line velocity and suggest that this, along with the $\Delta$m$_{15}$--p$_{\rm \ion{Si}{ii}}$ relationship, may be explained in the context of a delayed-detonation model. 
In contrast, we compared our observations to numerical predictions in the $\Delta$m$_{15}$--v$_{\rm \ion{Si}{ii}}$ plane and found a dichotomy in the polarization properties between Chandrasekhar and sub-Chandrasekhar mass explosions, which supports the possibility of two distinct explosion mechanisms.
A subsample of SNe display evolution of loops in the $q$--$u$ plane that suggests a more complex Si structure with depth. This insight, which could not be gleaned from total flux spectra, presents a new constraint on explosion models. Finally, we compared our statistical sample of the \ion{Si}{ii} polarization to quantitative predictions of the polarization levels for the double-detonation, delayed-detonation, and violent-merger models.
\end{abstract}


%%%%%%%%%%%%%%%%%%%%%%%%%%%%%%%%%%%%%%%%%%%%%%%%%%

%%%%%%%%%%%%%%%%% BODY OF PAPER %%%%%%%%%%%%%%%%%%

%\clearpage

\section{Introduction}
\label{sect_intro}

Type Ia Supernovae (SNe Ia) are used as standard candles (after applying proper scaling relations, see e.g., \citealt{1977AZh....54.1188P,1993ApJ...413L.105P}) to measure the expansion rate of the Universe \citep{1998AJ....116.1009R,1999ApJ...517..565P}. The ultimate future goal is to constrain the nature of dark energy, which requires accurate measurements of the equation-of-state parameter, \textit{w}.

Several upcoming sky surveys that are expected to discover SNe at a very high cadence will lead to an improvement in the statistics and advances in SN Ia cosmology. The Large Synoptic Survey Telescope (LSST) alone is expected to observe 50\,000 SNe Ia per year with an adequate time coverage for cosmological distance estimates \citep{2009arXiv0912.0201L}. Furthermore, future space missions will enable us to extend the SN\,Ia Hubble diagram to redshifts above z$\sim$2, and up to z$\sim$4 using the James Webb Space Telescope \citep{2013RSPTA.37120282H,2015arXiv150303757S,2017arXiv171007005W}. Even with a large statistical sample, it is demanding to accurately measure the equation-of-state parameter \textit{w}, due to systematic errors.
%, such as dust extinction, and evolutionary effects on the progenitor system. 
It is known that SNe Ia are exploding carbon/oxygen (C/O) white dwarfs close to the Chandrasekhar mass limit, however, their evolutionary path and the exact progenitor system are still not known. Identifying the SN\,Ia progenitors is important, because we need to understand the evolution of their luminosity with cosmic time, depending on e.g., metallicity, age, dust, etc. \citep{2006ApJ...648..884R}.

The three most popular SN Ia progenitor systems are: the single-degenerate, in which a white dwarf accretes material from a non-degenerate companion until it reaches the Chandrasekhar mass and explodes; the double-degenerate model in which two white dwarfs merge smoothly or violently; and the core-degenerate progenitor model \citep{2011MNRAS.417.1466K} in which a white dwarf merges with the core of a giant companion star (see \citealt{
2018PhR...736....1L} for a SNe Ia review). 

One approach to learn more about the progenitor system, is to investigate the environment of the SNe, because different progenitor models can imply different environments. Spectropolarimetry offers an independent method to the study inter/circum-stellar
dust properties (by observing the continuum polarization) and the analysis of the three-dimensional geometrical properties of unresolved sources (by observing the intrinsic continuum polarization and polarization across spectral lines). The latter provides insights on global and local asymmetries of SN explosions \citep[see e.g.][]{1982ApJ...263..902S,1991A&A...246..481H,1993PASAu..10..263S,1993ApJ...414L..21T,1996ApJ...467..435W}. These aspects are fundamental for our understanding of the phenomenon and are hardly approachable by any other observational technique \citep{2008ARA&A..46..433W,2017suex.book.....B}. \\

In this work we primarily focus on the linear polarization in the absorption lines, particularly the $\lambda$6355\,\AA\ silicon line, because it is the most prominent polarized line, along with the \ion{Ca}{ii}  triplet. However, this latter feature occurs in the low signal-to-noise ratio wavelength range of our observations. We use a statistical sample of 35 SNe Ia observed with FORS1 (before 2005) and FORS2, mounted at the Cassegrain focus of ESO's Very Large Telescope, to explore the polarization across the spectral lines, look for relationships between the \ion{Si}{ii} line velocity and polarization degree; and compare our polarization measurements with predictions from simulations with the main aim to constrain the SN Ia explosion mechanism and progenitor system. 
Additionally, because our sample includes a number of objects affected by low or negligible reddening, we derive an upper limit on the intrinsic continuum polarization.

To understand the polarization spectra observed toward SN\,Ia sightlines, it is important to understand the mechanisms of polarization, which are presented in Sect.~\ref{sect_polmechanisms}. In Sect.~\ref{sect_observations} we explain the instruments, observations, and our sample of SNe Ia, in Sect.~\ref{sect_methods} the methods used to reduce the data, in Sect.~\ref{sect_discussion} we show and discuss the results, and finally summarize the main results and conclude in Sect.~\ref{sect_summary}.

\section{Polarization mechanisms}
\label{sect_polmechanisms}

\subsection{Continuum polarization}
\label{sect_continuumpolmechanisms}

There are three continuum polarization mechanisms, relevant for the SN Ia observations, that we need to consider:

i) Polarization produced due to linear dichroism in non-spherical grains -- When light passes through the interstellar medium, or a cloud of non-spherical supramagnetic dust grains, which are aligned with the galactic magnetic field, the electric vector of the light wave parallel to the major axis of the dust grains will experience higher extinction than light waves parallel to the minor axis of the dust grain, and thus we will observe a net polarization \citep{1957lssp.book.....V,1974ApJ...187..461M,1975ApJ...201..151S}.

The Serkowski curve \citep{1975ApJ...196..261S} is an empirical wavelength dependence derived using a Milky Way sample in the optical \citep[see also][]{1980ApJ...235..905W,1982AJ.....87..695W,1992ApJ...386..562W}, which is used to characterize interstellar linear polarization curves:
\begin{equation}
\label{eq_serkowski}
\frac{p(\lambda)}{p_{\mathrm{\rm max}}} = \exp \left[-K \mathrm{ln}^2 \left( \frac{\lambda_{\mathrm{\rm max}}}{\lambda} \right) \right] ,
\end{equation}
where the wavelength of peak polarization, $\lambda_{\rm max}$, in general depends on the dust grain size distribution. For an enhanced abundance of small dust grains $\lambda_{\rm max}$ moves to shorter wavelengths and for an enhanced abundance of large dust grains to longer wavelengths. $p_{\rm max}$ is the peak degree of polarization, and $K$ depends on the width of the curve. Narrow curves will have large $K$ values, compared to wide curves that have small $K$ values.
\citet{1975ApJ...196..261S} also found that $R_V \approx 5.5 \lambda_{\rm max}$, where $\lambda_{\rm max}$ is in $\mu m$, and $R_V$ is the total-to-selective extinction ratio, $R_V$ = $A(V)/E(B-V)$.

ii) Polarization by scattering from nearby material -- Single scattering from nearby dust clouds or sheets produces polarized light perpendicular to the scattering plane \citep[see e.g.][]{1979ApJ...229..954W, 1979ApJ...230..116W, 1996ApJ...462L..27W}. The polarization curve produced by scattering is given by:
\begin{equation}
\label{eq_scattering}
p(\lambda) = c_R \times \lambda^{-4},
\end{equation}
where $c_R$ is the amplitude of the scattering (e.g., $c_R$=0.027 $\pm$ 0.002 per cent in \citealt{2013ApJ...775...84A}). The index of the power law is not well constrained, and is usually chosen to be $-$4, appropriate for both Rayleigh scattering from polarizable molecules and Mie scattering in the small grain limit \citep{2013ApJ...775...84A}. Note that Eq.~\ref{eq_scattering} is only valid in case of low optical depth. In case of large optical depths, the blue part is strongly depleted, because of multiple scattering, which is more probable in the blue part of the spectrum where scattering is more efficient \citep[see e.g.][]{2018ApJ...861....1N}.

\citet{2015A&A...577A..53P} found that highly reddened SNe with low total-to-selective extinction ratios, $R_V$, also have peculiar continuum polarization curves steeply rising toward the blue part of the spectrum, and with polarization peaks $\lambda_{\rm max} \lesssim 0.4 \mu m$. It is not fully understood whether such polarization curves are produced by small interstellar grains in their host galaxies or by scattering from nearby circumstellar material. For more information on interstellar (or circumstellar) continuum polarization toward SNe Ia, see \citet{2015A&A...577A..53P}, \citet{2017ApJ...836...88Z}, \citet{Cikota2017}, \citet{2017ApJ...836...13H}, \citet{2018A&A...615A..42C} and \citet{2018arXiv181005557H}.

iii) Polarization induced by electron scattering in globally aspherical photospheres -- In case of spherical photospheres, the intensity of scattered (linearly polarized) light from free electrons will be equal in orthogonal directions, and thus, we do not observe a net polarization (Fig.~\ref{fig:SNlinepol_photosphere}a). In the case of an aspherical photosphere, on the other hand, the net intensity of scattered light perpendicular to the major axis of the projected photosphere will be larger than the intensity of the scattered light perpendicular to the minor axis (Fig.~\ref{fig:SNlinepol_photosphere}b), and therefore, we will observe a net polarization \citep[see e.g.][]{1982ApJ...263..902S,1991A&A...246..481H}. 
The intrinsic continuum polarization in SNe\,Ia is typically $\lesssim$0.4 per cent, which is consistent with global asphericities at the $\sim$ 10 per cent level \citep{2008AJ....136.2227C}. 
The highest intrinsic continuum polarization was observed in the sub-luminous Type Ia SN\,1999by \citep{2001ApJ...556..302H}, which showed an intrinsic polarization of $\sim$0.8 per cent. \citet{2001ApJ...556..302H} found that the spectropolarimetric data could be modeled by an ellipsoid with an axis ratio of 1.17 seen equator-on.

Despite the Thomson scattering being wavelength--independent, in some cases the intrinsic continuum polarization of SNe Ia decreases toward the blue end of the spectrum \citep[see e.g. the sub-luminous SN\,1999by and SN\,2005ke,][]{2012A&A...545A...7P}. 
This is due to depolarization by a large number of bound--bound transitions, primarily of iron--peak elements \citep{2000ApJ...530..757P}, in the UV and blue part of the spectrum \citep{2008AJ....136.2227C}. 
This can be explained by the probability of the last scattering of the photons due to electrons versus the probability of the last scattering due to lines. If the opacity is dominated by bound-bound scattering rather than electron scattering, the polarization will be suppressed. For a Fe-rich ejecta, line opacity dominates, and the importance of electron scattering is suppressed.

\subsection{Line polarization}
\label{sect:line_polarization}

As explained in the previous subsection, SN Ia explosions are in general globally spherical explosions (Sect.~\ref{sect_continuumpolmechanisms}), which is also reflected in the spherical shapes of SN Ia remnants \citep[e.g.][]{2009ApJ...706L.106L}. Polarization across spectral lines (hereafter line polarization) can give us insight into the geometry of the ejecta, which depends on the explosion scenario and/or the progenitor system. In Sect.~\ref{sect:Bullasection}, we explain line polarization predictions for different explosion scenarios, based on simulations, whereas here we focus on how line polarization is produced in general.

\begin{figure}
\center
\includegraphics[trim=0mm 0mm 0mm 0mm, width=7.5cm, clip=true]{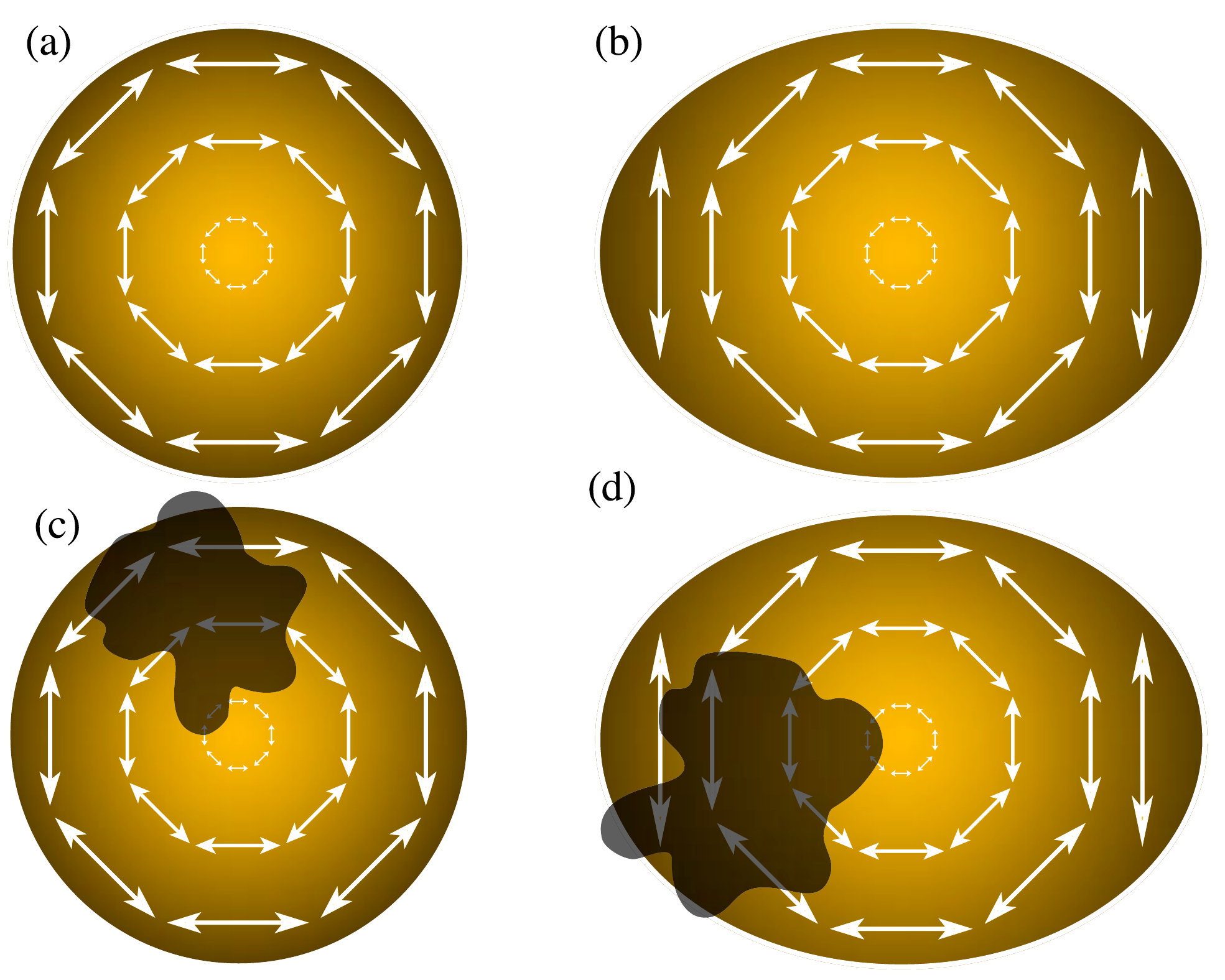}
\vspace{1mm}
\caption{Illustration of the mechanism by which polarization arises. (a) In case of a spherical SN photosphere, we will observe a net polarization of p=0. (b) In case of an aspherical photosphere, we will observe a net continuum polarization p > 0. (c) If there is an inhomogeneous distribution of material in front of the spherical photosphere, it will obscure part of the polarized light, and we will observe polarization across spectral lines only. (d) An aspherical photosphere with an inhomogeneous distribution of material in front of the photosphere would produce both continuum and line polarization.}
\label{fig:SNlinepol_photosphere}
\end{figure}

Figure~\ref{fig:SNlinepol_photosphere} shows a very simplified illustration of the mechanism by which polarization arises from the electron-scattering dominated photosphere. In the case of a spherical photosphere, there will be an equal amount of polarized light coming from all directions (Figure~\ref{fig:SNlinepol_photosphere}a), and therefore, we will observe a net polarization of p=0. However, if there is aspherically distributed material in front of the photosphere, it will obscure part of the polarized light from the photosphere, and thus the cancellation of the perpendicular intensity beams will be incomplete and we will observe a net polarization, p>0, in the polarization spectrum associated to the corresponding absorption line in the flux spectrum (Figures~\ref{fig:SNlinepol_photosphere}c and d, see also \citealt{2003ApJ...593..788K,2003ApJ...591.1110W}). 

%\citep{1984MNRAS.210..829M}

Depending on the geometry and velocity distribution of the blocking material, different properties in the Stokes $q$--$u$ plane\footnote{Stokes parameters are values that characterize the polarization of electromagnetic radiation (\citealt{1960ratr.book.....C}, see also \citealt{2008ARA&A..46..433W}).} will be observed. If the distribution of the absorbers (or clumps) is not spherically symmetric, does not have a common symmetry axis, and if there is a velocity dependent range of position angles, we will observe a smooth change in the polarization angle, i.e we will observe loops in the $q$--$u$ plane as function of wavelength. However, in case of an axisymmetric distribution of the ejecta (e.g., a torus, or bipolar ejecta), we will observe a line in the $q$--$u$ plane, but no loops, because in these configurations the geometry of the obscuring material shares the same symmetric axis as the photosphere (see \citealt{2001ApJ...550.1030W,2003ApJ...591.1110W,2003ApJ...593..788K,2008ARA&A..46..433W,2017ApJ...837..105T} for a more detailed explanation).

\subsubsection{Line polarization predictions for different SN Ia explosion models}
\label{sect:Bullasection}

\citet{2016MNRAS.455.1060B} ran simulations to predict polarization signatures for the violent-merger model of \citet{2012ApJ...747L..10P}, which is presumed to be highly asymmetric, hence describing what can be considered as a limiting case. They found that the polarization signal significantly varies with the viewing angle. In the equatorial plane of the explosion, the continuum polarization levels will be modest, $\sim$0.3 per cent, while at orientations out of the equatorial plane, where the departures from a dominant axis are higher, the degrees of polarization will be higher ($\sim$0.5-1 per cent). However, for all orientations, the polarization spectra display strong line polarization ($\sim$1-2 per cent). They suggest that the violent-merger scenario may explain highly polarized events such as SN\,2004dt \citep{2016MNRAS.455.1060B}. 

In a similar study, \citet{2016MNRAS.462.1039B} predict polarization signatures for the double-detonation (from \citealt{2010A&A...514A..53F}) and delayed-detonation (from \citealt{2013MNRAS.429.1156S}) models of SNe Ia. 

In the delayed-detonation model, a white dwarf near the Chandrasekhar mass, which accretes material from a non-degenerate companion, explodes after an episode of slow (subsonic) carbon burning (carbon deflagration) near the center \citep{1991A&A...245..114K}. In the double-detonation model the explosion in the core of a sub-Chandrasekhar white dwarf is triggered by a shock wave following a detonation of a helium layer at the white dwarf's surface, that has been accreted from a helium rich companion star \citep{2010A&A...514A..53F}.

For both explosion models, \citet{2016MNRAS.462.1039B} predict modest degrees of polarization ($\lesssim$ 1 per cent). The peak of the intrinsic continuum polarization of $\sim$0.1--0.3 per cent decreases after maximum brightness, and prominent polarization of the absorption features are present in the simulated spectra, in particular of the \ion{Si}{ii} $\lambda$6355\,\AA\ line. The simulations show low polarization across the \ion{O}{i} 7774\,\AA , which is consistent with the observed values in normal SNe Ia. The low polarization across \ion{O}{i} reflects the fact that the oxygen is not a product of the dynamical thermonuclear burning, so that oxygen in the spectra comes from the primordial oxygen of the white dwarf (which is supposed to be spherically distributed, \citealt{2006ApJ...653..490W,2009A&A...508..229P}). This is a very important result from polarimetry, placing a constraint on the explosion models \citep{2006NewAR..50..470H}. \\

\section{Observations and data}
\label{sect_observations}

\subsection{Instruments and observations}
\label{sect_instruments}

Our targets were observed with the FOcal Reducer and low dispersion Spectrograph (FORS) in spectropolarimetric mode (PMOS), mounted on the Cassegrain focus of the Very Large Telescope (VLT) at Cerro Paranal in Chile \citep{1998Msngr..94....1A}.

Two versions of FORS were built, and mounted to different VLT unit telescopes since the start of operations. FORS1 was installed on Antu (UT1) and commissioned in 1998, while FORS2 was installed on Kueyen (UT2) one year later. They are largely identical with a number of differences, particularly, the polarimetric capabilities were offered only on FORS1. In June 2004, FORS1 was moved to Kueyen (UT2) and FORS2 to Antu (UT1), and in August 2008 the polarimetric capabilities were transferred from FORS1 to FORS2.

FORS in PMOS mode is a dual-beam polarimeter, with a wavelength coverage from $\sim$330 -- 1100 nm. It contains a Wollaston prism, which splits an incoming beam into two beams with orthogonal directions of polarization, the ordinary (o) and extraordinary (e) beam. The observations of the SNe Ia in this work were obtained with FORS1 or FORS2, a 300V grism, with and/or without the order separating GG435 filter. The GG435 filter has cut-off at $\sim$ 435 $\mu m$, and is used to prevent second-order contamination. The effect of second-order contamination on spectropolarimetry is in most cases negligible, but can be significant for very blue objects \citep[see Appendix in][]{2010A&A...510A.108P}. The half-wave retarder plate was positioned at four angles of 0$^{\circ}$, 22.5$^{\circ}$, 45$^{\circ}$, and 67.5$^{\circ}$ per sequence. The ordinary and extraordinary beams were extracted using standard procedures in IRAF (as described in \citealt{2017MNRAS.464.4146C}). Wavelength calibration was achieved using He-Ne-Ar arc lamp exposures. The typical RMS accuracy is $\sim$\,0.3\,\AA. 
The data have been bias subtracted, but not flat field corrected; however, the detector artifacts get reduced by taking a redundant number of half-wave positions \citep[see][]{2006PASP..118..146P}.

\citet{2007ASPC..364..503F} analysed observations of standard stars for linear polarization obtained from 1999 to 2005 with FORS1, in imaging (IPOL) and spectropolarimetric (PMOS) mode. They found a good temporal stability and a small instrumental polarization in PMOS mode. In a similar study, \citet{2017MNRAS.464.4146C} tested the temporal stability of the PMOS mode of FORS2
since it was commissioned, using a sample of archival polarized and unpolarized standard stars, and found a good observational repeatability of total linear polarization measurements with an RMS $\lesssim$ 0.21 per cent. They also confirmed and parameterized the small ($\lesssim$ 0.1 per cent) instrumental polarization found by \citet{2007ASPC..364..503F}, that we correct by applying their linear functions to Stokes $q$ and $u$.

\subsection{Supernova sample}
\label{sect_sample}

In this work, we collected archival data\footnote{http://archive.eso.org} of 35 SNe Ia, observed with FORS1 and FORS2 in spectropolarimetry mode, between 2001 and 2015, at 127 epochs in total. A list of the targets with some basic properties is provided in Table~\ref{tab:SNsample}, while a full observing log is given in the supplementary material (online), Table~A1. The observations in Table~A1 are grouped into individual epochs, separated by a space.
%Table~\ref{table_obslog} 

\begin{figure*}
\includegraphics[trim=5mm 0mm 7mm 0mm, width=7.1 cm, clip=true]{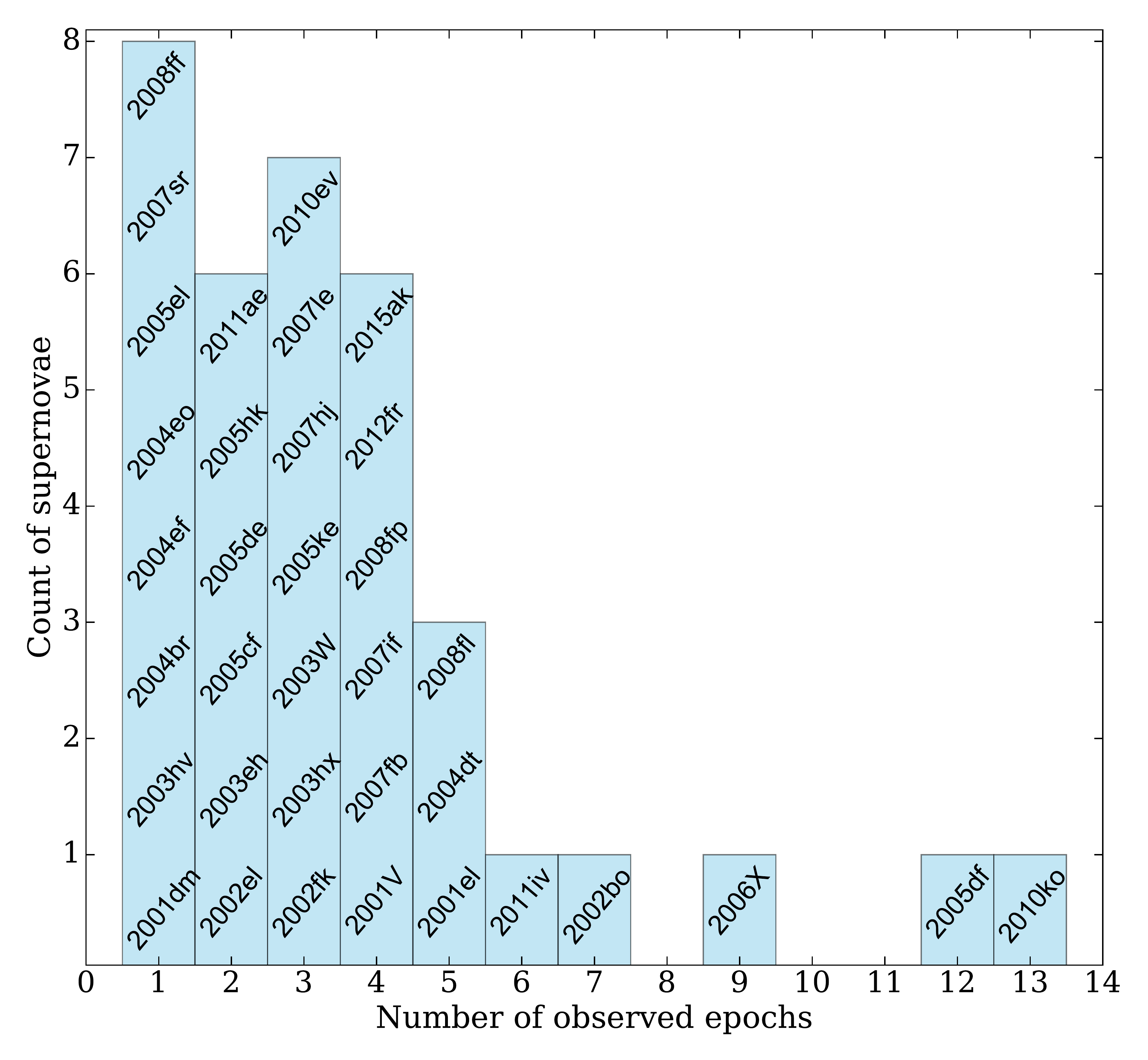}
\includegraphics[trim=5mm 0mm 7mm 0mm, width=10.5 cm, clip=true]{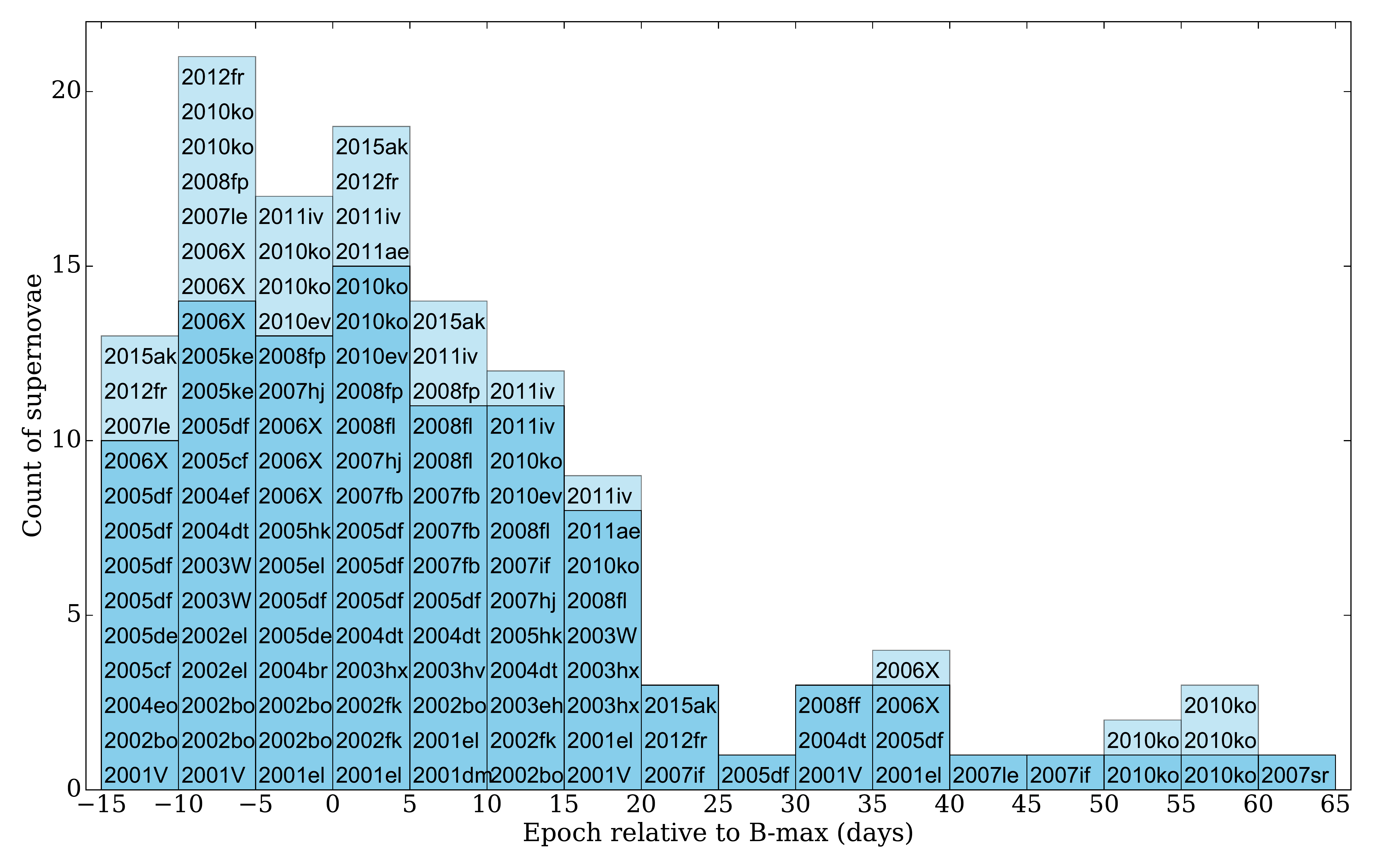}
\vspace{-5mm}
\caption{\textit{Left panel:} Distribution of number of epochs per SN. For example, SN\,2001el, SN\,2004dt and SN\,2008fl have been observed at 5 epochs. \textit{Right panel:} Distribution of observed epochs. For example, SN\,2007if, SN\,2012fr and SN\,2015ak have been observed between 20 and 25 days past peak brightness. The dark blue color is the number of unique supernovae observed per epoch bin (note that some SNe were observed at multiple epochs within the same epoch bin).}
\label{fig:epochshisto}
\end{figure*}

Figure~\ref{fig:epochshisto} shows the number of observed epochs per SN, and the distribution of observed epochs. Eight SNe have been observed at a single epoch, while the most frequently observed supernovae are SN\,2010ko (observed at 13 epochs), SN\,2005df (12 epochs), SN\,2006X (9 epochs) SN\,2002bo (7 epochs) and SN\,2011iv (6 epochs).

Our sample also contains a few spectroscopically peculiar sub-luminous objects \citep[see][for a review of different sub-types]{2017hsn..book..317T}. SN\,2005ke is a 91bg-like object \citep{2012A&A...545A...7P}; SN\,2005hk is a 2002cx-like object \citep{2010ApJ...722.1162M}, while SN\,2011iv and SN\,2004eo are transitional objects \citep[see][respectively]{2017arXiv170703823G, doi:10.1111/j.1365-2966.2007.11700.x}. Also a spectrum of SN\,2007hj near maximum light shows similarities to several sub-luminous SNe Ia \citep{2007CBET.1048....2B}.

{\small
\begin{table*}
\centering
\caption{SN Ia sample}
\label{tab:SNsample}
\begin{tabular}{lllc p{4.0 cm} l} % four columns, alignment for each
\hline
Name & z & $T_{\rm Bmax}$ (MJD) & $\Delta$m$_{15}$ (mag) & Epochs (days relative to $T_{\rm max}$) & References (z, $T_{\rm Bmax}$, $\Delta$m$_{15}$)\\
\hline
SN\,2001dm & 0.01455 & 52128.0 $\pm$ 10.0 & \dots & 6.3 & 1,  estimate , -- \\
SN\,2001V & 0.01502 & 51972.58 $\pm$ 0.09 & 0.73 $\pm$ 0.03 & $-$10.4, $-$6.4, 17.5, 30.6 & 2, SNooPy, 3 \\
SN\,2001el & 0.00364 & 52182.5 $\pm$ 0.5 & 1.13 $\pm$ 0.04 & $-$4.2, 0.7, 8.7, 17.7, 39.6 & 4, 5, 6 \\
SN\,2002bo & 0.00424 & 52356.5 $\pm$ 0.2 & 1.12 $\pm$ 0.02 & $-$11.4, $-$7.4, $-$6.4, $-$4.5, $-$1.4, 9.6, 12.6 & 7, 5, 6 \\
SN\,2002el & 0.02469 & 52508.76 $\pm$ 0.05 & 1.38 $\pm$ 0.05 & $-$8.6, $-$7.6 & 4, SNooPy, 3 \\
SN\,2002fk & 0.007125 & 52547.9 $\pm$ 0.3 & 1.02 $\pm$ 0.04 & 0.4, 4.4, 13.4 & 8, 5, 6 \\
%SN\,2002ic & 0.066 & None $\pm$ None & Nan $\pm$ Nan & 'nan' & {\citet{2002IAUC.8019....3B}}, {\citet{Nan}}, dm15ref \\
SN\,2003eh & 0.02539 & 52782.0 $\pm$ 10.0 & \dots & 0.0, 12.0 & 9, estimate, -- \\
SN\,2003hv & 0.005624 & 52891.2 $\pm$ 0.3 & 1.09 $\pm$ 0.02 & 6.1 & 4, 5, 6 \\
SN\,2003hx & 0.007152 & 52892.5 $\pm$ 1.0 & 1.17 $\pm$ 0.12 & 4.8, 16.8, 18.8 & 1, 10, 11 \\
SN\,2003W & 0.018107 & 52679.98 $\pm$ 0.11 & 1.3 $\pm$ 0.05 & $-$8.7, $-$6.7, 16.1 & 4, SNooPy, 3 \\
SN\,2004br & 0.019408 & 53147.9 $\pm$ 0.27 & 0.68 $\pm$ 0.15 & $-$3.9 & 4, SNooPy, 12 \\
SN\,2004dt & 0.01883 & 53239.98 $\pm$ 0.07 & 1.21 $\pm$ 0.05 & $-$9.7, 4.4, 5.2, 10.3, 33.2 & 13, SNooPy, 3 \\
SN\,2004ef & 0.028904 & 53264.4 $\pm$ 0.1 & 1.45 $\pm$ 0.01 & $-$5.3 & 4, 5, 6 \\
SN\,2004eo & 0.015701 & 53278.51 $\pm$ 0.03 & 1.32 $\pm$ 0.01 & $-$10.4 & 2, 14, 6 \\
SN\,2005cf & 0.006461 & 53533.94 $\pm$ 0.05 & 1.18 $\pm$ 0.01 & $-$11.9, $-$5.8 & 7, SNooPy, Wang (priv. comm.) \\
SN\,2005de & 0.015184 & 53598.89 $\pm$ 0.05 & 1.41 $\pm$ 0.06 & $-$10.9, $-$4.9 & 2, SNooPy, 3 \\
SN\,2005df & 0.004316 & 53599.18 $\pm$ 0.1 & 1.06 $\pm$ 0.02 & $-$10.8, $-$8.8, $-$7.8, $-$6.8, $-$2.8, 0.2, 4.2, 5.2, 8.2, 9.1, 29.2, 42.1 & 15, SNooPy, SNooPy \\
SN\,2005el & 0.01491 & 53647.0 $\pm$ 0.1 & 1.4 $\pm$ 0.01 & $-$2.7 & 13, 5, 6 \\
SN\,2005hk & 0.01306 & 53685.42 $\pm$ 0.14 & 1.47 $\pm$ 0.14 & $-$2.3, 11.7 & 16, SNooPy, 17 \\
SN\,2005ke & 0.00488 & 53699.16 $\pm$ 0.08 & 1.66 $\pm$ 0.14 & $-$9.1, $-$8.1, 75.9 & 13, SNooPy, 17 \\
SN\,2006X & 0.00524 & 53786.3 $\pm$ 0.1 & 1.09 $\pm$ 0.03 & $-$10.9, $-$9.0, $-$8.1, $-$7.1, $-$4.0, $-$3.0, $-$1.9, 37.9, 38.9 & 13, 5, 6 \\
SN\,2007fb & 0.018026 & 54288.41 $\pm$ 0.17 & 1.37 $\pm$ 0.01 & 3.0, 6.0, 6.9, 9.9 & 18, SNooPy, SNooPy \\
SN\,2007hj & 0.01289 & 54350.23 $\pm$ 0.1 & 1.95 $\pm$ 0.06 & $-$1.1, 4.9, 10.9 & 13, SNooPy, 12 \\
SN\,2007if & 0.073092 & 54343.1 $\pm$ 0.6 & 1.07 $\pm$ 0.03 & 13.1, 20.1, 45.0, 46.0 & 4, 5, 6 \\
SN\,2007le & 0.005522 & 54399.3 $\pm$ 0.1 & 1.03 $\pm$ 0.02 & $-$10.3, $-$5.1, 40.7 & 4, 5, 6 \\
SN\,2007sr & 0.005417 & 54447.82 $\pm$ 0.24 & 1.05 $\pm$ 0.07 & 63.4 & 18, SNooPy, 12 \\
SN\,2008ff & 0.0192 & 54704.21 $\pm$ 0.63 & 0.90 $\pm$ 0.06 & 31.0 & 19, 20, 21 \\
SN\,2008fl & 0.0199 & 54720.79 $\pm$ 0.86 & 1.35 $\pm$ 0.07 & 2.2, 8.3, 9.2, 11.3, 15.3 & 22, 20, 21 \\
SN\,2008fp & 0.005664 & 54730.9 $\pm$ 0.1 & 1.05 $\pm$ 0.01 & $-$5.6, -0.5, 1.4, 5.4 & 4, 5, 6 \\
SN\,2010ev & 0.009211 & 55385.09 $\pm$ 0.14 & 1.12 $\pm$ 0.02 & $-$1.1, 2.9, 11.9 & 16, SNooPy, 23 \\
SN\,2010ko & 0.0104 & 55545.23 $\pm$ 0.23 & 1.56 $\pm$ 0.05 & $-$7.1, $-$6.0, $-$2.0, $-$1.2, 1.9, 3.9, 10.9, 15.9, 50.9, 51.8, 57.8, 58.8, 59.8 & 24, SNooPy, SNooPy \\
SN\,2011ae & 0.006046 & 55620.22 $\pm$ 0.39 & \dots & 4.0, 16.0 & 18, SNooPy, -- \\
SN\,2011iv & 0.006494 & 55905.6 $\pm$ 0.05 & 1.77 $\pm$ 0.01 & -0.4, 2.6, 5.6, 11.5, 12.5, 19.5 & 16, 25, 26 \\
SN\,2012fr & 0.0054 & 56244.19 $\pm$ 0.0 & 0.8 $\pm$ 0.01 & $-$12.1, $-$6.9, 1.0, 23.1 & 27, SNooPy, SNooPy \\
SN\,2015ak & 0.01 & 57268.13 $\pm$ 0.1 & 0.95 $\pm$ 0.01 & $-$13.1, 4.9, 6.9, 23.9 & spec. fit, SNooPy, SNooPy \\
\hline
\multicolumn{6}{p{15.5 cm}}{References. SNooPy: see Sect.~\ref{sect_lightcurvefitting};
1: {\citet{2012MNRAS.425.1789S}}; 
2: {\citet{2013MNRAS.433.2240G}}; 
3: {\citet{2007Sci...315..212W}}; 
4: {\citet{2016A&A...594A..13P}}; 
5: {\citet{Dhawan2015}}; 
6: {\citet{Dhawan2015}}; 
7: {\citet{2011ApJ...731..120M}}; 
8: {\citet{2010ApJ...716..712A}}; 
9: {\citet{2008ApJ...673..999P}}; 
10: {\citet{2008MNRAS.389..706M}}; 
11: {\citet{2008MNRAS.389..706M}}; 
12: {\citet{2010ApJS..190..418G}}; 
13: {\citet{2013ApJ...773...53F}}; 
14: {\citet{Dhawan2015}}; 
15: {\citet{2012PASP..124..668Y}}; 
16: {\citet{2016ApJ...821..119C}}; 
17: {\citet{2009ApJ...700..331H}}; 
18: {\citet{2015ApJS..220....9F}}; 
19: {\citet{2008CBET.1488....1T}}; 
20: {\citet{2017AJ....154..211K}}; 
21: {\citet{2017AJ....154..211K}}; 
22: {\citet{2008CBET.1498....1P}}; 
23: {\citet{2016A&A...590A...5G}}; 
24: {\citet{2010CBET.2569....1L}}; 
25: {\citet{2017arXiv170703823G}}; 
26: {\citet{2017arXiv170703823G}}; 
27: {\citet{2012CBET.3277....3B}}. }\\
\end{tabular}
\end{table*}
}

\section{Methods}
\label{sect_methods}

\subsection{Stokes parameters, polarization degree and polarization angle}
\label{sect_stokes}

The normalized Stokes parameters $q$ and $u$ were derived following the recipe in the FORS2 User Manual (\citealt{FORS2manual}):
\begin{equation}
\begin{array}{l}
q = \frac{2}{N} \sum_{i=0}^{N-1} F(\theta_i)\cos(4\theta_i) \\ 
u = \frac{2}{N} \sum_{i=0}^{N-1} F(\theta_i)\sin(4\theta_i)
\end{array}
\end{equation}
where $F(\theta_i)$ are the normalized flux differences between the ordinary ($f^o$) and extra-ordinary ($f^e$) beams:
\begin{equation}
\label{eqF}
F(\theta_i) = \frac{f^o (\theta_i) - f^e (\theta_i)}{f^o (\theta_i) + f^e (\theta_i)}
\end{equation}
at N different half-wave retarder plate position angles $\theta_i = i * 22.5^{\circ}$ (0 $\leq i \leq$ 3).

We correct the retardance chromatism of the super-achromatic half wave plate (HWP) through a rotation of the Stokes parameters:
\begin{equation}
\begin{array}{l}
q_0 = q \cos 2\Delta\theta(\lambda) - u \sin 2\Delta\theta(\lambda) \\
u_0 = q \sin 2\Delta\theta(\lambda) + u \cos 2\Delta\theta(\lambda),
\end{array}
\end{equation}
using the wavelength dependent retardance offset ($\Delta\theta(\lambda)$), tabulated in the FORS2 User Manual.

Finally we calculated the polarization: 
\begin{equation}
\label{eqP}
p=\sqrt{q_0^2+u_0^2}
\end{equation}
and the polarization angle:
\begin{equation}
\label{eqtheta}
\theta_0 = \frac{1}{2}\arctan(u_0/q_0).
\end{equation}

The uncertainties on the normalized flux differences, the Stokes parameters, and the polarization and polarization angle were calculated by propagating the flux uncertainty ("sigma spectrum") of the ordinary and extra-ordinary beams, which were extracted, along with the beams, using the IRAF task apextract.apall.

\subsection{Wavelet decomposition and continuum removal}
\label{sect_wavelet}

In this paper we mainly focus on line polarization, and do not investigate interstellar, circumstellar or intrinsic continuum polarization. Therefore, we remove the whole continuum polarization without distinguishing between the three components (see Sect.~\ref{sect_continuumpolmechanisms}). 

There are different approaches to remove the interstellar continuum polarization (ISP). For instance, \citet{2009A&A...508..229P} fit a third order polynomial to the degree of linear polarization for the last epoch of SN\,2006X, and estimate the interstellar polarization Stokes parameters as follows:
\begin{equation}
\label{eqISPnando}
\begin{array}{l}
q_{ISP}(\lambda) =  p_{ISP}(\lambda) cos(2 \theta_{ISP}) \\
u_{ISP}(\lambda) =  p_{ISP}(\lambda) sin(2 \theta_{ISP}),
\end{array}
\end{equation}
where they compute $\theta_{ISP}$ as the average within some line-free regions (note that $\theta_{ISP}$ is wavelength independent). Then they calculate the intrinsic SN polarization at all epochs by subtracting $q_{ISP}$ and $u_{ISP}$ from the initial Stokes $q$ and $u$. 
SN\,2006X has a strong ISP contribution, so that any intrinsic polarization was negligible. Also, \citet{2009A&A...508..229P} use this method for the purpose of removing the ISP, but not all the continuum polarization.

For the purpose of a more general approach we preferred a different method, which doesn't require the $\theta_{ISP}$ to be determined and is also applicable to small levels of ISP.
Therefore, to determine and subtract the continuum, we first perform an $\rm \grave{a}$ trous wavelet decomposition \citep{1989wtfm.conf..286H} of the individual ordinary and extraordinary beams, which allows us to distinguish between the continuum spectra and the line spectra in a systematic way. 

The wavelet decomposition is a method to decompose a function into a set of $J$ scales, by convolving the function with a convolution mask with an increasing size. Assuming a convolution mask (e.g., the commonly used Mexican hat or simply a triangle), the first convolution is performed on the initial function $c_0(k)$, to generate $c_1(k)$. The difference $c_0(k) - c_1(k)$ is the first wavelet scale $w_1(k)$. The algorithm is then reapplied $j$ times, using a double sized convolution mask, until scale $J$ is reached \citep[see also][]{2010ApJ...711..711W}.
 
The sum of all wavelet scales reproduces the original function (i.e., spectrum):
\begin{equation}
\label{eq:waveletsum}
c_0(k)= c_J(k) + \sum_{j=1}^{J-1} w_j(k).
\end{equation}
 
As a convolution mask, we follow \citet{2010ApJ...711..711W}, and use a five-bin symmetric triangle function with weights: 3/8, 1/4, and 1/16.

\begin{figure}
\center
\includegraphics[trim=0mm 0mm 0mm 0mm, width=8.5cm, clip=true]{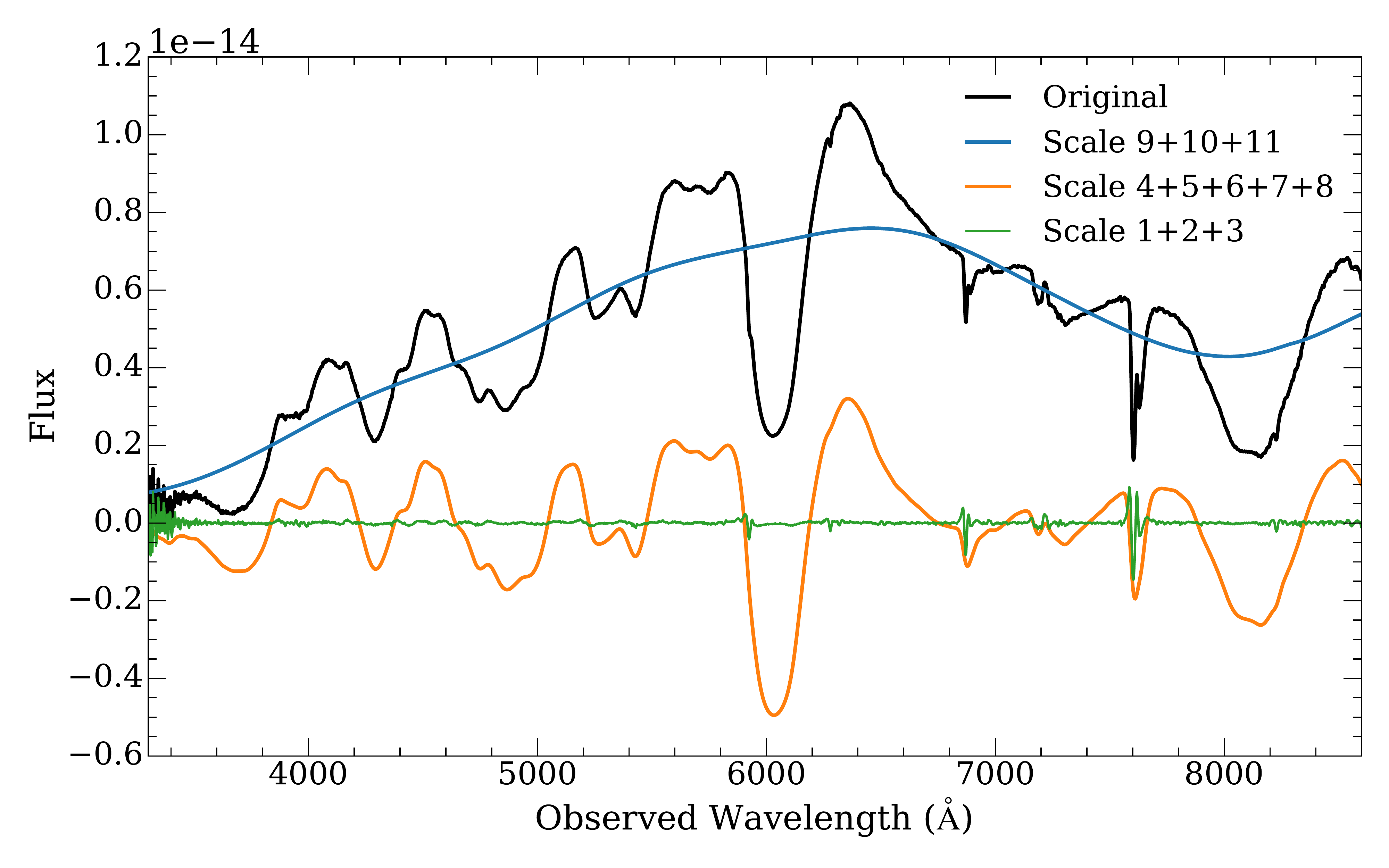}
\vspace{-7mm}
\caption{Example of an $\rm \grave{a}$ trous wavelet decomposition. The black line is an original spectrum, blue, orange, and green curves are the sum of the last three (9+10+11), first three (1+2+3) and middle 5 wavelet scales (4+5+6+7+8), respectively.}
\label{fig:atrous}
\end{figure}

Figure~\ref{fig:atrous} shows an example of a wavelet decomposition of a flux spectrum into a ``continuum" (sum of scales 9+10+11), ``noise" (scales 1+2+3) and the remaining spectrum (scales 4+5+6+7+8). Note that the ``continuum" does not necessarily coincide with the physical continuum of the SN. 
To determine the continuum subtracted polarization spectra, we first calculate the total Stokes $q$ and $u$ by summing all wavelet scales of the individual ordinary and extraordinary flux spectra, except the first three in order to reduce the noise, and following the equations in Sect~\ref{sect_stokes}. Then we calculate the continuum Stokes $q$ and $u$, using the ``continuum" spectra of the individual ordinary and extraordinary flux spectra, i.e. using the sum of the last three wavelet scales only. Finally, we subtract the continuum Stokes $q$ and $u$ from the total Stokes $q$ and $u$. 
Note that, in general, the normalized Stokes parameters q and u are not additive (in contrast to unnormalized Stokes Q and U). However, the ISP cannot be expressed as a regular Stokes vector (I,Q,U,V), because the ISP does not carry any intensity, I$_{\rm ISP}$. When a beam of radiation traverses the ISP, the Stokes vector is transformed and the process can be described as the product between the incoming Stokes vector and a Mueller matrix, which approximates a partial polarizer (see Appendix B in \citealt{2010A&A...510A.108P}). In the case of weak incoming polarization, this corresponds to a vectorial sum between S$_{\rm SN}$(I,Q,U) and S$_{\rm ISP}$(0,Q$_{\rm ISP}$,U$_{\rm ISP}$), where I$_{\rm out}$ is attenuated by the extinction \citep{2010A&A...510A.108P}.

\begin{figure}
\center
\includegraphics[trim=15mm 15mm 40mm 45mm, width=8.0cm, clip=true]{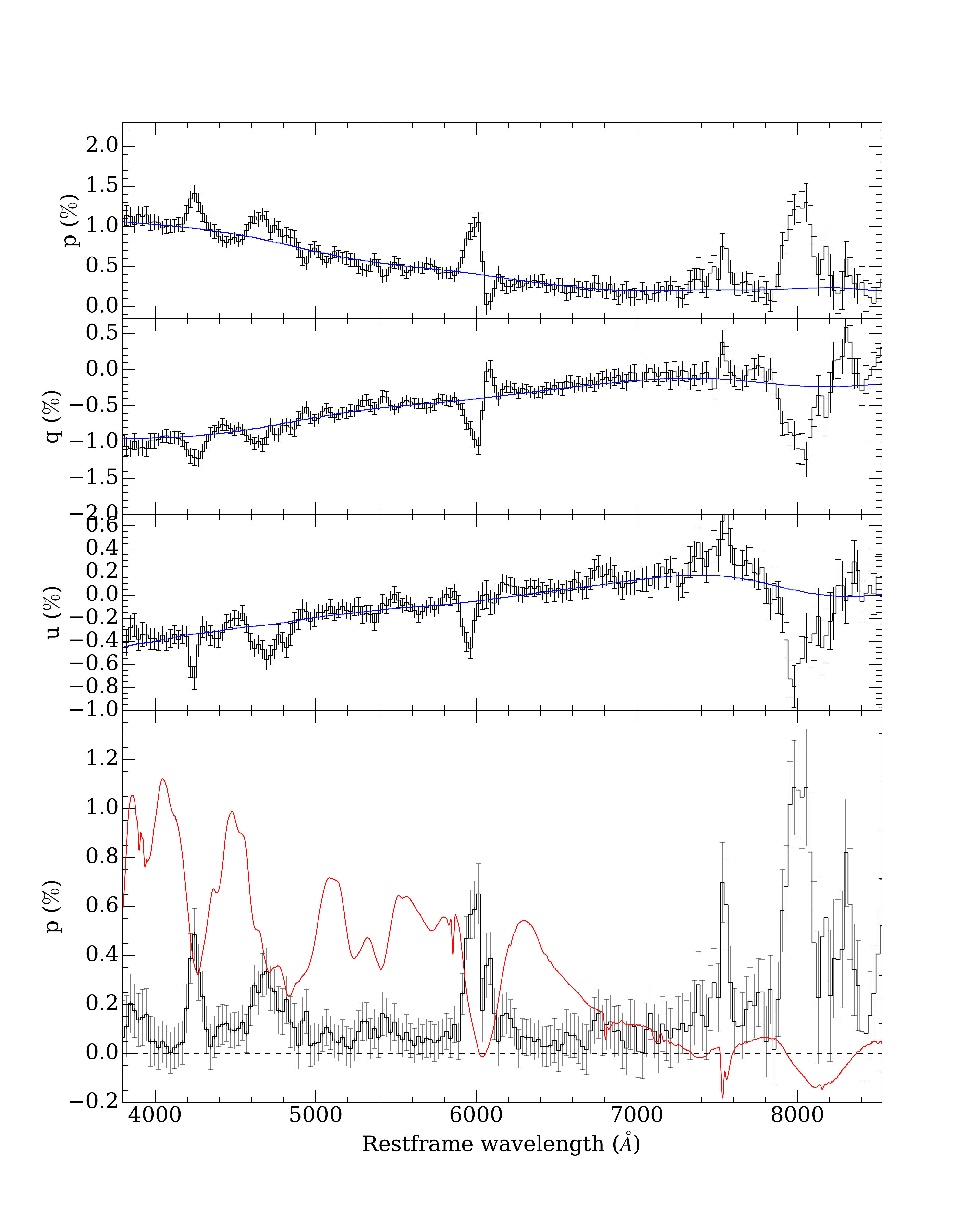}
\vspace{-4mm}
\caption{Example of continuum subtraction from a polarization spectrum of SN\,2002bo at $-$1 day relative to peak brightness. The top panel shows the total degree of polarization (black curve) and the middle panels are Stokes $q$ and $u$. The blue curves display the continuum polarization Stokes $q$ and $u$ determined with wavelet decomposition (middle panels), and the calculated continuum polarization from the Stokes parameters (top panel). The bottom panel shows the continuum subtracted polarization spectrum, compared to the flux spectrum (red curve). The $\lambda$6355\,\AA\ \ion{Si}{ii} line, at $\sim$ 6000\,\AA, displays two peaks, related to the high velocity and photospheric silicon lines, respectively. The bin size of the polarization spectra is 25\,\AA , and the error bars represent 1$\sigma$ uncertainties.}
\label{fig:2002bo_atrous}
\end{figure}

This also instantly removes the continuum polarization from the line photons (which may still contain some  polarization due to interstellar or circumstellar scattering), assuming that the continuum polarization crossing the absorption lines is smoothly following the continuum polarization determined by wavelet decomposition of the flux spectra.
An example of continuum subtraction from a polarization spectrum is shown in Fig.~\ref{fig:2002bo_atrous}. The blue curves in the middle panels display the continuum Stokes $q$ and $u$ determined as described.

\subsection{Light curve fitting}
\label{sect_lightcurvefitting}

For a number of SNe Ia, for which we did not find the date of peak brightness, $T_{\rm max}$ or the light-curve decline rate, $\Delta$m$_{15}$ \citep{1993ApJ...413L.105P} in the literature, we downloaded light curves from the Open Supernova Catalog\footnote{https://sne.space}, which is an online collection of observations and metadata for SNe \citep{2017ApJ...835...64G} and determined the missing parameters. 

\begin{figure}
\includegraphics[trim=0mm 0mm 0mm 10mm, width=8.5cm, clip=true]{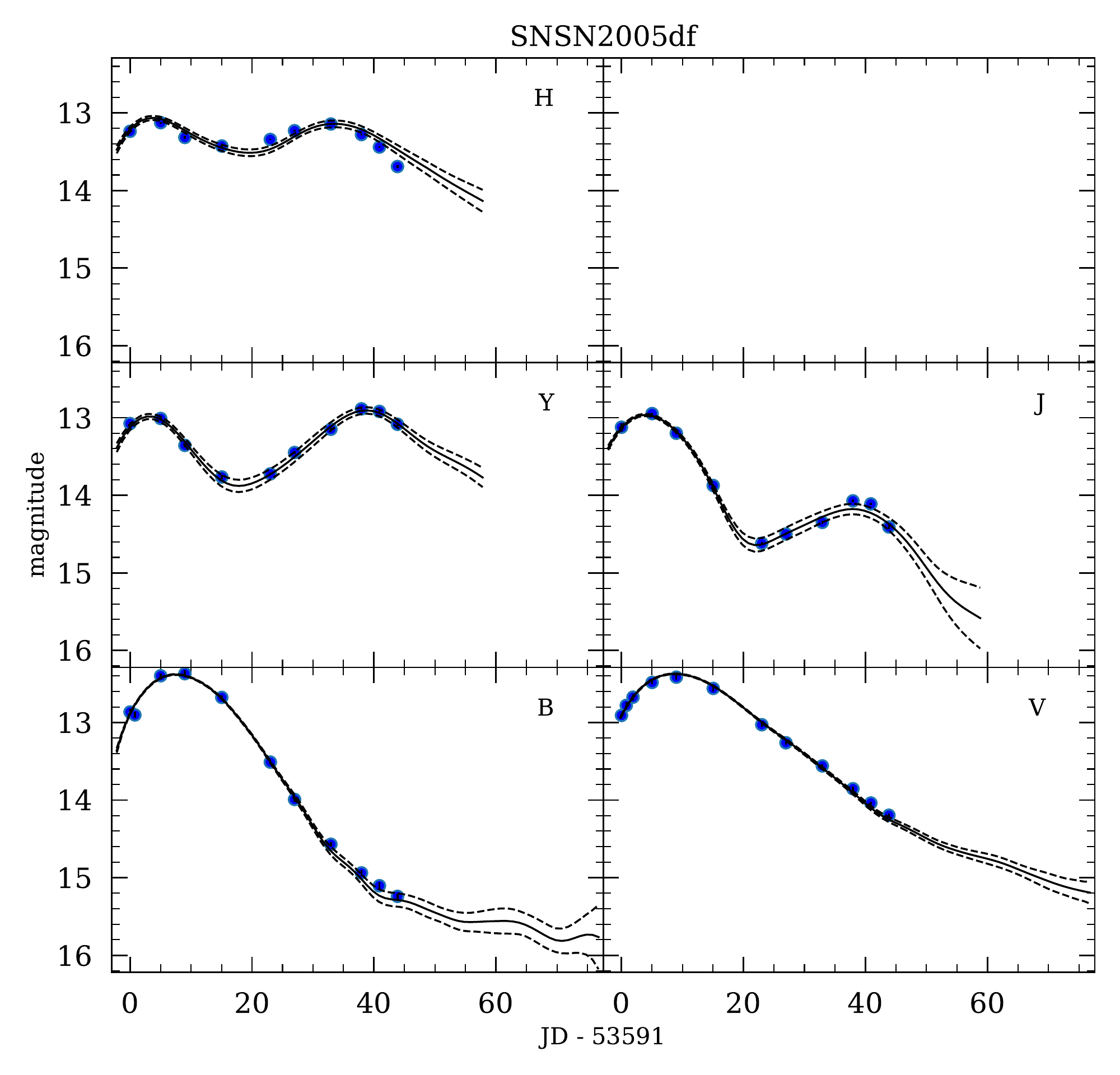}
\vspace{-7mm}
\caption{An example of light curve fitting with SNooPy. Shown is a multi-color light curve fit to SN\,2005df observations in 5 passband filters \citep{2017RNAAS...1...36K}. The derived light curve parameters are T$_{\rm max}$ = 53599.2 $\pm$ 0.1 MJD, $E(B-V)_{\rm Host}$ = 0.024 $\pm$ 0.012 mag, and $\Delta$m$_{15}$ = 1.06 $\pm$ 0.02 mag, respectively. The date on the x-axis is in units of days relative to an arbitrary zero-point. The solid line is the best fit line, and the dashed lines are the 1$\sigma$ uncertainties. Note that the \textit{B} and \textit{V} measurements at 53700 MJD \citep{2017RNAAS...1...36K} are outside of the axes.}
\label{fig:snoopySN2005df}
\end{figure}

For this purpose, we fit the light curves using the SNooPy package \citep{2011AJ....141...19B,2015ascl.soft05023B}, in order to determine the time of peak brightness, $T_{\rm max}$, and $\Delta$m$_{15}$. SNooPy contains tools for the analysis of SN Ia photometry, and includes four models to fit the light curves. For our purpose, we used the ``EBV\_model", which is based on the empirical method for fitting multicolor light curves developed by \citet{2006ApJ...647..501P}. Figure~\ref{fig:snoopySN2005df} shows an example fit of SN\,2005df observations.

A list of SNe Ia for which we fit their light curves, including the number of photometric observations used per passband filter and the corresponding references for the observations, is given in Table~A2 (online supplementary material). The results, i.e., $T_{\rm max}$ and $\Delta$m$_{15}$ are listed in Table~\ref{tab:SNsample}.
%%%\ref{tab:LCfittingsample)

%2017RNAAS...1...36K

% print 'SN', snname, round(s.EBVhost,3),round(s.e_EBVhost,3), round(s.Tmax,3), round(s.e_Tmax,3), round(s.DM,3), round(s.e_DM,3), round(s.dm15,3), round(s.e_dm15,3) 
% SN2005df , 0.024 , 0.012 , 53599.183 , 0.104 , 31.397 , 0.019 , 1.055 , 0.02

\subsection{Expansion velocities deduced from absorption lines}
\label{sect_expvel}

One of our main aims is the study of possible correlations between polarization (used as a proxy for asymmetries in the explosion) and other spectrophotometric properties that characterize the events. In particular, we focus here on the relationship between the line polarization and the expansion velocities. To measure the ejecta photospheric velocities, we use spectra from the Open Supernova Catalog, supplementary to our flux spectra (calculated by summing the ordinary and extraordinary beams) observed with VLT/FORS. A complete list of spectra taken from the Open Supernova Catalog is given in Table~A3 (online supplementary material).
%%\ref{tab:jsonspectrasample}.

\begin{figure}
\center
\includegraphics[trim=0mm 0mm 0mm 0mm, width=8.5cm, clip=true]{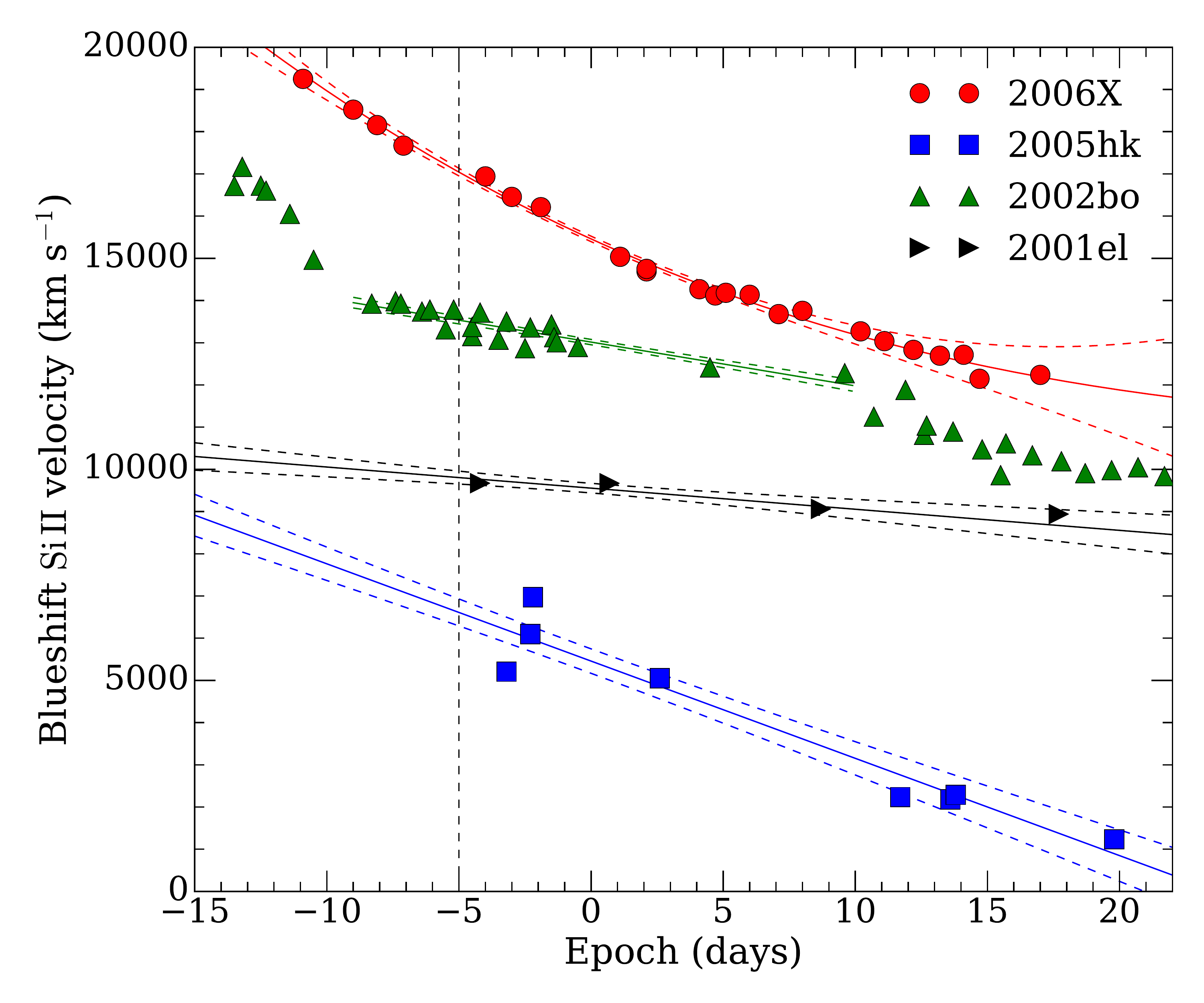}
\vspace{-6mm}
\caption{A third order polynomial fit (solid lines), and the 1$\sigma$ uncertainties (dashed lines) of the \ion{Si}{ii} $\lambda$6355\,\AA\ velocities for SN\,2005hk, SN\,2002bo, SN\,2001el, and SN\,2006X. From the polynomial fits we determined the velocity at $-$5 days relative to peak brightness (vertical black line). The individual velocity measurements are typically accurate within $\sim$ 75 km s$^{-1}$ (the uncertainties are not depicted).}
\label{fig:SiII-epoch-fit}
\end{figure}

We calculated the \ion{Si}{ii} $\lambda$6355\,\AA\ line velocities by smoothing the spectra and measuring the rest-frame wavelengths of the absorption minima. The uncertainty of the measurements mainly  depends on the spectral resolution, which is typically $\sim$ 75 km s$^{-1}$. To determine the velocity at $-$5 days relative to peak brightness, we fit a low-order polynomial function to the measured velocities at different epochs. 
The velocity is determined at $-$5 days relative to peak brightness, because the \ion{Si}{ii} $\lambda$6355\,\AA\ line polarization is expected to be highest before peak brightness \citep{2007Sci...315..212W}. 
Figure~\ref{fig:SiII-epoch-fit} shows an example fit to the \ion{Si}{ii} $\lambda$6355\,\AA\ line velocities at different epochs, for SN\,2005hk (the SN with the lowest \ion{Si}{ii} expansion velocity in our sample), SN\,2002bo, SN\,2001el, and SN\,2006X (the SN with the highest \ion{Si}{ii} expansion velocity in our sample). The results are given in Table~A4 (online supplementary material).
%\ref{tab:SiII_velocity}.

\subsection{Measuring of the \ion{Si}{ii} line polarization}
%%% and P(epoch) fitting
\label{sect_line_pol}

Measuring the linear polarization of lines is challenging, because the peak polarization depends on the binning, and there are sometimes multiple peaks related to one absorption feature, e.g., the high velocity and photospheric \ion{Si}{ii} $\lambda$6355\,\AA\ line components \citep{2003ApJ...591.1110W,2003ApJ...593..788K,2005PASP..117..545B,2005ApJ...623L..37M,2015MNRAS.451.1973S}.

To measure the peak polarization of a specific line, we first manually determine the lower and upper wavelength edges containing the line of interest from a plot with all VLT flux and polarization spectra.

For each polarization spectrum, we use three sets with different bin sizes of 25\,\AA , 50\,\AA\ and 100\,\AA , and apply the polarization bias correction. Since the polarization is by definition a positive quantity (see Eq.~\ref{eqP}), the presence of noise introduces a biased, overestimated, value. We correct the polarization bias following the equation given in \citet{1997ApJ...476L..27W}: 
\begin{equation}
p = (p_{\rm obs} - \sigma_p^2/p_{\rm obs}) \times h(p_{\rm obs} - \sigma_p),
\end{equation}
where $h$ is the Heaviside function, p$_{\rm obs}$ is the observed polarization and $\sigma_p$ is the 1$\sigma$ error on the polarization.

\begin{figure}
\center
\includegraphics[trim=0mm 0mm 0mm 0mm, width=8.5cm, clip=true]{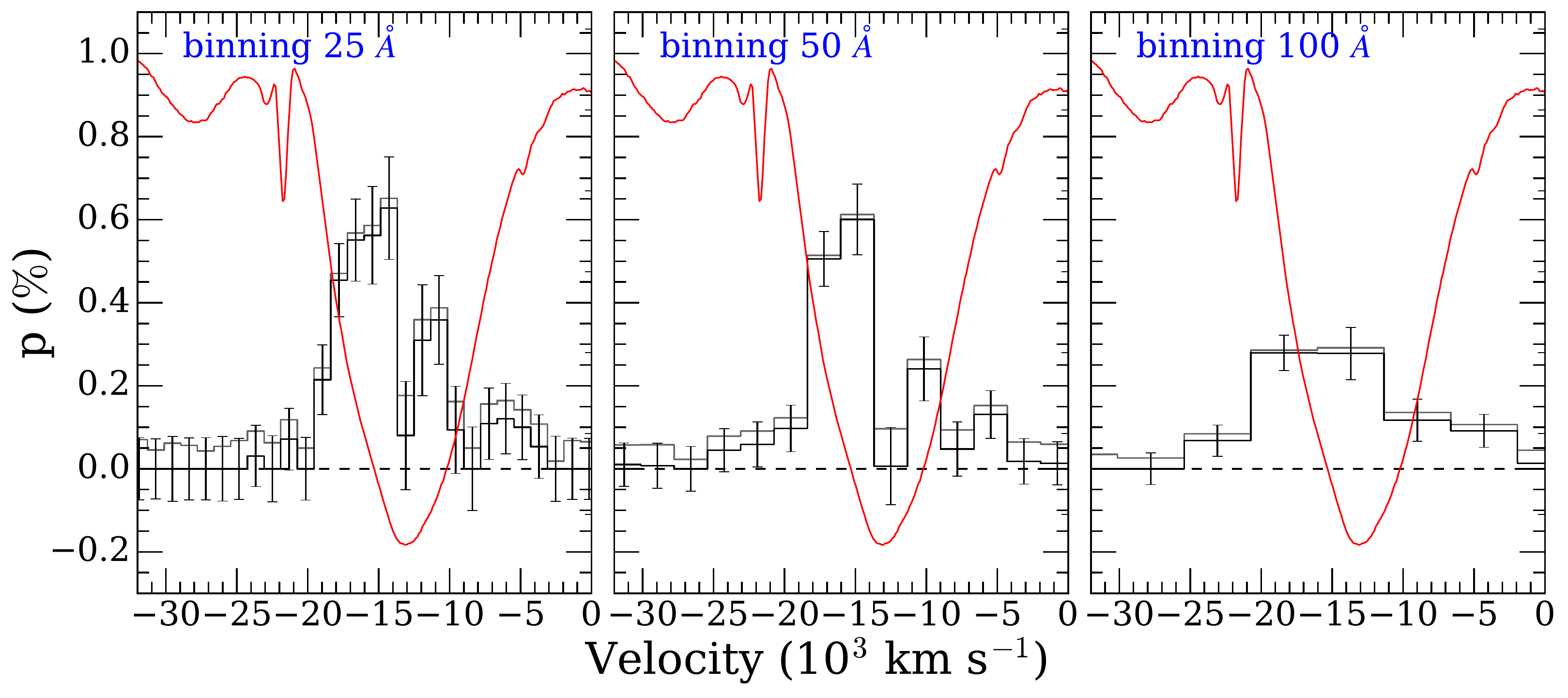}
\vspace{-5mm}
\caption{The linear polarization of the \ion{Si}{ii} $\lambda$6355\,\AA\ line in SN\,2002bo at $-$1 day relative to peak brightness. The three panels show the polarization degree derived with bin sizes of 25\,\AA\ (left), 50\,\AA\ (middle) and 100\,\AA\ (right). The error bars represent 1$\sigma$ uncertainties. The gray line is the polarization spectrum not corrected for bias, and the red line is the flux spectrum of the \ion{Si}{ii} line.}
\label{fig:binncomparison}
\end{figure}

Finally we measure the peak polarization value within the borders of the line. Fig.~\ref{fig:binncomparison} shows the \ion{Si}{ii} $\lambda$6355\,\AA\ line of SN\,2002bo at $-$1 day relative to peak brightness, at three different bin sizes. Note that at this epoch, there are two peaks visible in the 25\,\AA\ and 50\,\AA\ binned spectra. The peaks correspond to the photospheric (with the lower velocity) and the high velocity component of the \ion{Si}{ii} $\lambda$6355\,\AA\ line. In this case, we measure the polarization of the higher peak. In the 100\,\AA\ binned data the two peaks are blended.\\

\begin{figure}
\center
\includegraphics[trim=0mm 0mm 0mm 0mm, width=8.5cm, clip=true]{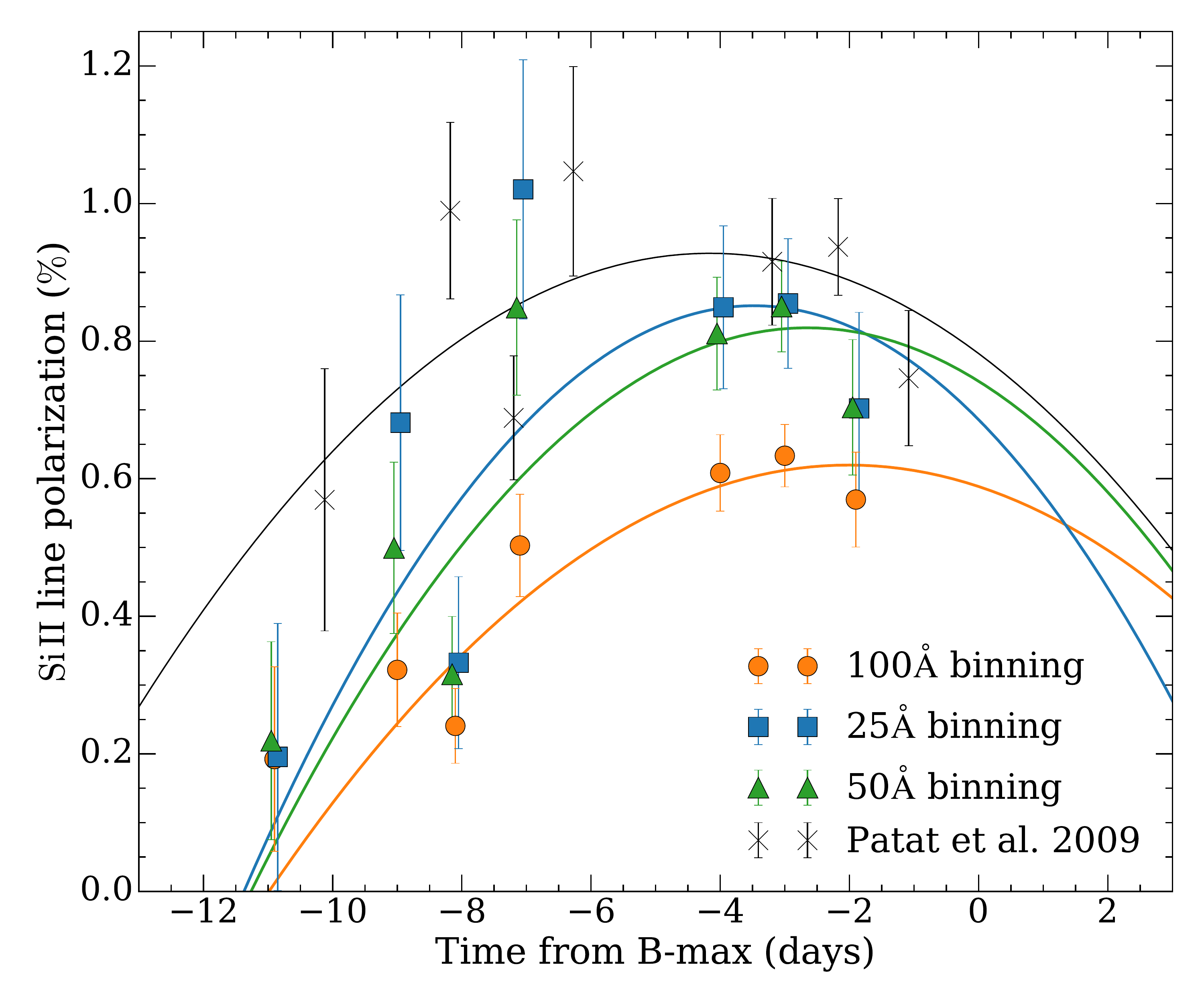}
\vspace{-7mm}
\caption{Peak polarization of the \ion{Si}{ii} $\lambda$6355\,\AA\ line in SN\,2006X, as a function of time, for three different bin sizes, compared to the measurements in \citet{2009A&A...508..229P}, who binned their data in $\sim$26\,\AA\ wide bins. The error bars represent 1$\sigma$ uncertainties. Note that the time of peak brightness in \citet{2009A&A...508..229P} is $\sim$1 day earlier than used in this work.}
\label{fig:epoch-SiII_2006X}
\end{figure}

After the data reduction and the measurements, we can now investigate the evolution of the line polarization as a function of time. Fig.~\ref{fig:epoch-SiII_2006X} shows the peak polarization of the \ion{Si}{ii} $\lambda$6355\,\AA\ line in SN\,2006X, as a function of time, for three different bin sizes of 25\,\AA , 50\,\AA , and 100\,\AA , compared to the measurements in \citet{2009A&A...508..229P}, who binned their data in $\sim$26\,\AA\ wide bins. This clearly shows that the measurements in highly binned data (100\,\AA ) exhibit less variability, and have lower values, but when the polarization is measured in data with smaller bin sizes (25\,\AA\ and 50\,\AA ) the measured peak polarization is higher, but so is the scatter. The choice of the appropriate bin size is a compromise between retaining spectral resolution and increasing the signal-to-noise. Broader bins increase the signal-to-noise ratio (SNR) per bin, but smooth the polarization features. In our analysis, we select the bin size depending on the necessity, and keep them consistent throughout the analysis. For instance, to search for distinct peaks in the \ion{Si}{ii} $\lambda$6355\,\AA\ line, we require a high spectral resolution and use data with a bin size of 25\,\AA ; to study the $q$--$u$ loops (Sect.~\ref{sect_loops}) we use 50\,\AA\ which is a good compromise between resolution and SNR; and to study different polarization relations, we use 100\,\AA\ in order to increase the SNR.

A list of all peak polarization measurements of the \ion{Si}{ii} $\lambda$6355\,\AA\ line, for different bin sizes, is given in Table~A5 (online supplementary material) and plots of the line polarization and flux spectra of all SNe are presented in the online supplementary material (Appendix~B). Note that at later epochs (i.e. $\gtrsim$ 25 days past peak brightness), the measured degrees of polarization around \ion{Si}{ii} $\lambda$6355\,\AA\ are no longer representing the polarization of the \ion{Si}{ii} feature, because at late epochs the absorption feature becomes increasingly contaminated by emerging lines from different ions \citep[see e.g.][]{2012AJ....143..126B}.
%%\ref{tab:SiII_pol_lines}
%\ref{sect_appendix_tabs_figs}

To fit the \ion{Si}{ii} polarization relations (e.g. in Fig.~\ref{fig:epoch-SiII_2006X}), we adopt the form p$_{\rm \ion{Si}{ii}}$ = a$\times$(t-t$_{\rm max}$)$^2$+p$_{\rm max}$, and use a nested sampling algorithm \citep{2004AIPC..735..395S}. We assume that the prior on the epoch of peak polarization, $t_{\rm max}$, is a uniform distribution between $-$10 and 0 days, the prior on the peak polarization p$_{\rm max}$ variable is a uniform distribution between 0 and 5 per cent, and the prior on the parameter $\rm a$ is a uniform distribution between $-$1 and 0.

\subsection{Analysis of the $q$--$u$ loops}
\label{sect_loops}

Depending on the geometry of the obscuring material, polarized lines sometimes display loops in the Stokes $q$--$u$ plane \citep[Sect.~\ref{sect:line_polarization}, see also][]{2008ARA&A..46..433W}. 
These loops result from variations of the polarization degree as function of wavelength, i.e. velocity, and the polarization angle, due to non-axisymmetric distribution of the ejecta. The shape and orientation of the loops depends on the projected distribution of the absorbers in front of the photosphere \citep{2003ApJ...593..788K,2008ARA&A..46..433W,2017suex.book.....B}.
Multi-epoch spectropolarimetry enables us to study the evolution of the 3D structure of the ejecta, and observe the change of the orientations of different chemical elements with epoch (see e.g. polar diagrams in \citealt{2009ApJ...705.1139M}).
However, detailed analysis of individual SNe Ia is out of the scope of this work and in order to quantify the evolution of the loops in a systematic way, we will consider the loop area only, which preserves useful geometrical information about the scattering region.
%%%%%%

\begin{figure*}
\includegraphics[trim=0mm 0mm 0mm 0mm, width=16.0cm, clip=true]{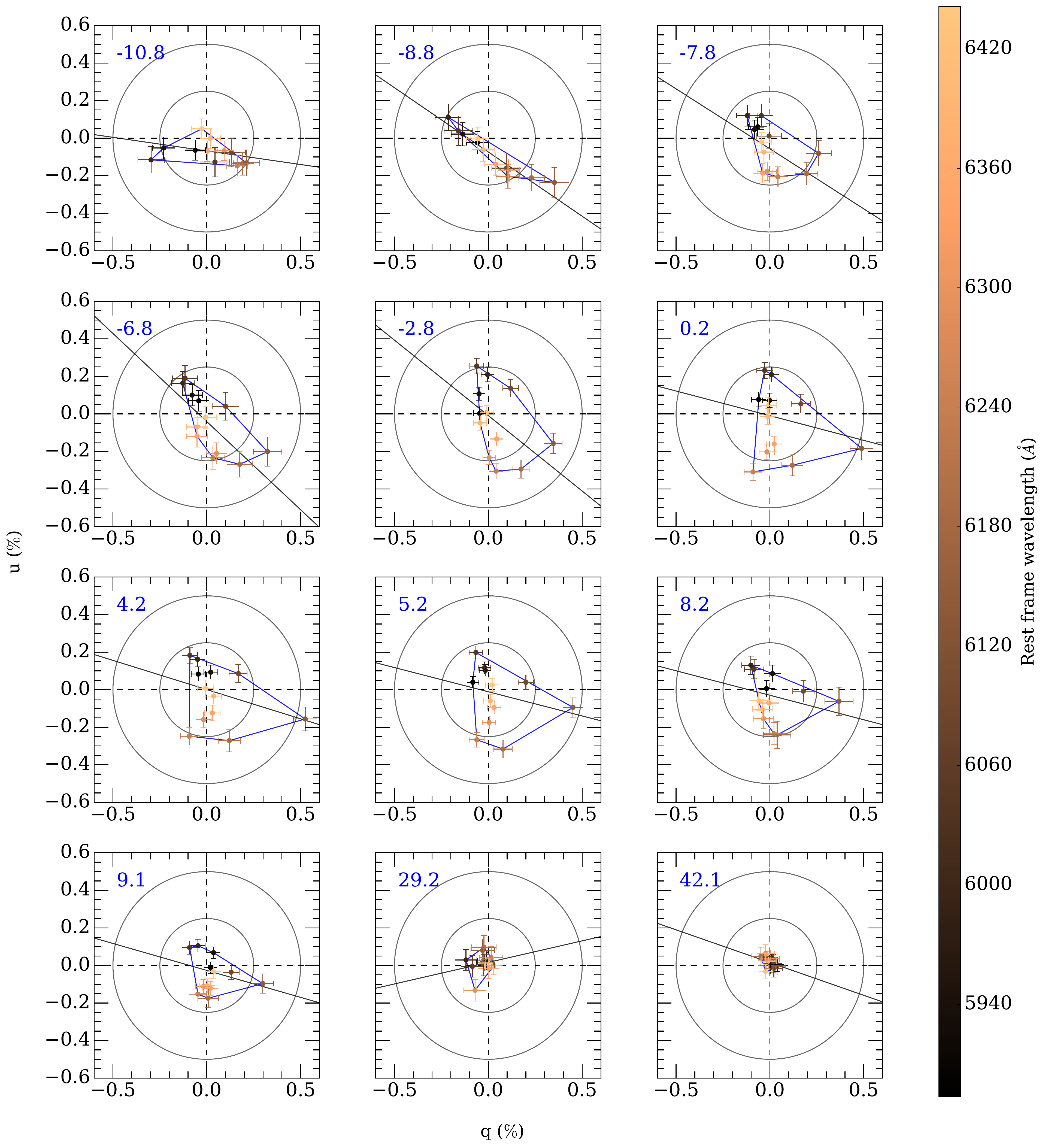}
\vspace{-0mm}
\caption{Evolution of the \ion{Si}{ii} $\lambda$6355\,\AA\ line polarization in the $q$--$u$ plane for SN\,2005df. The bin size of the used data is 50\,\AA , and the wavelengths are color coded. The error bars represent 1$\sigma$ uncertainties. The panels show different epochs relative to peak brightness (indicated in the upper left corner of each insert). The blue lines trace the convex hull used to calculate the area, and the black lines trace the dominant axes computed using the displayed $q$, $u$ data. The inner and outer concentric circles represent a polarization degree of 0.25 and 0.5 per cent, respectively.} 
\label{fig:QU_SN2005df}
\end{figure*}

Examples of loops are shown in Fig.~\ref{fig:QU_SN2005df}. To analyze the loops, we use polarization values calculated with bins of 50\,\AA .
%
%We parametrize those loops by determining the dominant axis \citep[see][]{2008ARA&A..46..433W}, and its slope, and calculate the enclosed area within the loops. To calculate the area, we first construct a convex hull from the ($Q$,$U$) tuples
We parametrize those loops by calculating the enclosed area within the loops. We first construct a convex hull from the ($q$,$u$) tuples. A convex hull of a set of points is the smallest convex set that contains the points \citep{Barber96thequickhull}. Examples of convex hulls, enclosing the loops, are shown in Fig.~\ref{fig:QU_SN2005df} (blue lines). Then we calculate the area of the convex hull using the Shoelace algorithm \citep{ShoelaceAlg}. The area of the loops is given in ``units" of (per cent)$^2$, because it was determined in the $q$--$u$ plane, with the values of $q$ and $u$ given in per cent. For instance, if the loop is a square with 1 per cent side, its area would be 1 (per cent)$^2$. To estimate the error of the area, we run a Monte Carlo simulation by introducing a Gaussian error to the ($q$,$u$) tuples, 10000 times for each epoch, and get a non-Gaussian distribution of the area values. We finally calculate the 25$^{\rm th}$ and 75$^{\rm th}$ percentile of the distribution to estimate the semi-interquartile range.

\section{Results \& discussion}
\label{sect_discussion}

Line polarization has implications for the progenitor model and the explosion mechanism \citep{2008ARA&A..46..433W,2015MNRAS.450..967B}. The most prominent line polarization is observed in the \ion{Si}{ii} $\lambda$6355\,\AA\ line and the near-IR \ion{Ca}{ii} triplet \citep[see][and this work]{2008ARA&A..46..433W}. The maximum degree of polarization of the \ion{Si}{ii} is typically $\sim$1 per cent, reached a few days before peak brightness. In this section we show and discuss the evolution of the \ion{Si}{ii} polarization, and the evolution of the loops in the $q$--$u$ plane, the $\Delta$m$_{15}$--\ion{Si}{ii} polarization relationship, investigate and interpret the \ion{Si}{ii} velocity--polarization relationship, and finally compare our sample to simulations.

\subsection{Time evolution of the \ion{Si}{ii} polarization}
\label{sect_results_epoch-P}

\begin{figure}
\center
\includegraphics[trim=0mm 0mm 0mm 0mm, width=8.5cm, clip=true]{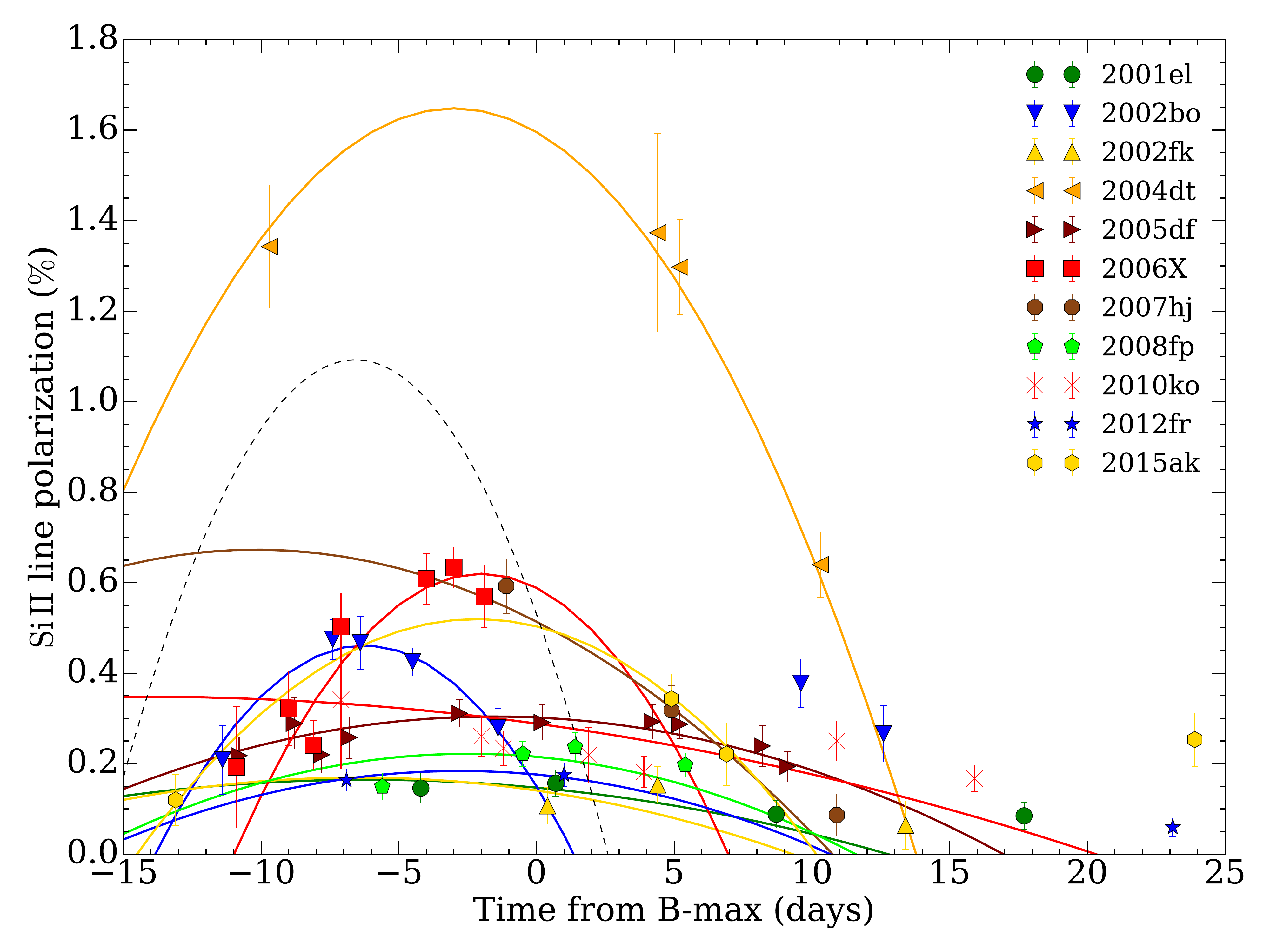}
\vspace{-6mm}
\caption{Time evolution of the peak polarization of the \ion{Si}{ii} $\lambda$6355\,\AA\ line, for a sample of SNe Ia. For comparison, the dotted line shows the p$_{\rm \ion{Si}{ii}}$--epoch fit from \citet{2007Sci...315..212W}. Note that our polarization measurements are systematically lower compared to the time dependence deduced by \citet{2007Sci...315..212W}, because we use 100\,\AA\ bins. 
To fit the polarization relations we adopt the form p$_{\rm \ion{Si}{ii}}$ = a$\times$(t$-$t$_{\rm max}$)$^2$+p$_{\rm max}$ (see Sect.~\ref{sect_line_pol}).}
\label{fig:PSIII-epoch}
\end{figure}

Figure~\ref{fig:PSIII-epoch} shows the time evolution of the peak polarization of the \ion{Si}{ii} $\lambda$6355\,\AA\ line for a selected subset of SNe Ia, that are well sampled in terms of time coverage, measured with a binning of 100\,\AA . The evolution was discussed in \citet{2007Sci...315..212W}, who fit a time dependence for the degree of polarization with a second order polynomial: p$_{\rm \ion{Si}{ii}}$ = 0.65$-$0.041(t$-$5)-0.013(t+5)$^2$, where t is the time in days relative to the peak brightness in the B-band. 
However, as \citet{2016ApJ...828...24P} already noticed, many events are strikingly different in their behavior compared to the polarization-epoch dependence given in \citet{2007Sci...315..212W}. There are different peak polarization values, at a range of different epochs. The \ion{Si}{ii} peak polarization values range from $\sim$0.1 per cent (in the case of, e.g., SN\,2002fk and SN\,2001el) to $\sim$0.6 per cent in the case of SN\,2006X. SN\,2004dt is an exception and has a peak \ion{Si}{ii} polarization greater than $\sim$ 1.4 per cent. The epochs of the peak polarization values range from $\sim$10 days before peak brightness to the time of peak brightness. 

Inspecting the high resolution polarization spectra, binned to 25\,\AA , we noticed that in some cases two distinct \ion{Si}{ii} $\lambda$6355\,\AA\ peaks are visible. The two peaks correspond to a high velocity and a lower velocity photospheric component of the \ion{Si}{ii} line \citep{2005ApJ...623L..37M}. These supernovae are listed in Table~\ref{tab:SNe_with_2peaks} (see also Appendix~B in the online supplementary material). Both polarization peaks are prominent at the listed epochs in Table~\ref{tab:SNe_with_2peaks}, although the two velocity components can often be identified also at other epochs.

% SN\,2002bo at epochs $-$4.5, $-$1.4 and 9.6; SN\,2002el at $-$7.6 days; SN\,2003W at all three observed epochs from $-$8.7 to 16.1 days,  SN\,2005df particularly at epochs $-$8.8 and $-$6.8; SN\,2005el at $-$2.7 (the only observed epoch); SN\,2007fb at 6 days past maximum; SN\,2007hj at 4.9 days; SN\,2008fp at -0.5 and 1.4; and SN\,2010ev at $-$1.1 and 2.9 days. 

\begin{table} %%[h!]  %% dear editor, please move table here.
\centering
\caption{SNe Ia with both high- and photospheric- velocity \ion{Si}{ii} $\lambda$6355\,\AA\ polarization peaks.}
\label{tab:SNe_with_2peaks}
\begin{tabular}{p{1.2cm} p{6.3cm}} 
\hline
Name & Epochs (relative to peak brightness) \\
\hline
SN\,2002bo & $-$4.5, $-$1.4 and 9.6 days\\
SN\,2002el & $-$7.6 days \\
SN\,2003eh & 0.0 days\\
SN\,2003W & all three epochs from $-$8.7 to 16.1 days\\
SN\,2005df & 8 epochs from $-$10.8 and $-$5.2 days\\
SN\,2005el & $-$2.7 days (the only observed epoch)\\
SN\,2007fb & 6.0 days\\
SN\,2007hj & 4.9 days\\
SN\,2008fp & $-$0.5 and 1.4 days\\
SN\,2010ev & $-$1.1 and 2.9 days\\
SN\,2012fr & $-$6.9 and 1.0 days\\
\hline
%\multicolumn{2}{p{11cm}}{text}\\
\end{tabular}
\end{table}

\begin{figure}
\center
\includegraphics[trim=0mm 0mm 0mm 0mm, width=8.5cm, clip=true]{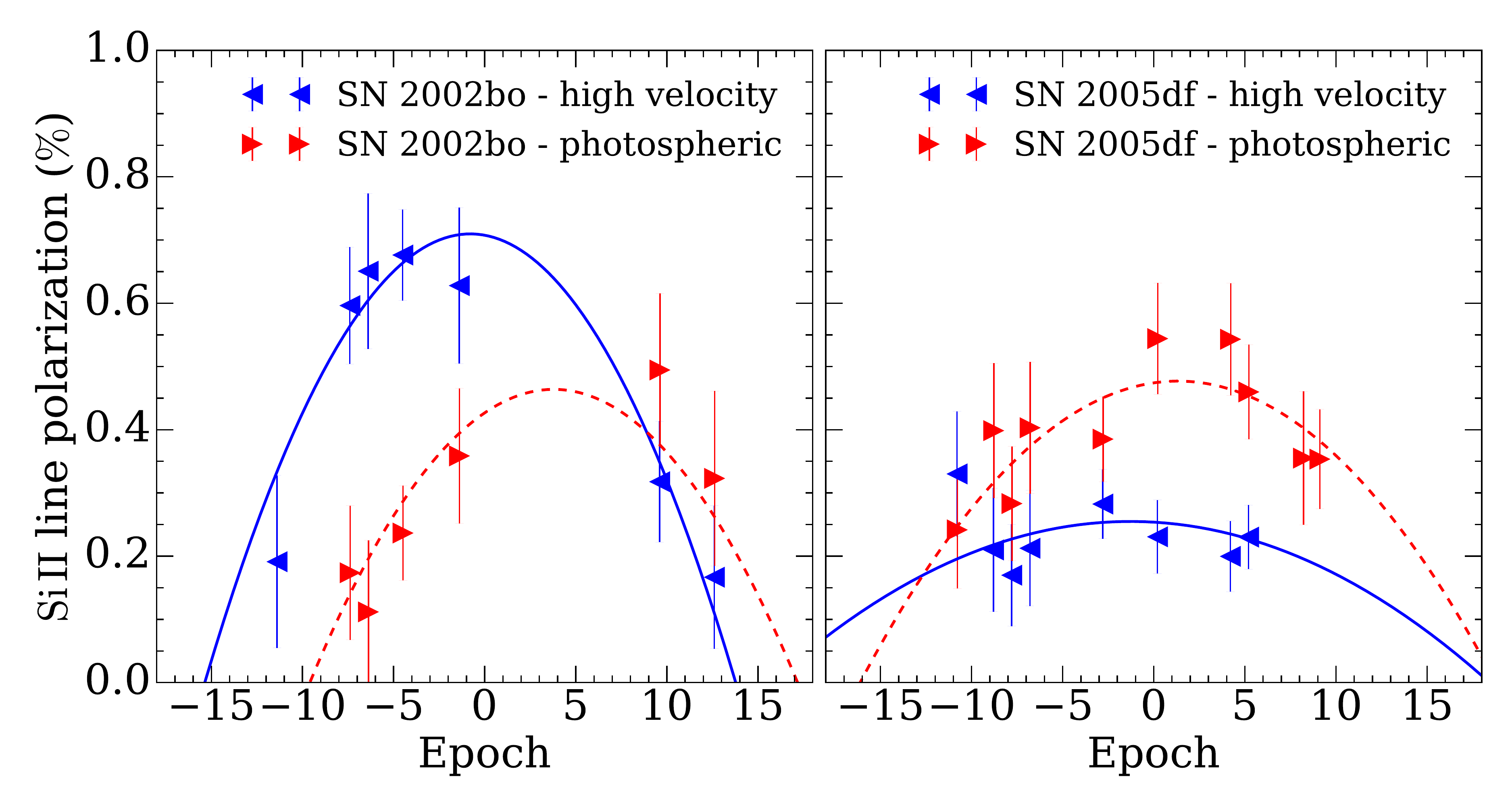}
\vspace{-6mm}
\caption{Time evolution of the peak polarization of the high velocity (blue symbols) and photospheric (red symbols) \ion{Si}{ii} $\lambda$6355\,\AA\ component, for SN\,2002bo (left panel) and SN\,2005df (right panel). The peak polarization of the two components was measured on polarization spectra with 25\,\AA\ bin size. To fit the polarization relations we adopt the form p$_{\rm \ion{Si}{ii}}$ = a$\times$(t$-$t$_{\rm max}$)$^2$+p$_{\rm max}$ (see Sect.~\ref{sect_line_pol}).}
\label{fig:PSiII_hv_phot-epoch}
\end{figure}

For the purpose of our statistical analysis we always measure the highest peak of the \ion{Si}{ii} $\lambda$6355 \,\AA\ line (Sect.~\ref{sect_line_pol}), except in the case of SN\,2002bo and SN\,2005df, which are well sampled (observed at 7 and 12 epochs, respectively) and display distinct high velocity and photospheric components in the \ion{Si}{ii} $\lambda$6355\,\AA\ polarization feature. 
Although it is beyond the scope of the paper to study individual objects in detail, for these two SNe we measured the polarization of the two velocity components separately with the aim of studying their evolution.
To identify the polarization peaks that correspond to the two velocity components, we compared the polarization spectra at different epochs and manually selected the bins that correspond to the high velocity and photospheric velocity component, respectively, if they were visible. 
 The time evolution of the high velocity and photospheric \ion{Si}{ii} $\lambda$6355\,\AA\ component, for SN\,2002bo and SN\,2005df is presented in Fig.~\ref{fig:PSiII_hv_phot-epoch}. 
In SN\,2002bo, between epochs $-$1.4 and 9.6, the polarization of the photospheric component becomes (and stays) more prominent than the high velocity component. The same behaviour is observed in SN\,2007fb between epochs 3.0 and 6.0, as well as in the case of SN\,2010ev between 2.9 and 11.9 days. On the other hand, in SN\,2005df, the high velocity component is more prominent only at the earliest observed epoch of $-$10.8 days and is relatively constant, while the photospheric component is more prominent between epochs $-$10.8 and 9.1 days.

We note that, although the velocity of the \ion{Si}{ii} $\lambda$6355\,\AA\ absorption minima at peak brightness of SN\,2002bo (v$_{\rm Si II}$ = $-$12854 $\pm$ 59 km s$^{-1}$) is higher compared to the \ion{Si}{ii} $\lambda$6355\,\AA\ absorption minima velocity of SN\,2005df (v$_{\rm Si II}$ = $-$9495 $\pm$ 37 km s$^{-1}$), 
the velocities of the high velocity polarization peaks of both SNe are comparable and consistent with the Si velocities of the the "HV" subclass of SNe Ia \citep[v$_{\rm Si} \geqslant$ 12000 km s$^{-1}$,][]{2009ApJ...699L.139W}. The velocities of the high velocity polarization peaks of SN\,2002bo and SN\,2005df are $\sim$ $-$14$\times$ 10$^3$ and $-$15$\times$ 10$^3$ km s$^{-1}$, respectively.

\subsection{${\rm \Delta}$m$_{15}$ -- p$_{\rm Si II}$ relationship}
\label{sect_results_dm15-P}

\citet{2007Sci...315..212W} presented peak polarization measurements of the \ion{Si}{ii} $\lambda$6355\,\AA\ line for a sample of 17 SNe Ia. They found a correlation between the degree of the peak polarization of the \ion{Si}{ii} line and the light-curve decline rate, $\Delta$m$_{15}$:
p$_{\rm \ion{Si}{ii}}$ = 0.48(03) + 1.33(15)($\Delta$m$_{15}$ - 1.1), where $\Delta$m$_{15}$ is the decline in B-magnitude from peak brightness to the brightness 15 days after peak \citep{1993ApJ...413L.105P}, and p$_{\rm \ion{Si}{ii}}$ is in percent.
They argue that this trend provides strong support for delayed-detonation models, as the dimmer SNe, which burn less material to thermonuclear equilibrium, are expected to have larger chemical irregularities (thus larger polarization). This is due to the fact that more complete burning tends to erase chemically clumpy structures, and produces less chemically-asymmetric explosions \citep[see also][]{2008ARA&A..46..433W}. 

\begin{figure*}
\center
\includegraphics[trim=0mm 0mm 0mm 0mm, width=16.5cm, clip=true]{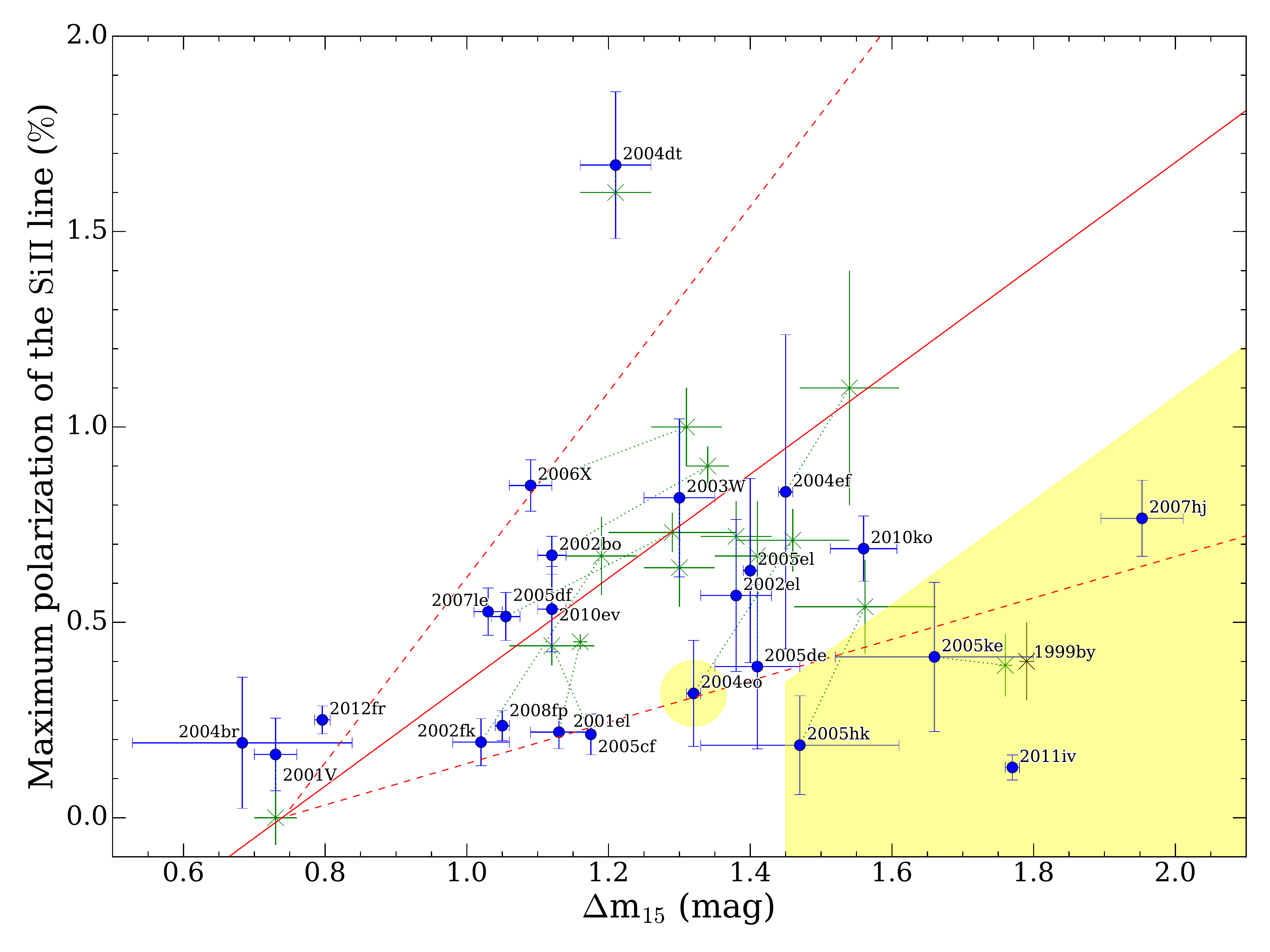}  %% 50AA
\vspace{-5mm}
\caption{Peak polarization of the \ion{Si}{ii} $\lambda$6355\,\AA\ line, measured at 50\,\AA\ binned data, between epochs $-$10 and 0, as a function of $\Delta$m$_{15}$ (blue dots). The red full line is the $\Delta$m$_{15}$--p$_{\rm \ion{Si}{ii}}$ relationship determined by \citet{2007Sci...315..212W}, and the red dashed lines indicate the 1$\sigma$ level of the most likely distribution of polarizations from their Monte Carlo simulation. 
For comparison we show measurements of the same SNe by other authors (green 'x' symbols connected with dotted lines). SN\,2006X, SN\,2005hk and SN\,2005ke have been measured by \citet{2009A&A...508..229P}, \citet{2010ApJ...722.1162M} and  \citet{2012A&A...545A...7P} respectively. All other green 'x' symbols represent the measurements by \citet{2007Sci...315..212W}. The yellow area includes subluminous and transitional objects. For comparison we additionally included the subluminous SN\,1999by \citep[black 'x', ][]{2001ApJ...556..302H,2007Sci...315..212W}. Also note that SN\,2004eo is classified as a transitional supernova.}
\label{fig:SiII-dm15}
\end{figure*}

As part of our analysis, we studied the $\Delta$m$_{15}$-p$_{\rm \ion{Si}{ii}}$ relationship using our larger sample and compared our \ion{Si}{ii} polarization measurements to the literature. Figure~\ref{fig:SiII-dm15} shows peak linear polarization measurements of the \ion{Si}{ii} $\lambda$6355\,\AA\ line, measured from 50\,\AA\ binned data, compared to the $\Delta$m$_{15}$--p$_{\rm \ion{Si}{ii}}$ relationship determined by \citet{2007Sci...315..212W} and measurements by other authors. 

Our linear polarization measurements of the \ion{Si}{ii} line in SN\,2002fk, SN\,2001el, SN\,2005cf, and SN\,2004eo are systematically lower compared to the literature, independent of the binning size, whereas our measured \ion{Si}{ii} polarization in SN\,2004ef is lower in the case of the 100\,\AA\ binned data, and consistent with the literature in the case of 25\,\AA\ and 50\,\AA\ binned data. The SNe in the yellow region of Figure~\ref{fig:SiII-dm15} are subluminous and transitional objects. The 91bg-like SN\,1999by \citep{2001ApJ...556..302H}, the 2002cx-like SN\,2005hk \citep{2010ApJ...722.1162M} and the 1991bg-like SN\,2005ke \citep{2012A&A...545A...7P} are well known subluminous events. SN\,2011iv is a transitional object \citep{2017arXiv170703823G} which has photometric and spectroscopic properties between a normal SN Ia and a subluminous 1991bg-like SN \citep[see also][]{2018MNRAS.476.2905M,2018MNRAS.tmp..611A}. A near-maximum spectrum of SN\,2007hj showed that it is similar to several sub-luminous SNe Ia, characterized by a strong \ion{Si}{ii} absorption feature at 580 nm \citep{2007CBET.1048....2B}. \citet{2012AJ....143..126B} classified SN\,2007hj as CL ("Cool"), which is characterized by deeper \ion{Si}{ii} $\lambda$5972\,\AA\ absorption, often associated with low-luminosity objects. 
Furthermore, \citet{doi:10.1111/j.1365-2966.2007.11700.x} suggest that SN\,2004eo should be considered as a transitional supernova, and \citet{2012AJ....143..126B} classified it as CL.

SN\,2004dt has an exceptionally high peak polarization of the \ion{Si}{ii} line \citep{2006ApJ...653..490W}. Predictions of polarization in the violent merger scenario are in good agreement with such observations \citep{2016MNRAS.455.1060B}, however, the explosion scenario is still debated \citep[see also][]{2007A&A...475..585A}. In addition, SN\,2003eh also shows high peak polarization of the \ion{Si}{ii} line ($\sim$ 0.8 per cent), however, it is not shown in Figure~\ref{fig:SiII-dm15}, because the lack of photometric data prevented us from deriving the decline rate $\Delta$m$_{15}$.\\

After removing the peculiar objects, the remaining data follow the $\Delta$m$_{15}$--p$_{\rm \ion{Si}{ii}}$ relationship, although the scatter is larger than for the measurements presented in \citet{2007Sci...315..212W}. The Pearson correlation coefficient for our sample, $\rho$ = 0.63 (p-value = 0.005), is lower than the coefficient determined by \citet{2007Sci...315..212W}, $\rho$~=~0.87. Note that in Fig.~\ref{fig:SiII-dm15}, we plot the maximum polarization within a range of $-$10 to 0 days relative to peak brightness, but we do not correct the polarization to a specific epoch (e.g., \citealt{2007Sci...315..212W} corrected their observations to -5 days relative to peak brightness using a deduced time dependence of the degree of polarization), because, as discussed in the previous section, the polarization evolves in a variety of different ways and, despite the similarity of some polarization-epoch curves, we did not identify a common pattern that could be used to normalize all of them. Furthermore, we measured the peak polarization of the \ion{Si}{ii} line using a constant bin size, while the binning used in \citet{2007Sci...315..212W} is not given. They collected the peak polarization measurements of the \ion{Si}{ii} line from the literature for a subsample of objects and combined them with their own measurements. Therefore, the bin size used for the measurements presumably differs from object to object.

The correlation between $\Delta$m$_{15}$ and p$_{\rm \ion{Si}{ii}}$ is not expected to be necessarily high, and the dispersion of the polarization may be a valuable diagnostic tool. \citet{2007Sci...315..212W} used a toy model to constrain the number of clumps and the opacity of inhomogeneities based on the observed dispersion. Their model assumes that the clumps are randomly distributed above a spherical photosphere dominated by electron scattering. The size and number of clumps then determine the observable degree of polarization, which is also constrained by the depth of the \ion{Si}{ii} absorption line profile. In such a model, the degree of polarization has a probability density distribution with a peak of about 0.5 per cent for 20 clumps covering the surface of the photosphere. Reducing the number of clumps introduces higher degrees of polarization and a larger dispersion of the degree of polarization. 
The results of their Monte-Carlo simulation of this effect are shown in Figure 2 in \citet{2007Sci...315..212W}, and for comparison in our Fig.~\ref{fig:SiII-dm15}. Our sample of normal SNe Ia is within the 1$\sigma$ level of the most likely distribution of polarizations from their simulation. The large dispersion of the SNe may imply that there is a small number of large clumps (likely large scale plumes or globally asymmetric chemical distributions) in the SN ejecta.

\subsection{Si II Velocity -- polarization relationship}
\label{sect_results_Vel-P}

\begin{figure*}
\center
\includegraphics[trim=0mm 0mm 0mm 0mm, width=16.5cm, clip=true]{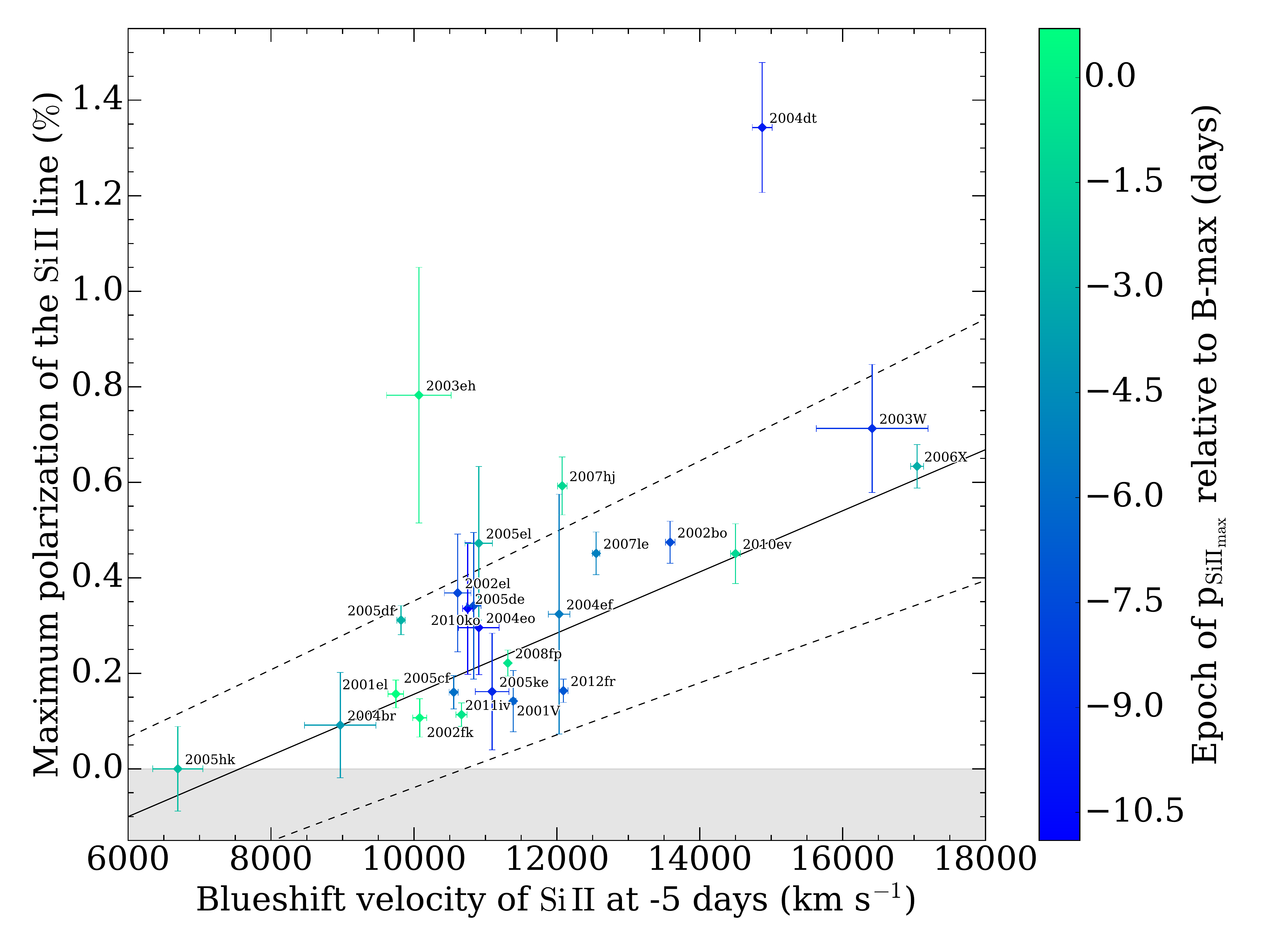}
\vspace{-3mm}
\caption{Maximum linear polarization of the \ion{Si}{ii} $\lambda$6355\,\AA\ line between $-$11 and 1 days relative to peak brightness (measured on polarization spectra of 100\AA\ bin size), as a function of the \ion{Si}{ii} $\lambda$6355\,\AA\ velocity at 5 days before peak brightness. The error bars represent 1$\sigma$ uncertainties. The colors indicate the epoch of the peak \ion{Si}{ii} $\lambda$6355\,\AA\ polarization relative to peak brightness. The black solid line is the linear least square fit to the data, and the dashed lines mark the 1$\sigma$ uncertainty. Note that the relationship is not valid for velocities $\lesssim$ 7500 km s$^{-1}$, because negative degrees of linear polarization are unphysical (gray area).}
\label{fig:SiII-vel}
\end{figure*}

Among the various possible correlations that we explored, we found a strong linear correlation between the degree of peak polarization and the expansion velocity of the \ion{Si}{ii} $\lambda$6355\,\AA\ line. 
Figure~\ref{fig:SiII-vel} shows a subset of our sample of SNe Ia that have at least one observation between $-$11.0 and 1.0 days relative to peak brightness. This selection produces a set of 23 SNe, observed within a relatively short range of time. The plot shows the maximum \ion{Si}{ii} $\lambda$6355\,\AA\ polarization in that period, as a function of the \ion{Si}{ii} $\lambda$6355\,\AA\ velocity at 5 days before peak brightness, v$_{\rm SiII-5}$. The \ion{Si}{ii} velocity has been determined from spectra (as described in Sect.~\ref{sect_expvel}) and the \ion{Si}{ii} polarization degree was measured from spectra of 100\,\AA\ bin size (see Sect.~\ref{sect_line_pol}).

There are two clear outliers: SN\,2004dt, which has been already studied by \citet{2006ApJ...653..490W} and may be the result of a violent merger \citep{2016MNRAS.455.1060B}, and SN\,2003eh. There is no photometry available in the literature for this object; therefore the $\Delta$m$_{15}$ value for this supernova is not known. However, we found spectra taken at six epochs in the literature (see Table~A3 in the online supplementary material), and obtained two spectropolarimetry observations at peak and 12 days past peak brightness. Therefore, the epoch of peak brightness has been estimated from spectra and has a large uncertainty of $\pm$10 days. %A3 \ref{tab:jsonspectrasample}
This object has been identified as a reddened SN Ia by \citet{2003IAUC.8134....2M}. We confirm the classification using our VLT observations and the Supernova Identification (SNID) code \citep{2007ApJ...666.1024B}. The spectrum of SN\,2003eh at peak brightness best matches the spectrum of SN 2008ar at 7 days past peak brightness. The total polarization spectra show a continuum, which rises from red to blue wavelengths, and highly polarized lines, similar to SN 2004dt. The continuum polarization reaches $\sim$1.5 per cent at 4000\,\AA , remains constant at both epochs, and is most likely produced by interstellar or circumstellar dust.

We test the \ion{Si}{ii} velocity -- polarization correlation for different epoch ranges. In the tests  the outliers (SN\,2004dt, SN\,2003eh) are excluded. The results are summarized in Table~\ref{tab:velocity_relaiton_results}. The correlation is in general strong for subsets before the peak brightness ($\rho$ $\gtrsim$ 0.7), and drops for subsets after peak brightness. The black line in Fig.~\ref{fig:SiII-vel} is the linear least squares fit to the data. The best-fit v$_{\rm SiII-5}$--p$_{\rm \ion{Si}{ii}}$ relation can be written as follows:
\begin{equation}
p_{\rm \ion{Si}{ii}} = (6.40\times10^{-5} \pm 1.28\times10^{-5})\times v_{\rm SiII-5} -0.484 \pm 0.147,
\label{sivelrelation}
\end{equation}
where p$_{\rm \ion{Si}{ii}}$ is the peak polarization of the \ion{Si}{ii} $\lambda$6355\,\AA\ line in percent, and $v_{\rm SiII-5}$ is the blueshift velocity of the \ion{Si}{ii} $\lambda$6355\,\AA\ line at $-$5 days relative to peak brightness in km s$^{-1}$. The Pearson correlation coefficient is $\rho$ = 0.8. Note that relationship (\ref{sivelrelation}) is valid only for velocities higher than $\sim$7500 km s$^{-1}$, because negative degrees of linear polarization are unphysical.

{\small
\center
\begin{table}
\caption{Linear regression test results for the p$_{\rm \ion{Si}{ii}}$ vs. \ion{Si}{ii} velocity relationship. No. is the number of included SNe in the Epoch range (days relative to peak brightness). $\rho$ and p-value are the Pearson correlation coefficient and the probability of an uncorrelated dataset, respectively. $\alpha$ and $\beta$ are the slope (in $\%$/mag) and the y-intercept (in $\%$) of a linear least squares fit, respectively.}
\label{tab:velocity_relaiton_results}
\begin{tabular}{lp{0.2cm}p{0.4cm}lll} % four columns, alignment for each
\hline
Epoch range & No. & $\rho$ & p-value  & $\alpha$ ($\times10^{-5}$) & $\beta$\\
\hline
$-$15.0 -- 0.0  & 23 & 0.738 & 5.77$\times10^{-5}$ & 5.93 $\pm$ 1.38 & $-$0.436 $\pm$ 0.161 \\
$-$11.0 -- 1.0  & 23 & 0.804 & 3.71$\times10^{-6}$ & 6.40 $\pm$ 1.28 & $-$0.484 $\pm$ 0.147 \\
$-$10.0 -- 0.0  & 21 & 0.803 & 1.17$\times10^{-5}$ & 6.39 $\pm$ 1.37 & $-$0.483 $\pm$ 0.159 \\
$-$10.0 -- $-$5.0 & 13 & 0.756 & 2.77$\times10^{-3}$  & 6.53 $\pm$ 2.16 & $-$0.536 $\pm$ 0.259 \\
$-$5.0 -- 0.0   & 13 & 0.812 & 7.41$\times10^{-4}$  & 6.34 $\pm$ 1.40 & $-$0.453 $\pm$ 0.163 \\
0.0 -- 5.0    & 13 & 0.391 & 0.187    & 2.45 $\pm$ 2.00 & $-$0.091 $\pm$ 0.223 \\
0.0 -- 10.0   & 15 & 0.375 & 0.169    & 3.98 $\pm$ 2.09 & $-$0.275 $\pm$ 0.233 \\
5.0 -- 10.0   & 9  & 0.404 & 0.281    & 4.21 $\pm$ 3.80 & $-$0.313 $\pm$ 0.411 \\
10.0 -- 20.0  & 13 & 0.806 & 0.87$\times10^{-3}$  & 5.38 $\pm$ 1.35 & $-$0.440 $\pm$ 0.149 \\
20.0 -- 30.0  & 4  & 0.560 & 0.44     & 1.19 $\pm$ 3.90 & $-$0.054 $\pm$ 0.460 \\
\hline
\end{tabular}
\end{table}
}

The interpretation of the velocity-polarization relation is not completely straightforward. In general, Fig.~\ref{fig:SiII-vel} connects the kinematics with ejecta asymmetry. If the relation is satisfactorily fit with one single linear function, it would imply that the geometric structure of the SNe is quite consistent and the differences in the polarization degree are caused largely by simple effects such as geometric projection towards the observer, or strong correlation between the observed ejecta velocity and absorption depth with ejecta asymmetry. If the dispersion of polarization across the fitted mean line of v$_{\rm \ion{Si}{ii}}$--p$_{\rm \ion{Si}{ii}}$ is low, it might suggest that the number of lumps covering the photosphere is either very small or so large that effectively only a global asymmetry matters.

% \citep{2009ApJ...699L.139W} velocity - EW relationship

\subsubsection{Off-center delayed-detonation model}

The velocity-polarization relationship may be explained in the context of a combination of an off-center delayed-detonation (DDT) model and brightness decline \citep{2006NewAR..50..470H}. This can be understood with the following considerations. If Chandrasekhar mass supernovae, for the range of normally bright to subluminous supernovae\footnote{Note that this statement is strongly debated, e.g., \citet{2018ApJ...852L..33G} suggest that only sub-Chandrasekhar mass explosions can reproduce the fast decline rates $\Delta$m$_{15}$>1.55, while all SN Ia models (Chandrasekhar, sub-Chandrasekhar and some super-Chandrasekhar mass white dwarf explosions) can reproduce the slow decline rates, 0.8<$\Delta$m$_{15}$<1.55.}, the peak brightness depends on the amount of $^{56}$Ni produced in the explosion \citep{2002ApJ...568..791H,2006NewAR..50..470H}, then intermediate mass elements (including silicon) are produced in larger quantities with decreasing peak brightness (i.e. increasing $\Delta$m$_{15}$).

A reasonable scenario involves the dynamical structure of the ejecta. For example, in the context of delay-detonation models, the \ion{Si}{ii} line polarization is formed when the underlying Thomson photosphere is covered by Si-rich matter with high optical depth. Silicon and sulfur are produced in the quasi-statistical equilibrium (QSE) group in oxygen burning, flanked by explosive carbon and elements produced in nuclear statistical equilibrium (NSE). Therefore, the Si polarization is formed after the photosphere passed the O/Ne/Mg -- Si/S interface. For normal SNe Ia this happens $\sim$ 7 to 5 days before peak brightness, while for sub-luminous supernovae the photosphere recedes faster in mass and the interface happens at higher density \citep{2002ApJ...568..791H}.

If for a given SN Ia explosion the Si line polarization is formed further out (higher velocity) with less mass involved, the polarization may be higher. The reason is that off-center delayed detonation in SNe Ia does not always occur at the same distance from the SN center. Larger off-center distances of DDT lead to more Si at high velocities \citep[see][]{2002ApJ...568..791H,2006NewAR..50..470H}.
Furthermore, as discussed in \citet{2007Sci...315..212W}, higher silicon polarization values suggest that there are more compositional irregularities left in the central region by the pure deflagration phase, and thus less material is burned to thermonuclear equilibrium.

Figure~\ref{fig:dm15-velSiII-pol} shows SNe Ia in the $\Delta$m$_{15}$ -- v$_{\rm \ion{Si}{ii}}$ plane. The color coded dots (depending on the \ion{Si}{ii} polarization) are SNe in our VLT sample. The \ion{Si}{ii} velocities correspond to the velocity at peak brightness\footnote{We decided to plot the \ion{Si}{ii} velocity at peak brightness instead of the velocity at 5 days before peak brightness, in order to be consistent with other works (see discussion below).}.
For comparison, we plot an additional sample of $\sim$ 100 SNe Ia from the Open Supernova Catalog \citep[gray dots,][]{2017ApJ...835...64G} that includes well observed SNe Ia with spectra close to day 0 and a sufficient number of photometric measurements.
We fit their light curves with SNooPy and determine $\Delta$m$_{15}$ as described in Sect.~\ref{sect_lightcurvefitting}, and measure the \ion{Si}{ii} velocity as explained in Sect.~\ref{sect_expvel}. SNe with normal \ion{Si}{ii} velocities ($\sim$ 9000-12000 km s$^{-1}$) are distributed across the whole range of $\Delta$m$_{15}$ values, while the SNe with high silicon velocities ($\gtrsim$ 12000 km s$^{-1}$) are clustered around $\Delta$m$_{15} \sim$ 1.1 mag (Fig.~\ref{fig:dm15-velSiII-pol}). This can be explained as a consequence of an asymmetric delayed detonation scenario. Models show that, for a given $\Delta$m$_{15}$, the asymmetric detonation pushes Si out aspherically (predominantly in one direction), and line polarization produced by Si starts to form in the regions with higher velocities. For example, at day $-$5 in normal and subluminous SNe we observe line polarization produced by the outermost material only, and not the entire ejected Si material (see off-center delayed detonation models, \citealt{fesen_2004}, \citealt{2006NewAR..50..470H}, and chemical structures in \citealt{2002ApJ...568..791H}). This is the high velocity end in the $\Delta$m$_{15}$ -- v$_{\rm \ion{Si}{ii}}$ relation. Furthermore, it is expected that the \ion{Si}{ii} velocity increases with decreasing $\Delta$m$_{15}$ \citep[see][]{2002ApJ...568..791H}.

\subsubsection{Double-detonation model}

\begin{figure*}
\center
\includegraphics[trim=0mm 0mm 0mm 0mm, width=16.5cm, clip=true]{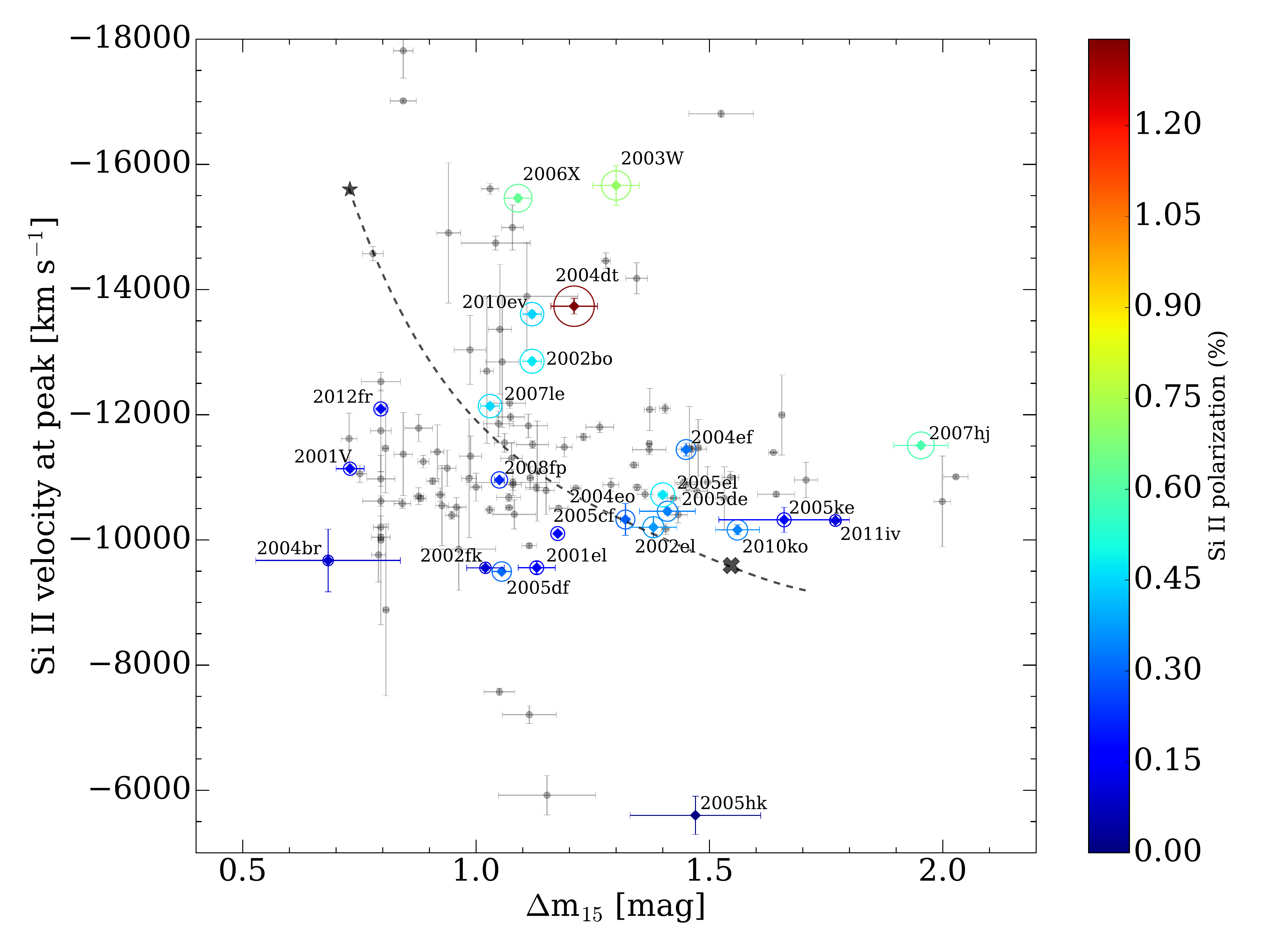}
\vspace{-3mm}
\caption{Supernovae Ia in the $\rm \Delta m_{15}-v_{\rm \ion{Si}{ii}}$ plane. The \ion{Si}{ii} velocity corresponds to the velocity at peak brightness. The colored dots represent the VLT sample. The color of the dots and circles indicates the maximum polarization degree of the \ion{Si}{ii} $\lambda$6355\,\AA\ line measured between $-$11 and 1 days relative to peak brightness, on polarization spectra of 100\AA\ bin size. The radius of the circles is proportional to the polarization degree. The dashed line corresponds to numerical predictions for sub-Chandrasekhar white dwarfs with masses from 0.9 M$_{\odot}$ ($\star$) to 1.2 M$_{\odot}$ ($\times$), and a thin 0.01 M$_{\odot}$ He shell \citep{2018arXiv181107127P}. A sample of archival SNe Ia (gray dots) from The Open Supernova Catalog is shown for comparison.}
\label{fig:dm15-velSiII-pol}
\end{figure*}

Although the off-center delayed detonation explosion scenario provides a good explanation of the observations, a double-detonation scenario cannot be excluded. \citet{2018arXiv181107127P} perform an analysis of the numerical parameter space of 1-D sub-Chandrasekhar mass white dwarf explosion models and found a correlation between mass, brightness and the \ion{Si}{ii} velocity exhibited in these double-detonation models with thin helium shells. 
\citet{2018arXiv181107127P} predicted the \ion{Si}{ii} line velocity as function of the absolute peak brightness in B-band, M$\rm _B$, for sub-Chandrasekhar white dwarfs with a thin 0.01 M$_{\odot}$ He shell and a range of white dwarf masses (see Fig.~11 in \citealt{2018arXiv181107127P}). 
In Fig.~\ref{fig:dm15-velSiII-pol} we included these numerical predictions\footnote{We used the \citet{1993ApJ...413L.105P} relationship to convert M$\rm _B$ from \citet{2018arXiv181107127P} to $\Delta$m$_{15}$.} \citep[dashed line, ][]{2018arXiv181107127P}. 
The dashed line traces the $\Delta$m$_{15}$--v$_{\rm \ion{Si}{ii}}$ predictions for different white dwarf masses: from 0.9 M$_{\odot}$ at the faint end (i.e. large $\Delta$m$_{15}$ values), to 1.2 M$_{\odot}$ at the bright end (see Fig.~11 in \citealt{2018arXiv181107127P}).

\begin{figure}
\center
\includegraphics[trim=0mm 0mm 0mm 0mm, width=8.5cm, clip=true]{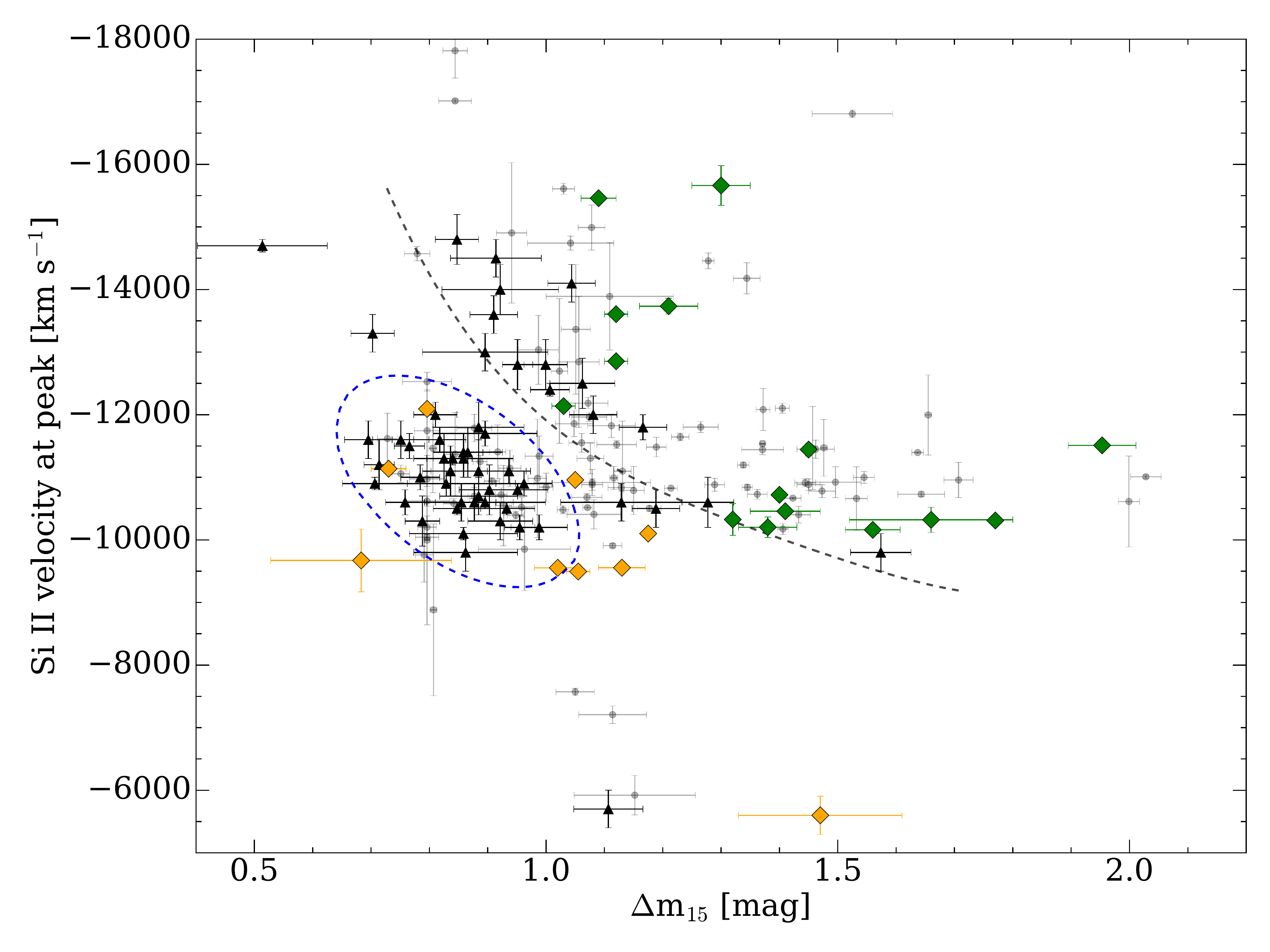}
\put(-30,150){(a)}  \\
\includegraphics[trim=0mm 0mm 0mm 0mm, width=8.5cm, clip=true]{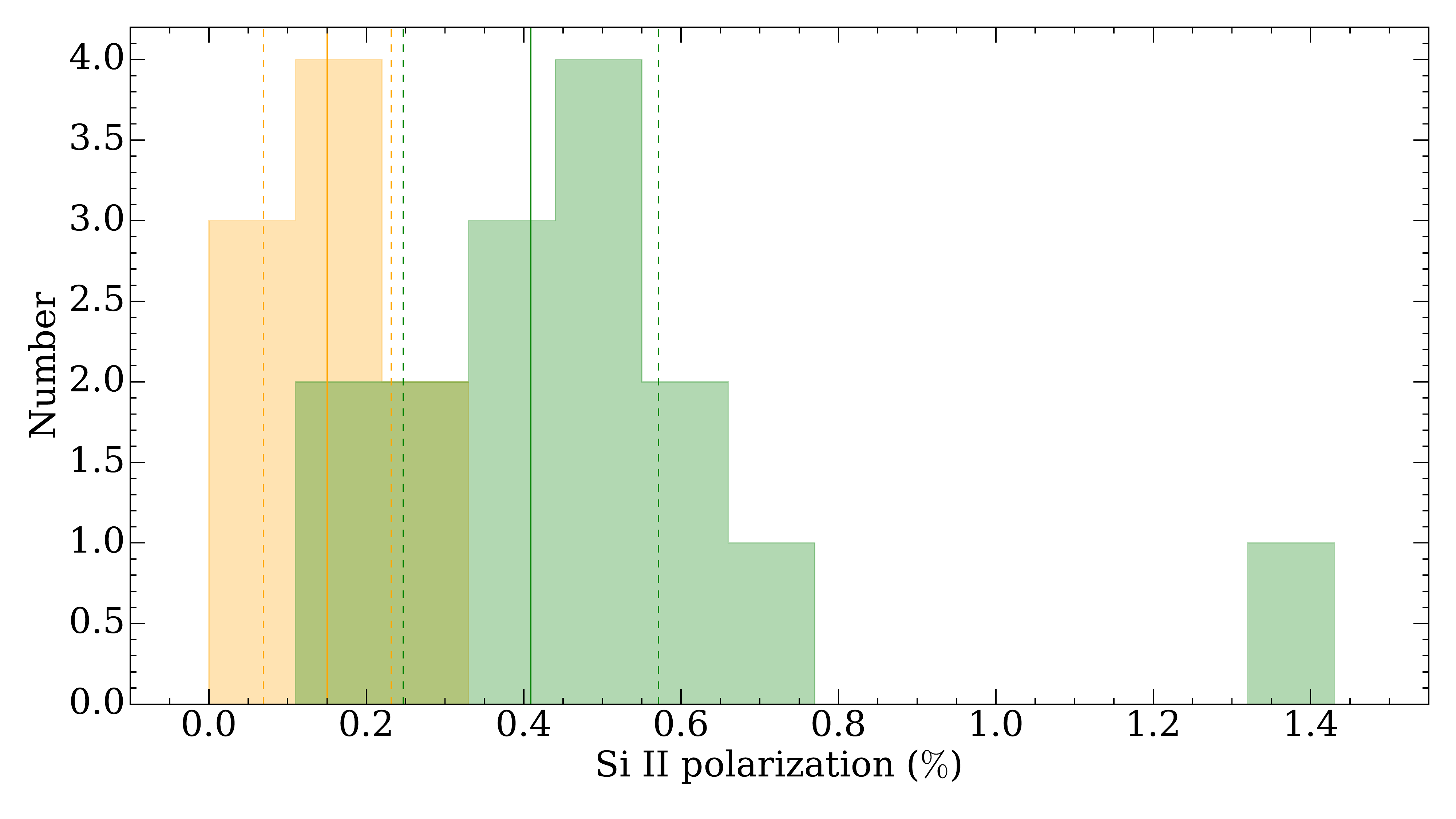}
 \put(-30,110){(b)} 
\vspace{-3mm}
\caption{\textit{Panel a:} SNe Ia in the $\rm \Delta m_{15}-v_{\rm \ion{Si}{ii}}$ plane. The orange and green diamonds represent the VLT sample, which is divided in Chandrasekhar-like (orange diamonds), and sub-Chandrasekhar SNe (green diamonds).
The sample from \citet{2018ApJ...858..104Z} (black dots) is consistent with the sample from the Open Supernova Catalog (gray dots). The dashed blue ellipse marks the ``cluster" 
of Chandrasekhar-like SNe, which may be a distinct population of SNe Ia, relative to the sub-Chandrasekhar SNe that are located along the numerical predictions for sub-Chandrasekhar white dwarfs with a thin He shell \citep[gray dashed line,][]{2018arXiv181107127P}.
\textit{Panel b:} Distribution of the peak polarization of the \ion{Si}{ii} line for the VLT sample in Fig.~\ref{fig:histo_SiII-pol}a located above the $\Delta$m$_{15}$--v$_{\rm \ion{Si}{ii}}$ prediction of \citet{2018arXiv181107127P} (green color), compared to the sub-sample in the ``cluster" (orange color), under the numerical prediction. The average polarization of the SNe in the cluster is p$_{\rm \ion{Si}{ii}}$ = 0.15 $\pm$ 0.08 per cent, and the average polarization of the SNe above the $\Delta$m$_{15}$--v$_{\rm \ion{Si}{ii}}$ prediction is p$_{\rm \ion{Si}{ii}}$ = 0.41 $\pm$ 0.16 per cent (after excluding SN\,2004dt). }
\label{fig:histo_SiII-pol}
\end{figure}

\citet{2018arXiv181107127P} compared their numerical predictions to a sample of well observed SNe \citep{2018ApJ...858..104Z} and found evidence of two distinct populations in the \ion{Si}{ii} velocity -- M$\rm _B$ plane, either having masses close to the Chandrasekhar, or sub-Chandrasekhar masses. Fig.~\ref{fig:histo_SiII-pol}a shows a direct comparison between the \citet{2018ApJ...858..104Z}, the Open Supernova Catalog and our VLT sample in the $\Delta$m$_{15}$--v$_{\rm \ion{Si}{ii}}$ plane. 
The blue ellipse denotes the cluster of SNe that are suggested to be exploding Chandrasekhar mass white dwarfs, and the other SNe that fall uniformly above the numerical predictions are suggested to be sub-Chandrasekhar mass white dwarfs \citep{2018arXiv181107127P}. 
The three samples coincide with each other, except that our sample from the Open Supernova Catalog includes a broader brightness range of supernovae, from normal to subluminous supernovae. Therefore, the two possible populations of SNe do not appear as distinct in the Open Supernova Catalog sample (gray dots), as they do in the \citet{2018ApJ...858..104Z} sample (black dots). 

We note that neither of the samples perfectly traces the $\Delta$m$_{15}$--v$_{\rm \ion{Si}{ii}}$ numerical prediction. However, following \citet{2018arXiv181107127P} we grouped supernovae into the ``cluster" of objects that have a normal peak brightness and normal \ion{Si}{ii} velocities (orange diamonds in Fig.~\ref{fig:histo_SiII-pol}a); and those that are suggested to be sub-Chandrasekhar mass white dwarfs (and fall above the prediction, green diamonds). 
Interestingly, the supernovae with higher polarization of the \ion{Si}{ii} line tend to lie above the numerical $\Delta$m$_{15}$--v$_{\rm \ion{Si}{ii}}$ predictions for sub-Chandrasekhar mass explosions by \citet{2018arXiv181107127P}, compared to supernovae in the ``cluster", below the $\Delta$m$_{15}$--v$_{\rm \ion{Si}{ii}}$ line (Fig.~\ref{fig:histo_SiII-pol}a), that have lower polarization degrees of the \ion{Si}{ii} line. In Fig.~\ref{fig:histo_SiII-pol}b, we show the \ion{Si}{ii} line polarization distribution for SNe of the two distinct groups. The average polarization of the SNe in the ``cluster" is p$_{\rm \ion{Si}{ii}}$ = 0.15 $\pm$ 0.08 per cent, and the average polarization of the sub-Chandrasekhar mass SNe is p$_{\rm \ion{Si}{ii}}$ = 0.41 $\pm$ 0.16 per cent (after excluding SN\,2004dt).

% 0.15 0.08
% 0.47 0.28
%0.41 0.16

\subsubsection{Other models}

Alternatively, the large scale asymmetry of the \ion{Si}{ii} line formation region as inferred from this dataset, may also be explained in the context of merging scenarios. It is not clear how this large scale asymmetry above the \ion{Si}{ii} layer is related to the rather spherical photosphere post optical maximum. \citet{2016MNRAS.455.1060B} investigated polarization signatures for the violent merger model by \citet{2012ApJ...747L..10P}. \citet{2016MNRAS.455.1060B} could not reproduce the low levels of \ion{Si}{ii} line polarization typically observed in normal SNe Ia. However, the ejecta morphologies for other merger scenarios, e.g. the tidal mergers involving an accretion phase \citep[e.g.][]{1990ApJ...348..647B} or violent head-on collisions (e.g. \citealt{2009ApJ...705L.128R}, see also \citealt{2018PhR...736....1L} for summary of merger scenarios) may produce different polarization signatures. These data provide new observational diagnostics for more theoretical understandings of SN\,Ia physics.

\subsection{Si II $q$--$u$ loops}
\label{sect_results_QUloops}

Some SNe included in our sample draw loops in the $q$--$u$ plane in the wavelength range around the \ion{Si}{ii} $\lambda$6355\,\AA\ line (see Appendix~B in the online supplementary material). %\ref{sect_appendix_tabs_figs}
These SNe are listed in Table~\ref{tab:SNe_with_loops}.
Other SNe may show a loop on at least one epoch, but they are below the noise level and cannot be analyzed in a statistically meaningful way.

\begin{table} %%[h!]  %% dear editor, please move table here.
\centering
\caption{SNe Ia with loops in the $q$--$u$ plane in the wavelength range around the \ion{Si}{ii} $\lambda$6355\,\AA\ line}
\label{tab:SNe_with_loops}
\begin{tabular}{p{1.2cm} p{6.3cm}} 
\hline
Name & Epochs (relative to peak brightness) \\
\hline
SN\,2005df & 8 epochs from $-$7.8 to 9.1 days \\
SN\,2006X & 5 epochs from $-$8.1 to $-$1.9 days\\
SN\,2002bo & 5 epochs from $-$7.4 to 9.6 days\\
SN\,2004dt & 4 epochs from 4.4 to 33.2 days\\
SN\,2007fb & 3 and 6 days\\
SN\,2010ko & 1.9 and 10.9 days\\
SN\,2015ak & clear loop at 4.9 days\\
SN\,2010ev & 11.9 days \\
SN\,2011ae & 4 days\\
SN\,2007hj & $-$1.1 days \\
SN\,2005ke & $-$9.1 days \\
SN\,2005cf & $-$5.8 days\\
SN\,2001el & clear but small loops at $-$4.2 and 0.7 days\\
\hline
\end{tabular}
\end{table}
\rm 

\begin{figure}
\center
\includegraphics[trim=0mm 0mm 0mm 0mm, width=8.5cm, clip=true]{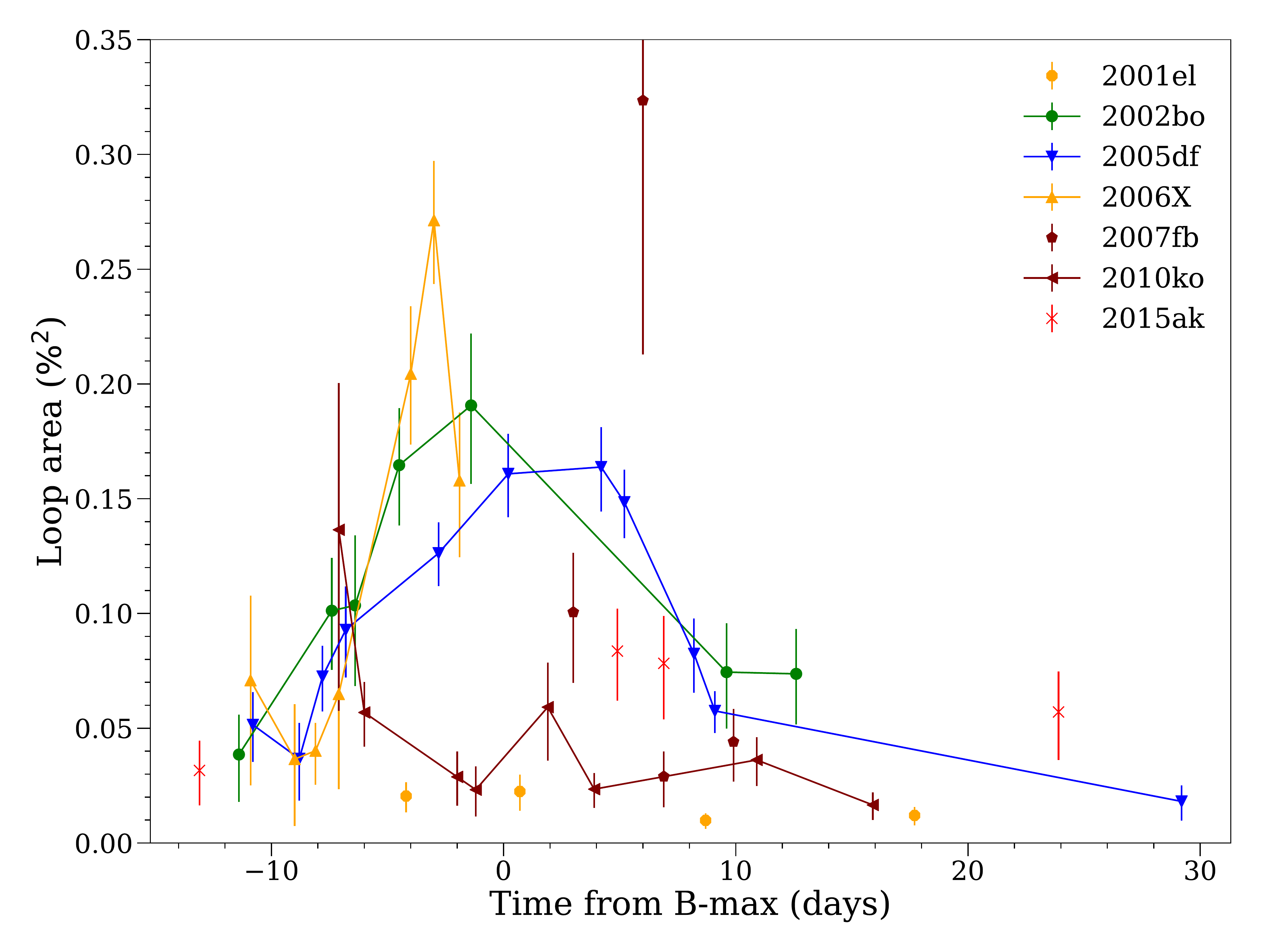}
\vspace{-3mm}
\caption{The area contained in $q$--$u$ loops as a function of epoch. Shown are all SNe that have been observed on at least four epochs (SN\,2004dt is off scale in area). The lines connect measurements of SNe that have been observed on five epochs or more. The error bars represent the semi-interquartile range.}
\label{fig:epoch-looparea}
\end{figure}

To characterize in a quantitative way the evolution of the loops we calculated the area inside of the loops, using polarization spectra binned to 50\,\AA, as described in Sect.~\ref{sect_loops}. Figure~\ref{fig:epoch-looparea} shows the area, A, as a function of epoch, for a subsample of SNe observed on at least 4 epochs before 30 days past peak brightness. The largest loops were observed in SN\,2006X, SN\,2002bo, and SN\,2005df. The area of the loops in these SNe smoothly evolves and is largest between approximately $-$5 and $\sim$5 days relative to the peak brightness. The peculiar SN\,2004dt is located outside of the axis limits (A$>$0.32 (per cent)$^2$, up to A$\sim$1.75\,(per cent)$^2$ at 4.4 days past peak).

%Mattia's simulations:\\
To explore the effect of clumps on the polarization spectra and $q$--$u$ plane we run simple simulations (see \citealt{2017PhDT.........1B} for more details). In the simulations we make the following assumptions: 

\begin{itemize}
 \itemsep0em
\item spherical ejecta
\item photosphere at 10000 km s$^{-1}$
\item roughly 0.5 M$_{\odot}$ above the photosphere
\item outer boundary: v$_{\rm outer}$ $\sim$ 20 000 km s$^{-1}$
\item density power law index = 7
\item density at the photosphere = 7$\times 10^{-14}$ g cm$^{-3}$
\item electron scattering optical depth: $\tau_{\rm es}$ $\sim$ 0.6 
\item line opacity outside clumps: $\tau_{\rm line}$ = 1
\item line opacity inside clumps: $\tau_{\rm line clumps}$ = 10
\end{itemize}

\begin{figure*}
\center
\includegraphics[trim=170mm 0mm 0mm 0mm, width=4.0cm, clip=true]{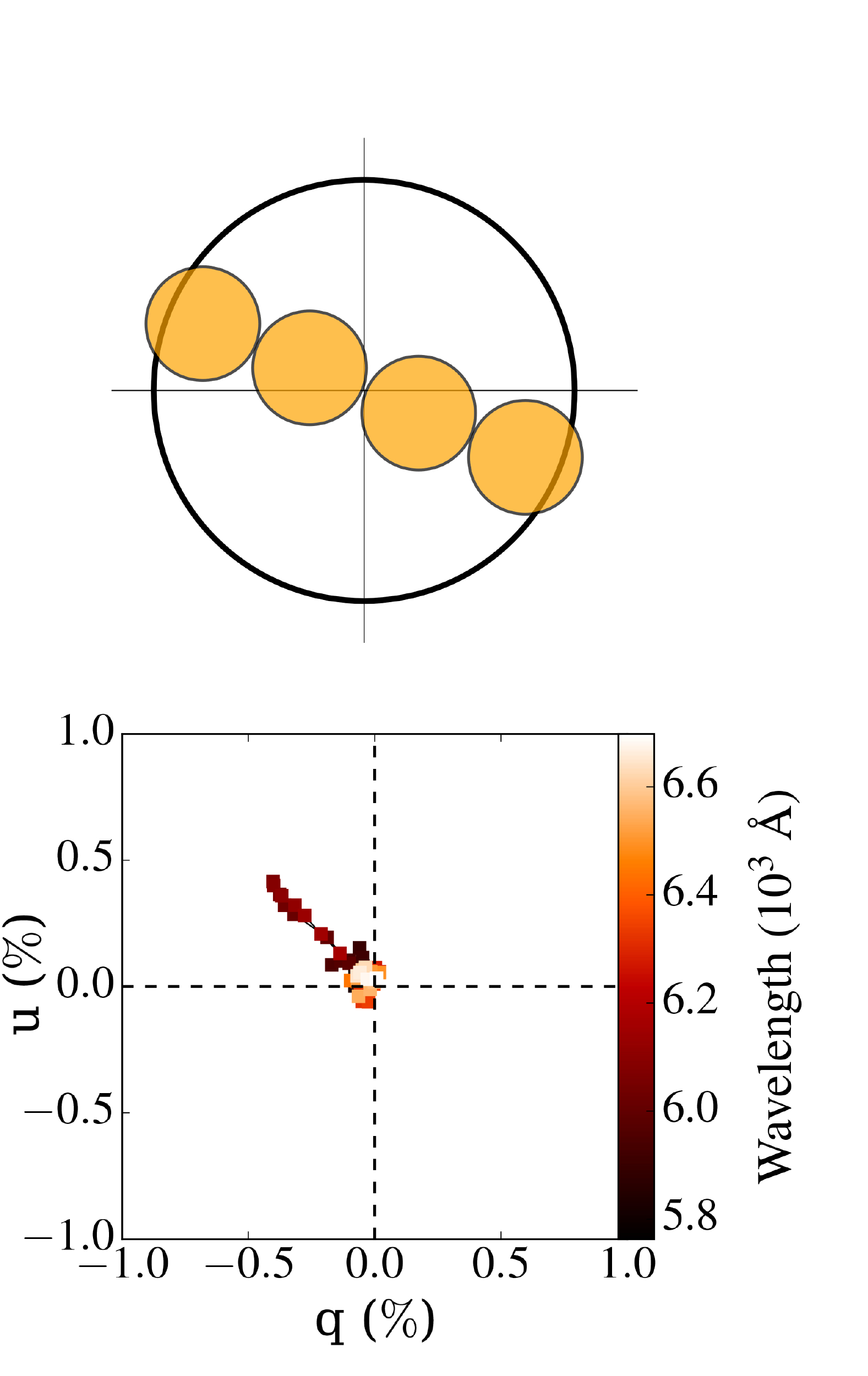}
\includegraphics[trim=170mm 0mm 0mm 0mm, width=4.0cm, clip=true]{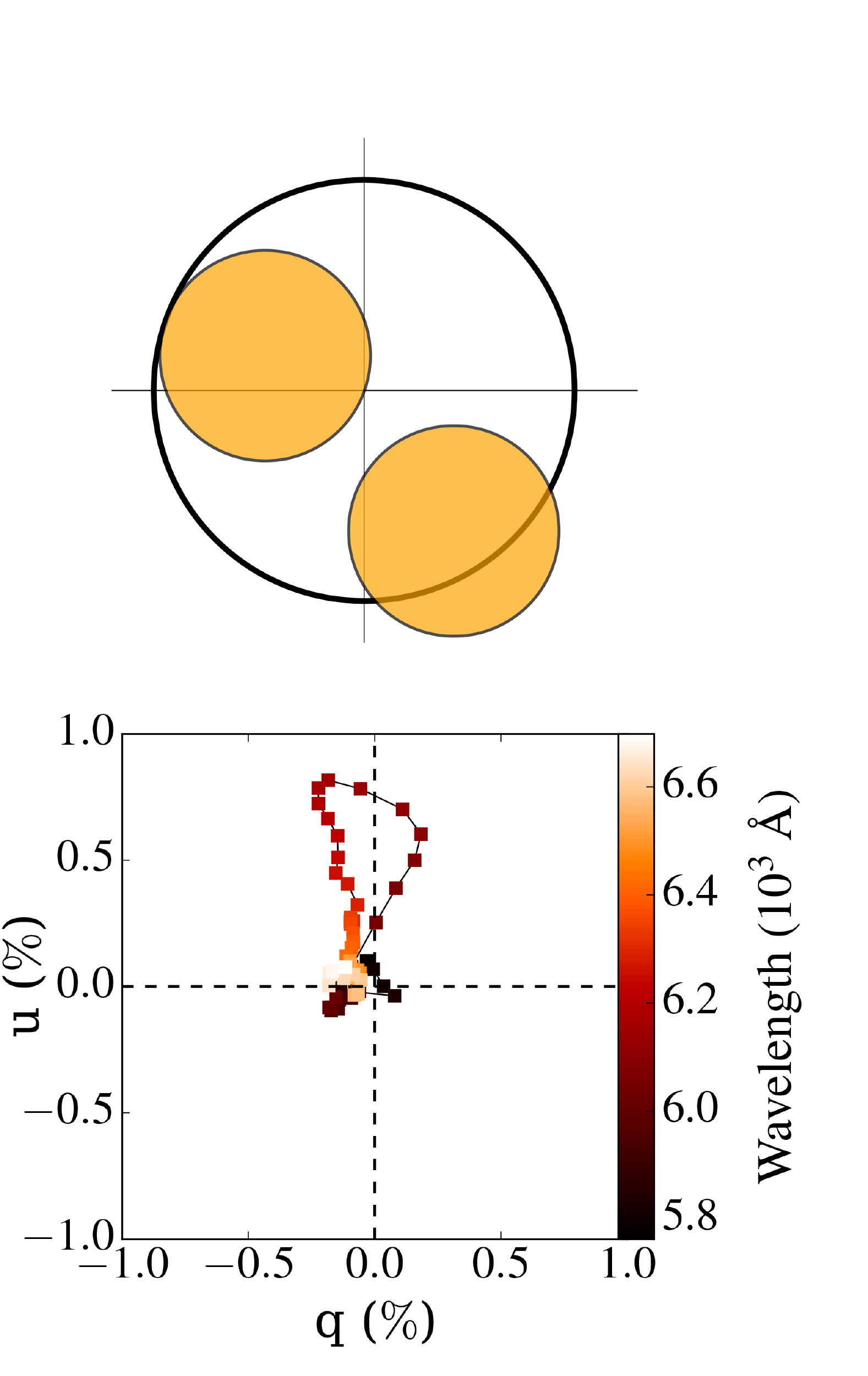}
\includegraphics[trim=170mm 0mm 0mm 0mm, width=4.0cm, clip=true]{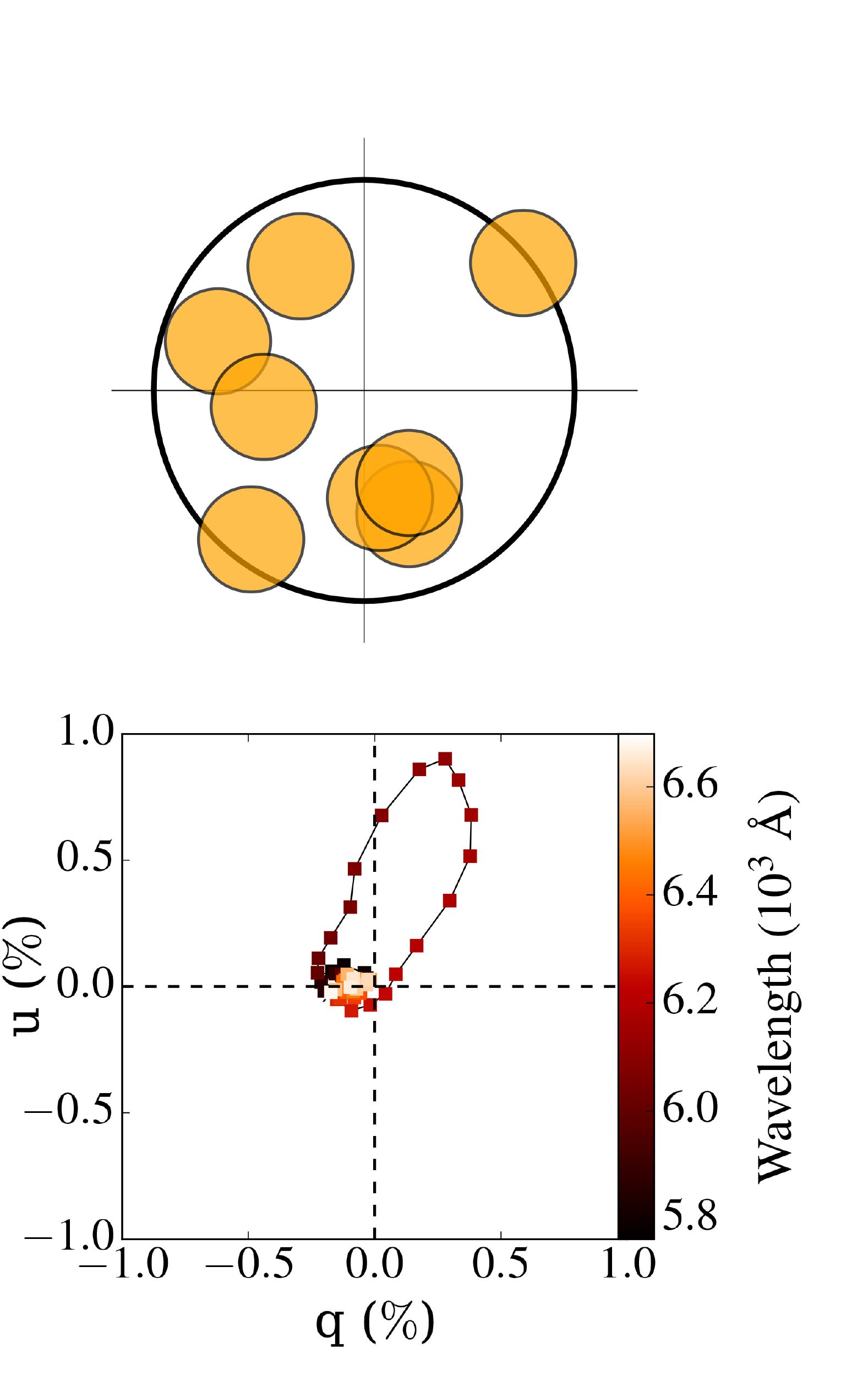}
\includegraphics[trim=170mm 0mm 0mm 0mm, width=4.0cm, clip=true]{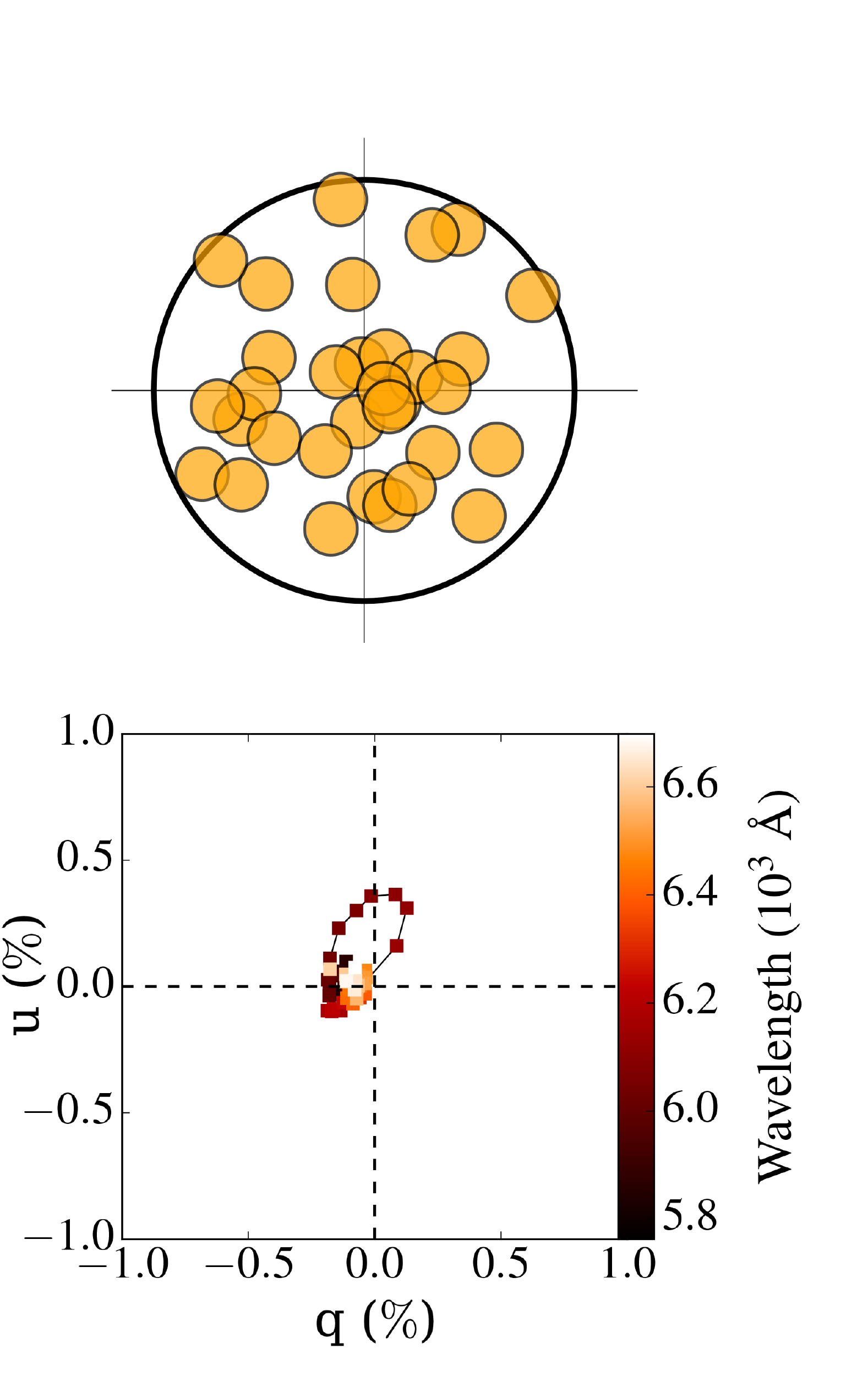}
\put(-450,160){(a)}
\put(-330,160){(b)}
\put(-220,160){(c)}
\put(-100,160){(d)}
\caption{Simulations of polarization induced by clumps, for four different scenarios. The upper panels illustrate the size and distribution of clumps (filled orange circles) in front of the photosphere (large circle), while the bottom panels show the resulting polarization in the $q$--$u$ plane. Panel (a) shows four clumps axisymmetrically distributed. It produces polarization, but no loops in the $q$--$u$ plane. The other three panels (b,c,d) illustrate randomly distributed clumps, of different sizes. The smallest clumps produce a small loop, compared to the clumps of intermediate and large size.}
\label{fig:loops_simulaitons}
\end{figure*}

The clump positions are randomly generated in regions of the ejecta moving toward the observer (blueshifted velocities). The signs of $q$ and $u$ are defined as in \citet{2015MNRAS.450..967B}.
Also, the clump centers are randomly generated between the inner (photosphere) and outer boundary. Depending on the sizes, some regions of the clumps could actually fall either inside the photosphere or outside the ejecta (see upper panels in Figure~\ref{fig:loops_simulaitons}). 

Figure~\ref{fig:loops_simulaitons} shows simulations for four different scenarios. The left panel presents the case of four clumps with an axisymmetric distribution, to simulate the effect of a torus-like geometry. The clump sizes (in velocity space) are v$_{\rm outer}$/4.0 = 5000 km s$^{-1}$.
The other three panels show randomly distributed clumps, from left to right: 2 clumps with radius v$_{\rm outer}$/2 = 10 000 km s$^{-1}$, 8 clumps with radius v$_{\rm outer}$/4 = 5000 km s$^{-1}$, and 32 clumps with radius v$_{\rm outer}$/8 = 2500 km s$^{-1}$. The total covering factor of the clumps would be equal in the three simulations if there was no overlapping. 
Non-zero polarization levels across the spectral line are found for all four scenarios. In the case of an axisymmetric distribution of the clumps, there are no loops in the $q$--$u$ plane, while for non-axisymmetric distributions, loops are produced. The loop areas are similar for the first two distributions, but smaller in the third case with small clumps. 
%Given the clump positions, the signs of $q$ and $u$ (as defined in Bulla+2015) appear reasonable to me.

Previous studies already suggested that the \ion{Si}{ii} line polarization is produced by large clumps, i.e. large-scale asymmetry. \citet{2010ApJ...725L.167M} report that the velocity evolution of the \ion{Si}{ii} $\lambda$6355\,\AA\ line, \.{v}$\rm _{\rm \ion{Si}{ii}}$, is linearly related to the polarization of \ion{Si}{ii}. They suggest that the velocity evolution is predominantly due to a general asymmetry and that the \ion{Si}{ii} polarization is unlikely to be produced by small randomly distributed clumps, but rather by a general asymmetry of the ejecta. Furthermore, \citet{2007Sci...315..212W} also suggest that the $\Delta$m$_{15}$ -- p$_{\rm \ion{Si}{ii}}$ relation is an indication of a large-scale asymmetry, for instance in form of large clumps above the photosphere in the outermost layers.

However, in this section we study the evolution of the \ion{Si}{ii} polarization in the $q$--$u$ plane and demonstrate the importance of high-cadence spectropolarimetric observations, which enables us to study the evolution of the ejected material and reveals a non-monotonic and complex behavior with time. In SN\,2005df, SN\,2006X, and SN\,2002bo, which have been observed at many epochs, we can observe the evolution of the polarization of the \ion{Si}{ii} line, from high polarization and no loops, over large loops, to small loops (see all plots in the online supplementary material, Appendix~B). By comparing to the simple simulations in Fig.~\ref{fig:loops_simulaitons}, the evolution of loops can be explained by the change of the silicon ejecta geometry and clumps size, from large axisymmetric structures, to large clumps, and finally to small clumps, as time evolves and the photosphere moves inwards.

\begin{figure}
\center
\includegraphics[trim=2mm 0mm 44mm 0mm, height=3.9cm, clip=true]{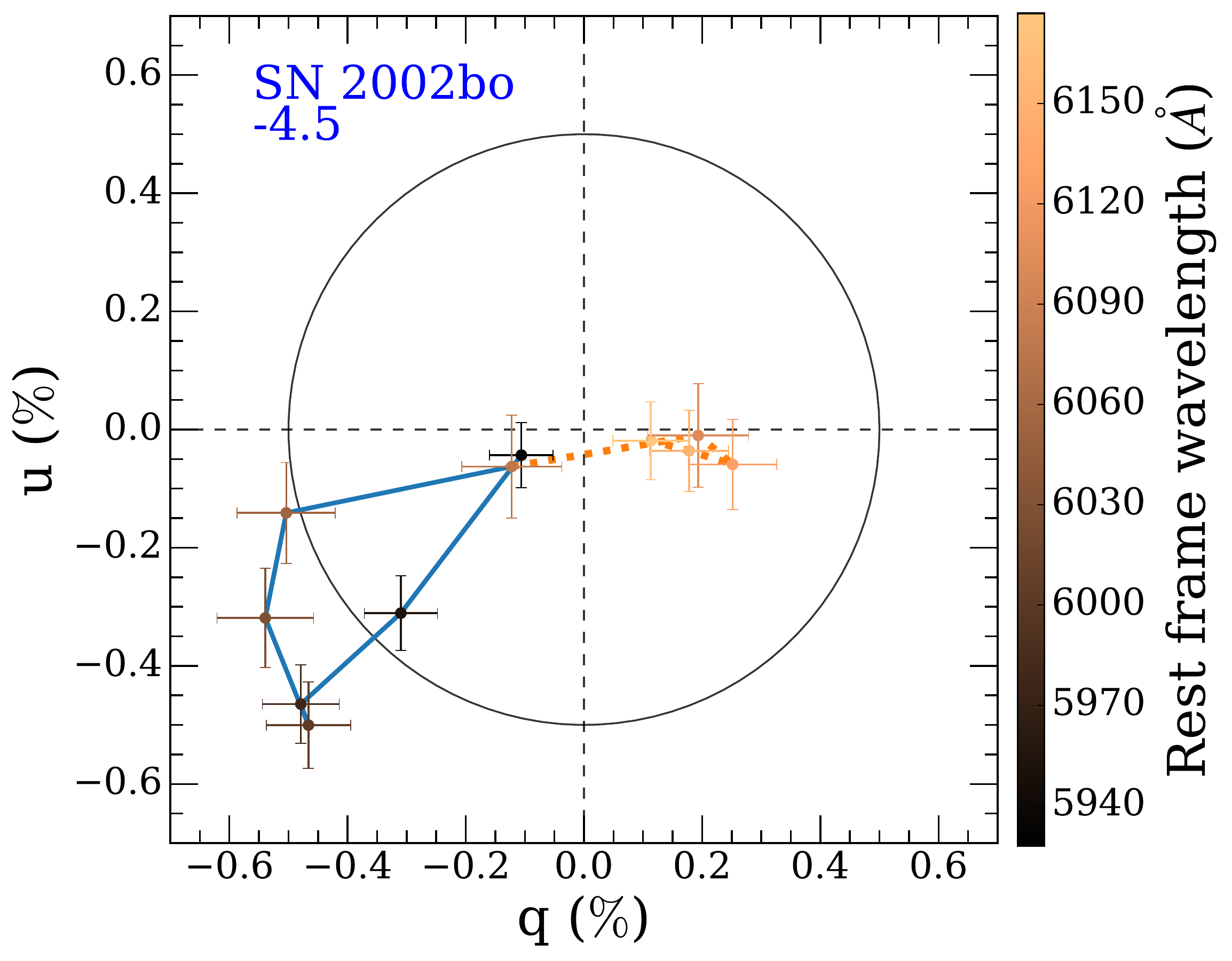}
\includegraphics[trim=32mm 0mm 3mm 0mm, height=3.9cm, clip=true]{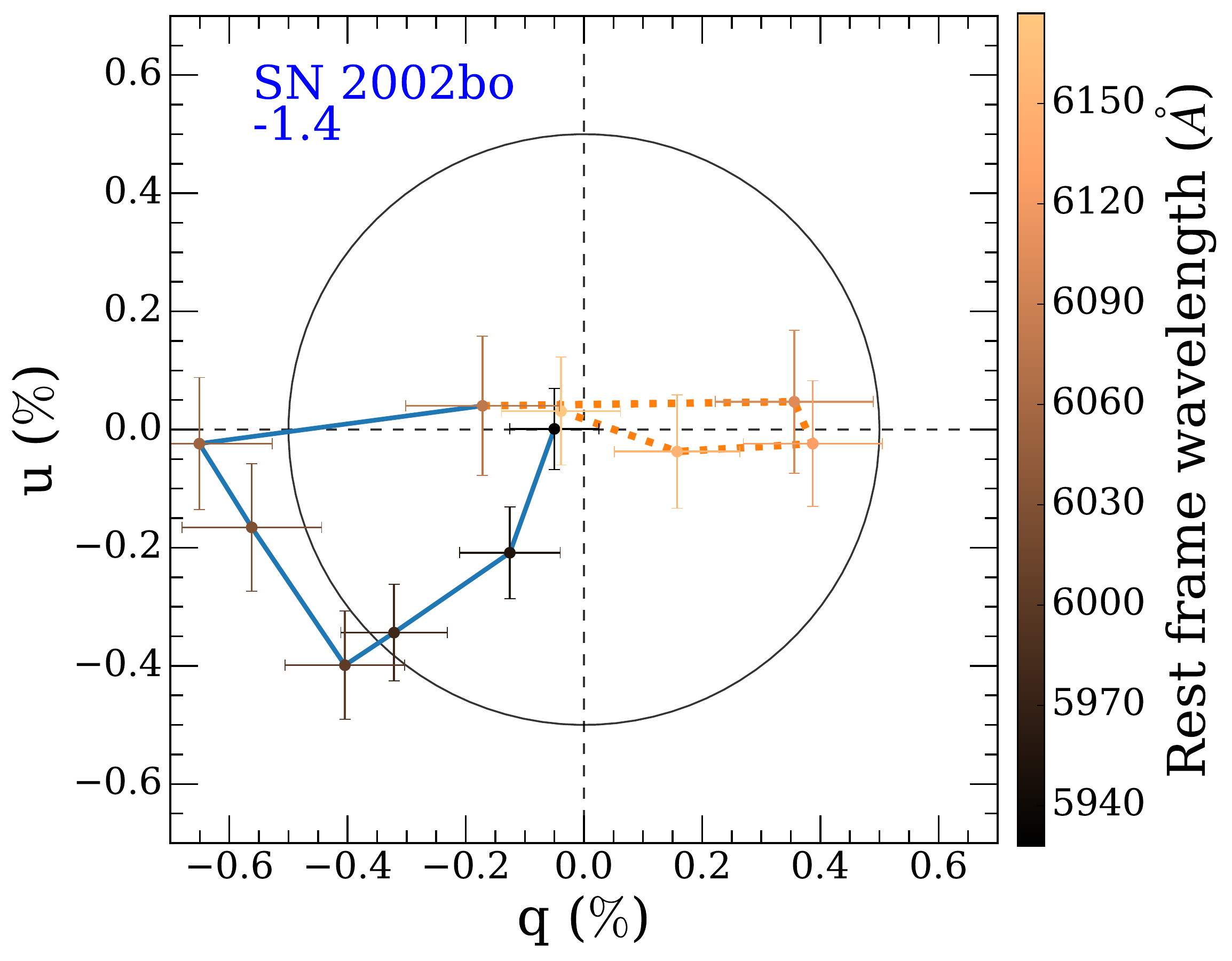} \\
\includegraphics[trim=2mm 0mm 44mm 0mm, height=3.9cm, clip=true]{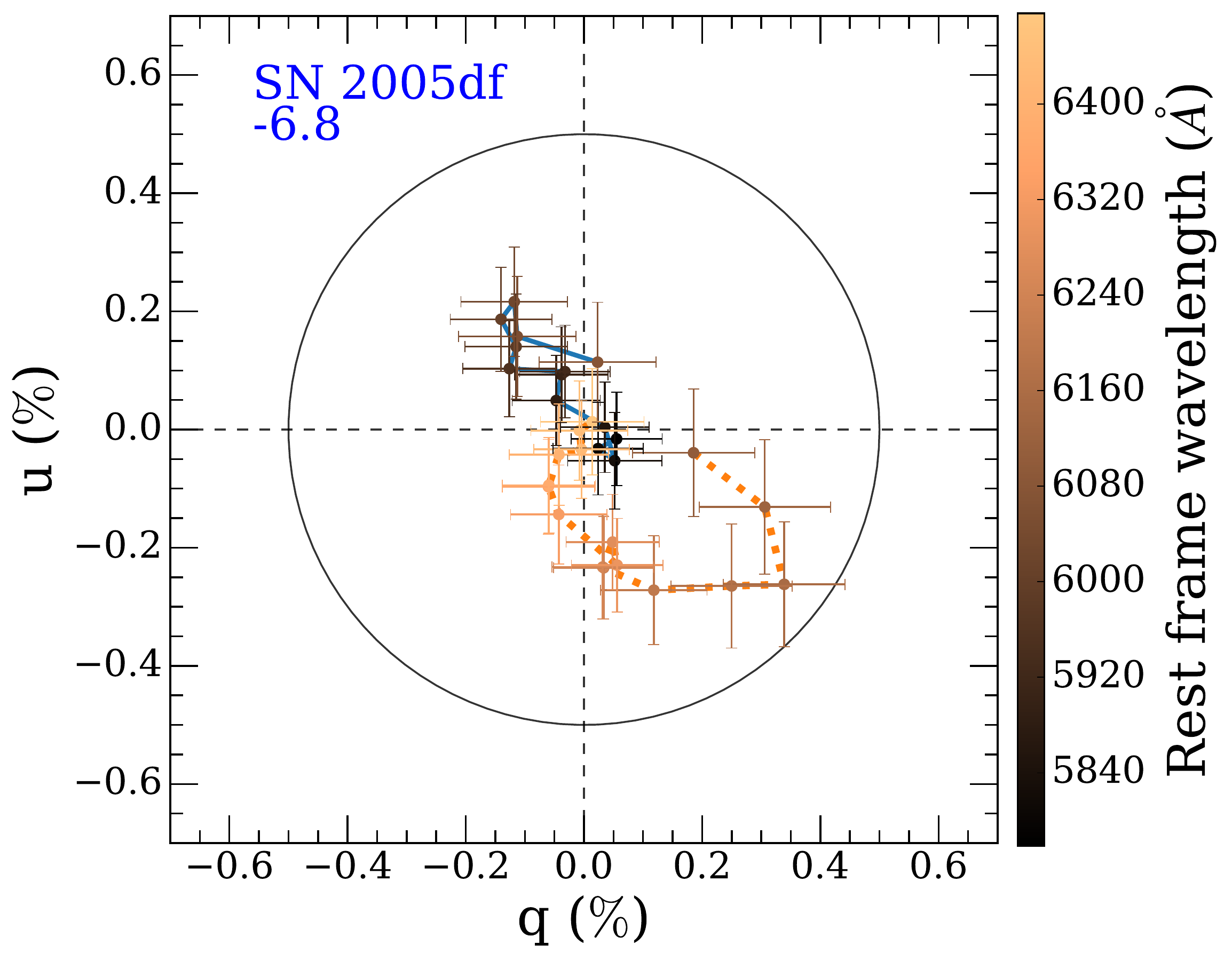}
\includegraphics[trim=32mm 0mm 3mm 0mm, height=3.9cm, clip=true]{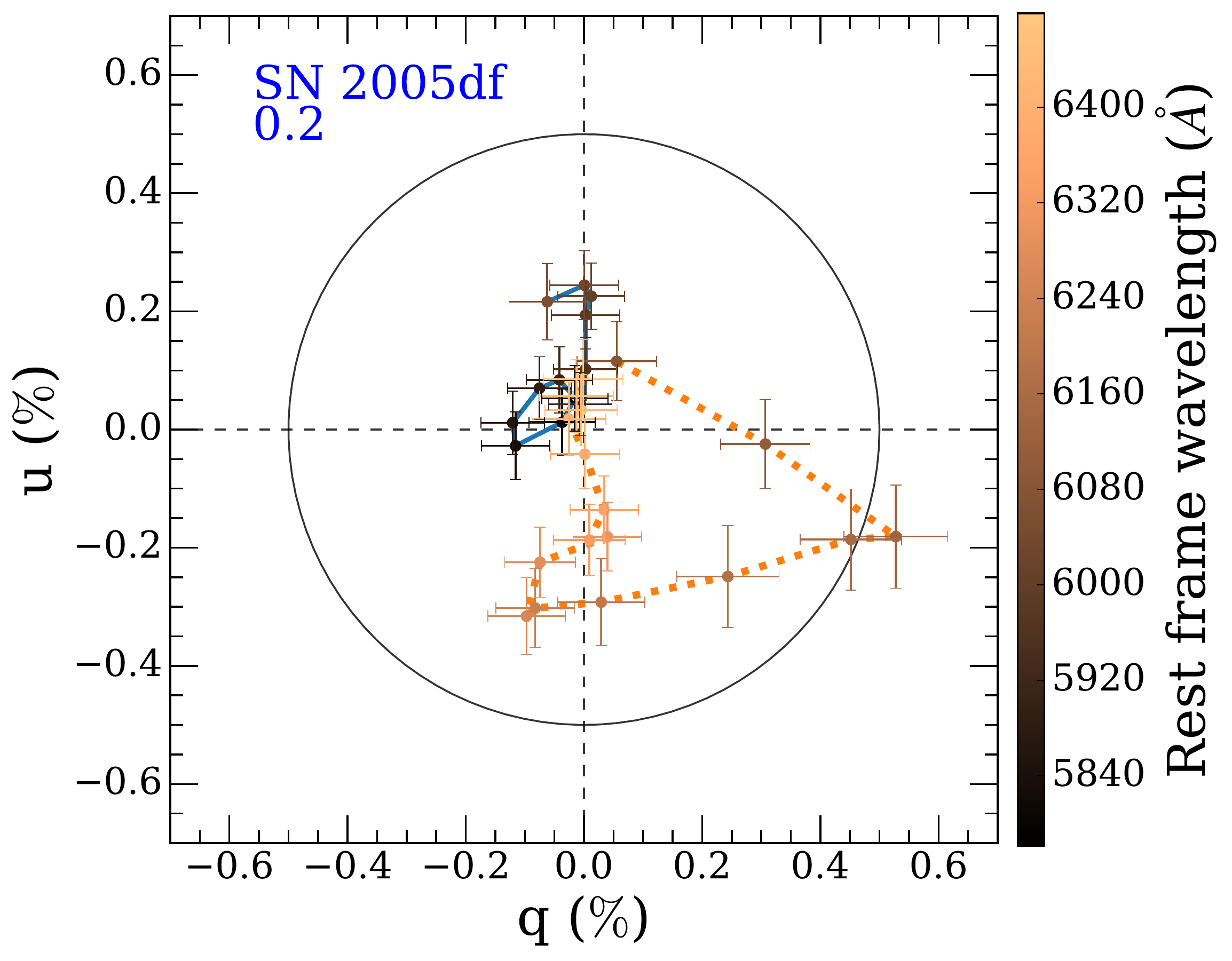}
\caption{Polarization of the \ion{Si}{ii} $\lambda$6355\,\AA\ line in the $q$--$u$ plane for SN\,2002bo (upper panels) and SN\,2005df (bottom panels) at two selected epochs. The color of the dots indicates the wavelength. The orange (dotted) and blue (solid) lines connect the dots that correspond to the photospheric and high velocity component, respectively. The circle represents a polarization degree of 0.5 per cent.}
\label{fig:loops_simulaitons_doublepeaks}
\end{figure}

We notice that in SN\,2002bo and SN\,2005df, which also display the \ion{Si}{ii} photospheric and high velocity components in the polarization spectra, the individual velocity components also display distinct $q$--$u$ loops at some epochs. 
In Fig.\ref{fig:loops_simulaitons_doublepeaks} the polarization of the photospheric and high velocity component is connected with the orange (dotted) and the blue (solid) line, respectively. Although the data at high resolution (with bin size of 25\,\AA ) have large uncertainties, and are not of sufficient quality to study the time-evolution of the distinct loops, these are visible in the case of SN\,2002bo, that has both, prominent photospheric and high velocity components, in contrast to SN\,2005df which displays prominent photospheric lines (and its corresponding loop), but has a relatively low polarized high velocity component (see Fig.~\ref{fig:PSIII-epoch}), and therefore displays a small loop at 0.2 days past peak brightness.

\subsection{Comparison to simulations}
\label{sect_results_simulations_comparison}

We compared our sample to simulations by \citet{2016MNRAS.455.1060B, 2016MNRAS.462.1039B}, who computed polarization spectra for different SN Ia explosion models. In particular, we examined the silicon line velocities and the \ion{Si}{ii} polarization in the polarization spectra of the delayed-detonation \citep[DDT, ][]{2013MNRAS.429.1156S}, double-detonation \citep{2010A&A...514A..53F} and violent merger \citep{2012ApJ...747L..10P} models, at $-$5, $-$2.5, 0, 2.5, 5, 7.5 days relative to B-band maximum. We determined the maximum \ion{Si}{ii} polarization and the wavelength of the absorption minimum to determine the photospheric expansion velocity.
These values are calculated for different viewing angles in each model. Specifically, we use only viewing angles for which high signal-to-noise calculations are available: 5 orientations for DDT and violent merger models and 3 orientations for double-detonation model. 
The choices of these angles are explained in \citet{2016MNRAS.455.1060B, 2016MNRAS.462.1039B}, and given in Table~A8 (online supplementary materials), along with the measurements from the simulations of the \ion{Si}{ii} velocities and polarization values.

The wavelength bin size in the simulated polarization spectra changes depending on the wavelength. Around the silicon line the size of one pixel is $\sim$10\,\AA , but the polarization spectra were Savitzky-Golay filtered using a first-order polynomial and a window of 3 pixels ($\sim$30\,\AA ). Therefore, we compare our data that were measured on polarization spectra of 25\,\AA\ bin size to the simulations.

\begin{figure}
\center
\includegraphics[trim=0mm 0mm 0mm 0mm, width=8.5cm, clip=true]{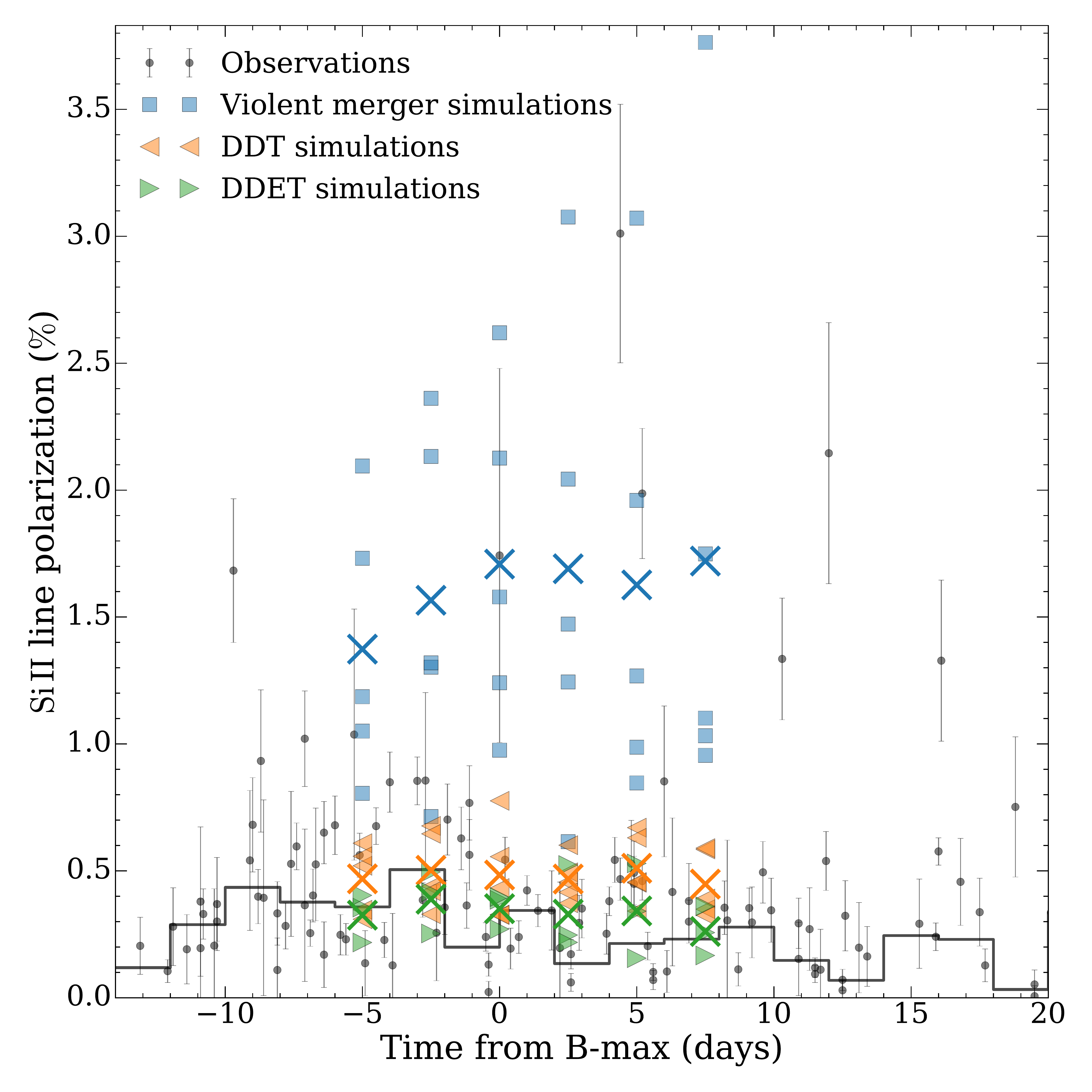}
\vspace{-4mm}
\caption{Observed peak polarization of the \ion{Si}{ii} $\lambda$6355\,\AA\ line for all epochs (black dots, measured on polarization spectra of 25\,\AA\ bin size), compared to predictions by \citet{2016MNRAS.455.1060B, 2016MNRAS.462.1039B} for the delayed-detonation (DDT, orange triangles pointing left), violent merger (blue squares), and double-detonation (DDET, green triangles pointing right) models. Simulated values for multiple orientations at $-$5, $-$2.5, 0, 2.5, 5, 7.5 days relative to B-max are shown. The crosses mark the values averaged over all orientations, and the full black line is the weighted mean of our observations.}
\label{fig:epoch-Sii_bullacomparison}
\end{figure}

Figure~\ref{fig:epoch-Sii_bullacomparison} compares our observations to the delayed-detonation (DDT), violent merger, and double-detonation (DDET) models. The simulations show that the polarization degree of the \ion{Si}{ii} line in the violent merger observations is higher compared to the delayed-detonation and double-detonation models. Furthermore, the delayed-detonation and double-detonation models have a comparable degree of polarization. Our observations of the peak polarization of the \ion{Si}{ii} line are consistent with predictions for the delayed-detonation and double-detonation models, while only SN\,2004dt has a comparable polarization to the predictions for the violent-merger model.

The ejecta in the DDT model is globally distributed spherically symmetrical and therefore the degree of polarization is similar regardless of the viewing direction. On the contrary, in the violent merger model, the higher polarization values typically correspond to viewing angles closer to the poles (away from the equatorial plane, see Table~A8). The DDET model is symmetric about the z-axis and therefore, three orientations were sufficient to properly sample this model. For the DDET model, the viewing angles closer to the equatorial plane typically display slightly higher polarization values, although the values are similar regardless of the viewing angle (Table~A8).

\begin{figure}
\center
\includegraphics[trim=0mm 0mm 0mm 0mm, width=8.5cm, clip=true]{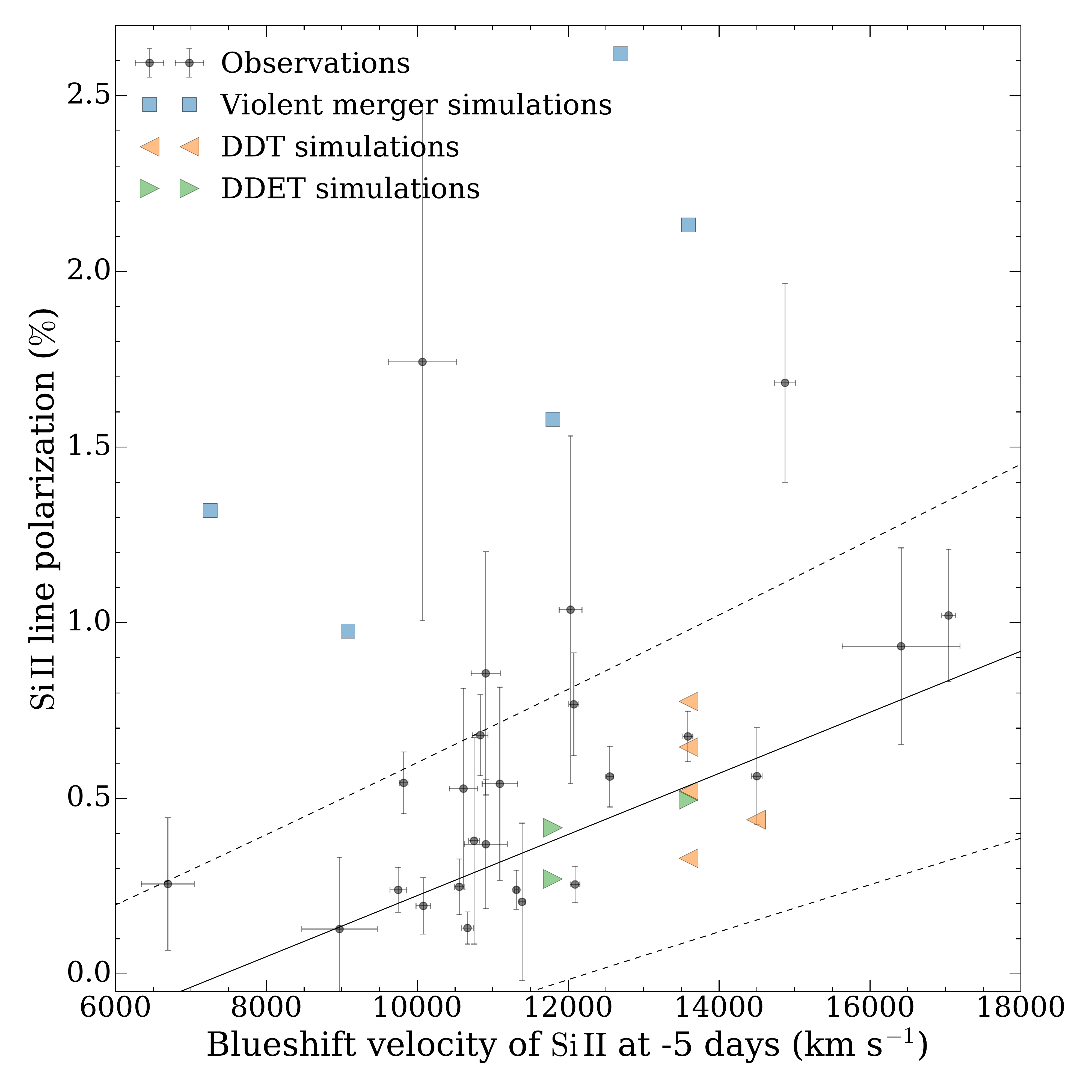}
\vspace{-6mm}
\caption{The observed maximum polarization of the \ion{Si}{ii} $\lambda$6355\,\AA\ line between $-$11 and 1 days relative to peak brightness (black dots, measured on polarization spectra of 25\,\AA\ bin size), versus the \ion{Si}{ii} velocity at 5 days before the peak brightness, compared to simulations. Red triangles pointing left mark the delayed-detonation (DDT), blue squares the violent merger and green triangles pointing right the double-detonation (DDET) model. The lines show the observed velocity-polarization relationship (solid line) and the 1$\sigma$ uncertainty (dashed line).}
\label{fig:vel_P_bullacomparison}
\end{figure}

Figure~\ref{fig:vel_P_bullacomparison} shows the comparison between simulations and observations of the observed maximum polarization of the \ion{Si}{ii} line between $-$11 and 1 days relative to peak brightness, against the \ion{Si}{ii} velocity at 5 days before the peak brightness. The DDT and DDET simulations results coincide with the observations, and the degree of \ion{Si}{ii} polarization in the violent merger simulation is offset from the DDT and DDET simulations. Although the observations display a velocity-polarization trend, the velocity range of the simulations is not sufficient to reliably confirm a relationship. Note that the polarization of the \ion{Si}{ii} line in this figure was measured on polarization spectra of 25\,\AA\ bin size, compared to Fig.~\ref{fig:SiII-vel}, which shows the velocity--polarization relationship using polarization measurements on spectra with bin size of 100\,\AA .

\subsection{Intrinsic continuum polarization}
\label{sect_results_continuumpolarization}

Although the analysis of this paper is based on continuum-subtracted spectra (which implicitly removes the problem of ISP correction for the continuum polarization), our sample includes a number of objects affected by low or negligible reddening. This allows us to derive an upper limit on the intrinsic continuum polarization. 

For this purpose, we determined the average continuum polarization from total polarization spectra with a bin size of 100\,\AA . The average continuum polarization was calculated in a region between 6400-7200\,\AA , which contains no strong spectral features \citep[see Sect. 5.4 in][]{2009A&A...508..229P}. The results are given in Table~A6 (online supplementary materials).
%table

Then we estimated the reddening, $E(B-V)$, by fitting the SN light curves from the Open Supernova catalog with SNooPy (see Sect.~\ref{sect_lightcurvefitting}), or took the $E(B-V)$ values from the literature. The $E(B-V)$ values with references are given in Table~A7 (online supplementary materials).

\begin{figure*}
\includegraphics[trim=0mm 0mm 0mm 0mm, width=8.5cm, clip=true]{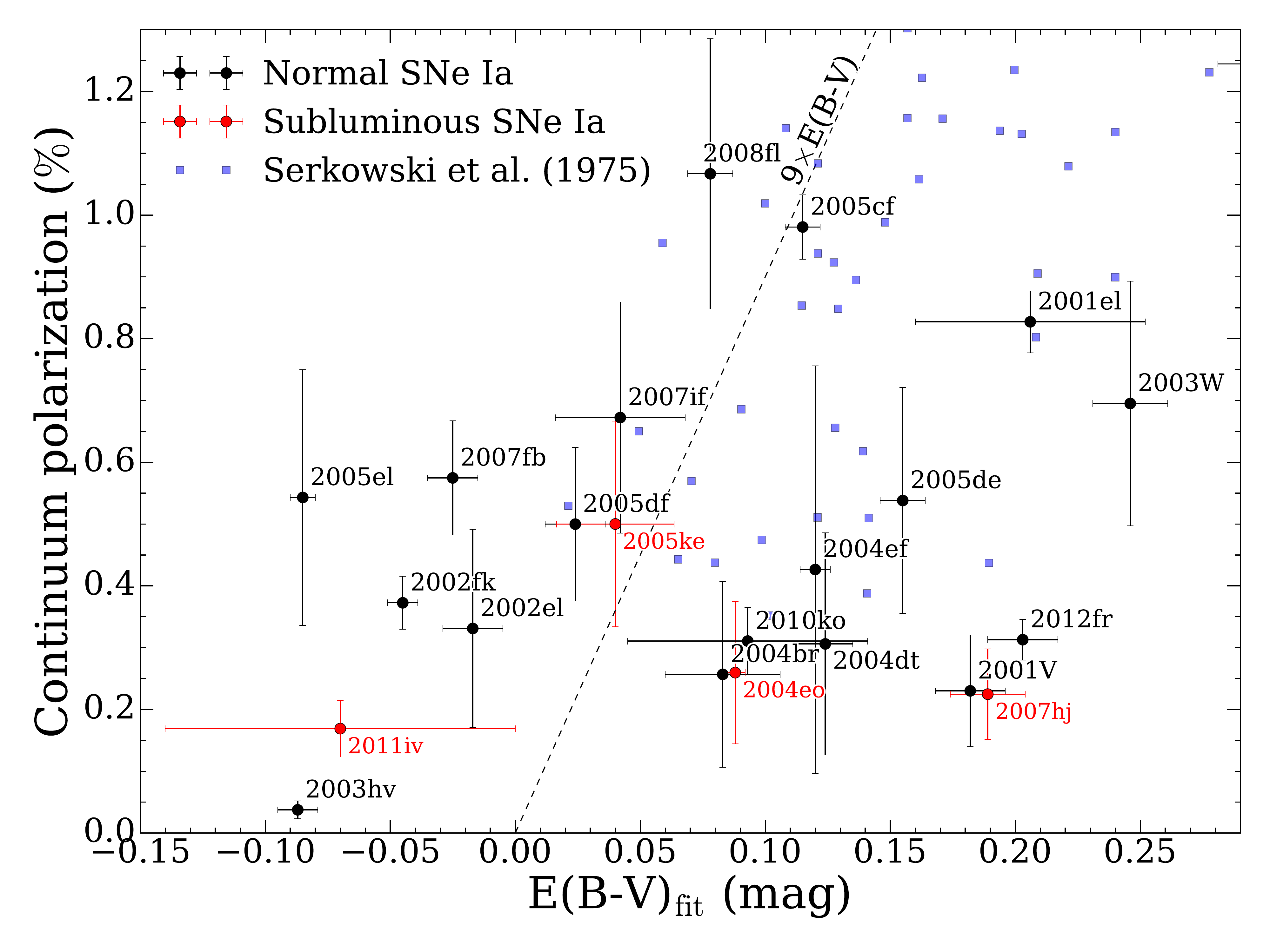}
\includegraphics[trim=0mm 0mm 0mm 0mm, width=8.5cm, clip=true]{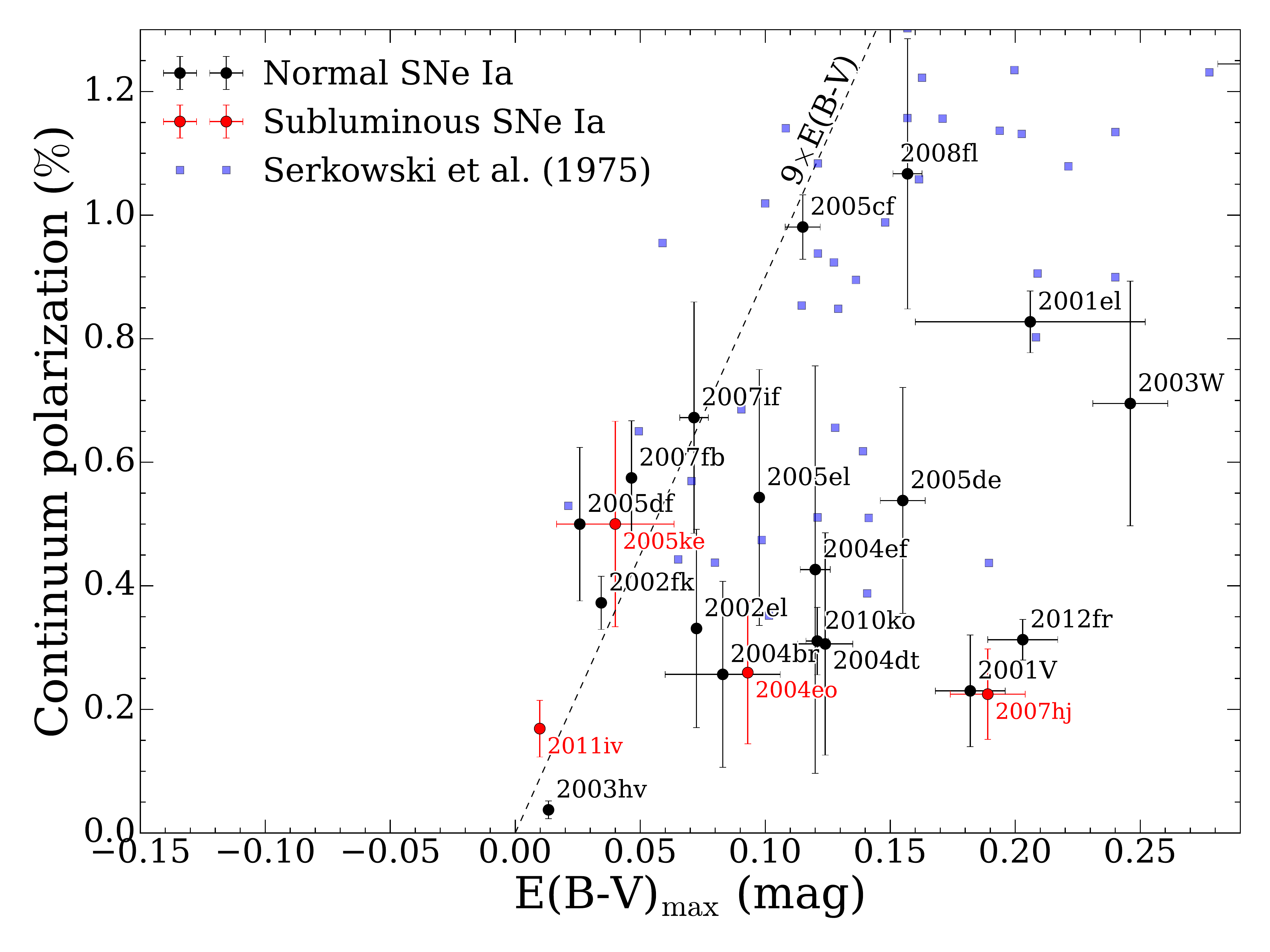}
\vspace{-3mm}
\caption{Continuum polarization as function of $E(B-V)$. The black and red dots depict normal SNe and subluminous/transitional SNe from our VLT sample, respectively. Milky Way stars from \citet{1975ApJ...196..261S} are included for comparison (grey squares). The dashed line marks the nominal upper ISP limit, 9$\times E(B-V)$, determined from Milky Way stars \citep{1975ApJ...196..261S}.  \textit{Left panel:} The maximum continuum polarization for each SN, measured before +15 days relative to peak brightness, as a function of $E(B-V)_{\rm fit}$ determined by fitting the SN light curves. \textit{Right panel:} Same as the left panel, but instead of $E(B-V)_{\rm fit}$, we use the larger reddening value of $E(B-V)_{\rm MW}$ and $E(B-V)_{\rm fit}$. A few SNe Ia have a higher polarization compared to the nominal upper ISP limit, which may be an indication of intrinsic SN continuum polarization, or can be due to potential ISP contribution from the host galaxy (see discussion in Sect.~\ref{sect_results_continuumpolarization}).}
\label{fig:EBV-P}
\end{figure*}

In the left panel in Fig.~\ref{fig:EBV-P} we show the maximum continuum polarization for each SN, measured before +15 days relative to peak brightness, as a function of reddening  determined by fitting the SN light curves, $E(B-V)_{\rm fit}$. Although the best fit B-V color of some SNe is bluer ($\sim$ -0.1 mag) compared to normal SN Ia templates, possibly because of errors in the photometric data or intrinsic differences, there is always some amount of Galactic dust present, which can produce a polarization of up to 9$\times E(B-V)$ per cent \citep[hereafter, upper ISP limit,][]{1975ApJ...196..261S}, and should not be neglected. 
Therefore, we extracted the Galactic dust reddening values for all SN locations from \citet{2011ApJ...737..103S} and compared the $E(B-V)_{\rm MW}$ values to the reddening determined from the light curves of SNe, $E(B-V)_{\rm fit}$. In the right panel in Fig.~\ref{fig:EBV-P} we present the continuum polarization as a function of the larger reddening value, $E(B-V)_{\rm max}$ = max[$E(B-V)_{\rm MW}$, $E(B-V)_{\rm fit}$].
This moves the blue SNe closer to the upper polarization limit, where the continuum polarization may be explained by ISM. A few SNe have a higher polarization compared to the ISP limit, which may be an indication of intrinsic SN continuum polarization, however, the confidence interval of the ISP upper limit is not given in the literature, and there are also some stars in the sample of \citet{1975ApJ...196..261S} that lie above the ISP limit. 
Another possible explanation why some SNe Ia lie above the ISP limit is a contribution of host galaxy dust to the observed polarization, and perhaps with a different polarization law \citep[e.g.][]{2002AJ....124.2506L,2016ApJ...828...24P}. Although the SNe Ia appear un-reddened, the contribution of host galaxy dust can not be excluded, as in the case of Galactic dust, which in contrast to the host galaxy dust, was independently measured by \citet{2011ApJ...737..103S}. On the contrary, note that polarization by dust in the host galaxy could also partially cancel out polarization by Galactic dust, depending on the relative position angles.

Finally, to constrain the upper limit on the intrinsic continuum polarization of SNe Ia, we select a subsample of six SNe with $E(B-V)_{\rm max}$<0.05 mag: SN\,2002fk, SN\,2003hv, SN\,2005df, SN\,2005ke, SN\,2007fb and SN\,20011iv. %Five of these SNe are exceeding the ISP upper limit. 
The mean continuum polarization of these six SNe with low reddening is 0.32 $\pm$ 0.15 per cent. The upper limit is 0.61 per cent, with a confidence interval of 95 per cent. These results are consistent with the literature \citep[][see also Sect.~\ref{sect_continuumpolmechanisms}\,iii]{2001ApJ...556..302H,2008AJ....136.2227C}. Assuming that there is no contribution from the ISP, and that the intrinsic continuum polarization is produced by to electron scattering in an aspherical photosphere, the upper limit of intrinsic polarization of 0.61 per cent corresponds to an oblate ellipsoid with an axis ratio of E=a/b$\approx$0.88 seen at an inclination angle of 90$^\circ$ \citep[see Fig. 4 in][]{1991A&A...246..481H}.

%% A few SNe have a higher polarization compared to the nominal upper ISP limit, which may be an indication of intrinsic SN continuum polarization, or can be due to potential ISP contribution from the host galaxy (see discussion in Sect.~\ref{sect_results_continuumpolarization}).
%% some
%SNe Ia lie above the ISP limit is a contribution of host galaxy dust to the observed polarization, and perhaps with a different polarization law
%(see Sect.~\ref{sect_results_continuumpolarization})
%discussion in the text (Section 5.6) of the potential ISP contribution from the host galaxy.
%\ref{sect_results_continuumpolarization}

\section{Summary and conclusions}
\label{sect_summary}

We reduced and examined, in a systematic way, archival linear spectropolarimetric data of a sample of 35 SNe Ia observed with the VLT's FORS1+2 between 2001 and 2015 at 127 epochs in total. To the best of our knowledge this is the largest polarimetric dataset of SNe Ia ever examined. Our main results can be summarized as follows:

\begin{enumerate}
\item The peak polarization of the \ion{Si}{ii} $\lambda$6355\,\AA\ line displays an evolution in time with a range of peak polarization degree from $\sim$ 0.1 to 1.7 per cent at epochs from $-$10 to 0 days relative to peak brightness (Fig.~\ref{fig:PSIII-epoch}). This differs from previous studies based on smaller samples \citep{2007Sci...315..212W}. Also the polarization peaks of the distinct \ion{Si}{ii} $\lambda$6355\,\AA\ velocity components behave differently (Fig.~\ref{fig:PSiII_hv_phot-epoch}).

\item We populated the $\Delta$m$_{15}$--p$_{\rm \ion{Si}{ii}}$ plane, and re-analyzed the 
$\Delta$m$_{15}$--p$_{\rm \ion{Si}{ii}}$ relationship proposed by \citet{2007Sci...315..212W}. Although the overall behavior is reproduced, our larger sample reveals a significantly larger scatter. We show that subluminous and transitional objects display lower \ion{Si}{ii} line polarization values and are located below the main $\Delta$m$_{15}$--p$_{\rm \ion{Si}{ii}}$ relationship (Fig.~\ref{fig:SiII-dm15}). This confirms previous results based on smaller samples \citep{2001ApJ...556..302H,2012A&A...545A...7P}.

\item We found a statistically significant linear relationship ($\rho \gtrsim 0.8$) between the degree of linear polarization of the \ion{Si}{ii} $\lambda$6355\,\AA\ line before maximum with the \ion{Si}{ii} $\lambda$6355\,\AA\ line velocity (Fig.~\ref{fig:SiII-vel}). We suggest that this relationship, along with the $\Delta$m$_{15}$--p$_{\rm \ion{Si}{ii}}$ relationship \citep{2007Sci...315..212W}, may be explained with the off-center delayed-detonation model \citep{2006NewAR..50..470H}.

\item However, a comparison of our sample in the $\Delta$m$_{15}$ -- $v_{\rm \ion{Si}{ii}}$ plane to numerical predictions for double-detonation explosions \citep{2018arXiv181107127P} revealed that there are two distinct subgroups. The SNe that fall uniformly above the sub-Chandrasekhar double-detonation explosion model predictions line display higher polarization degrees of the \ion{Si}{ii} line compared to a distinct ``cluster" of normal Chandrasekhar-like SNe (see Figs.~\ref{fig:dm15-velSiII-pol} and \ref{fig:histo_SiII-pol}). This dichotomy in the polarization may imply that there are indeed two distinct populations of SNe Ia, as suggested in \citet{2018arXiv181107127P}, with different explosion mechanisms. However, this tentative conclusion requires further theoretical investigations. 

\item We examined the evolution of the \ion{Si}{ii} line in the $q$--$u$ plane for a subsample of SNe, observed at multiple epochs, and ran simple simulations to explore the effect of clumps on the polarization spectra. In the cases of SN\,2005df, SN\,2006X, and SN\,2002bo, we observe the evolution of the formation of loops, their growth and subsequent shrinking (see all plots in the online supplementary material, Appendix~B), which may be explained by the evolution of the projected silicon ejecta size, from large axisymmetric structures to large clumps, and finally to small clumps as time evolves and the photosphere moves inwards. 

\item We compared our sample of the \ion{Si}{ii} polarization measurements to predictions calculated for the double-detonation, delayed detonation and violent merger models by \citet{2016MNRAS.455.1060B, 2016MNRAS.462.1039B}. Our observations of the peak polarization of the \ion{Si}{ii} line are consistent with predictions for the delayed-detonation and double-detonation models, which have a comparable degree of polarization. Only SN\,2004dt has a degree of polarization that is comparable to predictions for the violent-merger model.

\item We attempted to reproduce the \ion{Si}{ii} -- velocity relationship using simulations by \citet{2016MNRAS.462.1039B,2016MNRAS.455.1060B}. Although the calculations show a global velocity-polarization trend, the velocity range of the models available is not sufficient to stringently test the observed relationship.

\item Additionally, we determined an upper limit of the intrinsic continuum polarization. For a subsample of six objects with $E(B-V)$<0.05 mag, we estimate an upper limit of 0.61 per cent, with a confidence of 95 per cent. This result is consistent with the literature. Assuming that the polarization is produced due to electron scattering in an aspherical photosphere, and that there is no contribution from ISP, this degree of polarization corresponds to an oblate ellipsoid with an axis ratio of a/b $\gtrsim$ 0.88 seen at \textit{i}=90$^\circ$ \citep{1991A&A...246..481H}.

\end{enumerate}

This work shows that even typical SN\,Ia display a distribution of polarization properties that illustrates a richness in the geometry of the explosion that is a challenge to models and that cannot be revealed by total flux spectra. The evolution of the Si polarization amplitude and the size of loops in the $q$--$u$ plane suggest that the Si distribution gets more complex, ``clumpier," with depth. If this structure is the result of plumes resulting from the burning dynamics, the plumes have a fine structure at greater depth but merge to a few large features in outer layers. Again, the implied Si plume structure is a new constraint on multi-dimensional models that is not revealed in total flux spectra.
Furthermore, this work shows the power of increasing the sample size of SN with high-cadence spectropolarimetry. Only frequent observations enable the study of the evolution of the polarization of individual lines. Here, we have concentrated on the \ion{Si}{ii} line, but a similar
analysis should be done on other prominent lines. If the next generation of large ground-based telescopes is designed and build to support precision polarimetry, they will allow the study of the polarimetric evolution of different species, different velocity components (high velocity and photospheric), and $q$--$u$ loops with high accuracy. This will allow us to reconstruct the 3D structure of the explosion and to study the polarimetric tomography in great detail. It is critical that the new generation of large telescopes have the capacity to do spectropolarimetry.

\section*{Acknowledgements}

We would like to thank to Jason Spyromilio, Thiem Hoang, Ivo Seitenzahl, Ashley Ruiter, Wolfgang Kerzendorf, Stefan Taubenberger and Barna Barnab\'{a}s for helpful discussions. 
%HD~141318 was observed with ESO Telescopes at the Paranal Observatory under Programme ID 094.C-0686. 
A. Clocchiatti is supported by grant IC120009 (MAS) funded by the Chilean Ministry of Economy, Development and Tourism, and by grant Basal CATA PFB 06/09 from CONICYT. LW is supported by NSF 1817099. JCW is supported by NSF Grant 1813825. The research of Y. Yang is supported through a Benoziyo Prize Post-doctoral Fellowship.
This work is based on observations collected at the European Organisation for Astronomical Research in the Southern Hemisphere under ESO programmes 67.D-0517(A), 66.D-0328(A), 68.D-0571(A), 69.D-0438(A), 70.D-0111(A), 71.D-0141(A), 073.D-0771(A), 073.D-0565(A), 075.D-0628(A), 075.D-0213(A), 076.D-0178(A), 076.D-0177(A), 079.D-0090(A), 080.D-0108(A), 081.D-0557(A), 081.D-0558(A), 085.D-0731(A), 086.D-0262(A), 086.D-0262(B), 088.D-0502(A), 095.D-0848(A), 095.D-0848(B), 290.D$-$5009(A), 290.D$-$5009(B), 290.D$-$5009(C) and 290.D$-$5009(D); the execution in service mode of these observations by the VLT operations staff is gratefully acknowledged.

This work benefited from The Open Supernova Catalog \citep[http://sne.space,][]{2017ApJ...835...64G}, SNooPy \citep{2011AJ....141...19B}, SNID \citep{2007ApJ...666.1024B}, L.A.Cosmic \citep{2001PASP..113.1420V}, IRAF \citep{1986SPIE..627..733T}, PyRAF and PyFITS. PyRAF and PyFITS are products of the Space Telescope Science Institute, which is operated by AURA for NASA. We thank the authors for making their tools and services publicly available.

%%%%%%%%%%%%%%%%%%%%%%%%%%%%%%%%%%%%%%%%%%%%%%%%%%



\bibliographystyle{mnras}
\bibliography{references.bib} % if your bibtex file is called example.bib

\section*{Supporting material}

Additional supporting material may be found in the online
version of this article:\\
{\bf Table A1.} Observing log\\
{\bf Table A2.} ﻿SN Ia light curves from the Open Supernova Catalog\\
{\bf Table A3.} ﻿SN Ia spectra from the Open Supernova Catalog\\
{\bf Table A4.} \ion{Si}{ii} $\lambda$6355\,\AA\ blueshift velocity at $-$5 days relative to peak brightness.\\
{\bf Table A5.﻿} \ion{Si}{ii} line polarization measured from polarization spectra with different bin sizes\\
{\bf Table A6.} ﻿Average continuum polarization measured from 6400--7200\,\AA .\\
{\bf Table A7.} Reddening determined from SN light curves and Galactic reddening taken from \citet{2011ApJ...737..103S}.\\
{\bf Table A8.} Measurements of the \ion{Si}{ii} $\lambda$6355\,\AA\  line velocities and polarization values from  simulations of the delayed-detonation (DDT), double-detonation (DDET) and violent merger models \citep{2016MNRAS.455.1060B, 2016MNRAS.462.1039B}.\\
{\bf Figures B1-B35.} For all SNe: Polarization of the \ion{Si}{ii} $\lambda$6355\,\AA\ line at different epochs relative to peak brightness, and \ion{Si}{ii} polarization in the $q$--$u$ plane.\\
{\bf Figures C1-C127.} Individual polarization spectra for all epochs.\\

% Don't change these lines
\bsp	% typesetting comment
\label{lastpage}
\end{document}